\documentclass[12pt]{article}
\usepackage{amsfonts,amsmath,amssymb}
\usepackage{xcolor}
\usepackage{todonotes}
\usepackage{graphicx}
\usepackage{rotating}
\usepackage{multicol}
\usepackage{float}
\usepackage{multirow}
\usepackage{enumerate}
\usepackage{wasysym}
\usepackage{bbm}
\usepackage{subfigure}
\usepackage[normalem]{ulem}
\usepackage[flushleft]{threeparttable} 
\usepackage[authoryear,round,longnamesfirst]{natbib}
\usepackage[colorlinks, citecolor=blue, linkcolor=blue]{hyperref}
\usepackage{xr}
\usepackage{longtable}

\allowdisplaybreaks
\usepackage[width=17cm, height=22cm]{geometry}

\numberwithin{equation}{section}
\usepackage{ntheorem}
\theoremseparator{:}

\newtheorem{thm}{Theorem}[section]

\newtheorem{lm}{Lemma}[section]

\newtheorem{pp}{Proposition}

\newtheorem{as}{Assumption}[section]

\newtheorem{rem}{Remark}
\allowdisplaybreaks

\begin{document}
	\author{Nan Liu\footnote{\setlength{\baselineskip}{3.8mm}Nan Liu, Wang Yanan Institute for Studies in Economics (WISE), Department of Statistics \& Data Science, School of Economics, Xiamen University.} \and Yanbo Liu\footnote{Yanbo Liu, School of Economics, Shandong University.} \and Yuya Sasaki\footnote{Yuya Sasaki, Department of Economics, Vanderbilt University.} \and Yuanyuan Wan\footnote{Yuanyuan Wan, Department of Economics, University of Toronto.}}
	\title{{\Large \textbf{Nonparametric Uniform Inference in \\Binary Classification and Policy Values}\thanks{\setlength{\baselineskip}{3.8mm} We are grateful to Toru Kitagawa and Whitney Newey for very intensive discussions, and to numerous participants in the seminars at Aarhus University, Kyoto University, Syracuse University, UCLA, UCSC, and University of Virginia; 2025 JER/SNSF Workshop on Recent Advances in Econometrics, and SETA 2025 for their valuable comments. All remaining errors are our own. Yuya Sasaki gratefully acknowledges the generous financial support of Brian and Charlotte Grove. Yuanyuan Wan gratefully acknowledges the support from the SSHRC Insight Grant \#43520240418.}}}

\date{}
	\maketitle

\begin{abstract}\setlength{\baselineskip}{5.9mm} 
We develop methods for nonparametric uniform inference in cost-sensitive binary classification, a framework that encompasses maximum score estimation, predicting utility maximizing actions, and policy learning. These problems are well known for slow convergence rates and non-standard limiting behavior, even under point identified parametric frameworks. In nonparametric settings, they may further suffer from failures of identification. 
To address these challenges, we introduce a strictly convex surrogate loss that point-identifies a representative nonparametric policy function. 
We then estimate this representative policy function to conduct inference on both the optimal classification policy and the optimal policy value. This approach enables Gaussian inference, substantially simplifying empirical implementation relative to working directly with the original classification problem. 
In particular, we establish root-$n$ asymptotic normality for the optimal policy value and derive a Gaussian approximation for the optimal classification policy at the standard nonparametric rate. Extensive simulation studies corroborate the theoretical findings. We apply our method to the National JTPA Study to conduct inference on the optimal treatment assignment policy and its associated welfare.

\vskip0.4cm
		
		
\vskip1cm
		
\baselineskip=15pt
\end{abstract}

\newpage
\section{Introduction}

Cost-sensitive binary classification problems encompass important classes of econometric models: e.g., the maximum score estimator \citep{manski1975maximum, manski1985semiparametric}; utility maximization with a binary action \citep{elliott2013predicting}; and a recent strand of research on treatment choice problems \citep{manski2000identification, manski2004statistical, stoye2009minimax, stoye2012minimax, hirano2009asymptotics, bhattacharya2012inferring, tetenov2012statistical, kitagawa2018should, athey2021policy, mbakop2021model, manski2021econometrics, viviano2025policy}. See \citet{hirano2020asymptotic} for a review of statistical decision analysis in econometrics.

A large majority of the existing literature focuses on analyzing learning performance, such as consistency, convergence rates, limit distribution, regret bounds, rather than on statistical inference. Inference, however, is well known to be a challenging problem. For $M$-estimation with a discontinuous objective, \citet{kim1990cube} show that the estimated policy parameters exhibit a nonstandard (non-Gaussian) limit distribution with a cube-root-$n$ rate of convergence. Moreover, a naive bootstrap is not valid in this setting \citep[see][]{abrevaya2005bootstrap, leger2006bootstrap}. Consequently, statistical inference for cost-sensitive binary classification problems remains challenging in practice, even when one considers a parametrized policy class.\footnote{With this said, it is worth emphasizing that there have been some developments toward practical inference in the parametric case, such as the smoothing approach \citep{horowitz1992smoothed}, the subsampling approach \citep{delgado2001subsampling, seo2018local}, the $m$-out-of-$n$ bootstrap approach \citep{lee2006m, bickel2011resampling}, and a modified bootstrap approach \citep{cattaneo2020bootstrap, cheng2024inference}.}

If a researcher further wishes to allow for a nonparametric policy class to accommodate more flexible classifiation policies, the problem becomes even more severe. Parametric problems are still point-identified after a suitable normalization \citep[e.g.,][]{manski1975maximum, manski1985semiparametric}. By contrast, there is no obvious normalization that guarantees point identification when the policy function is nonparametric. Statistical inference under partially identified parameters poses significant challenges, and the development of valid and practical inference methods in this setting remains an open question.

This paper proposes a novel method of Gaussian inference for cost-sensitive binary classification problems with a nonparametric class of classification policy functions. The framework encompasses the aforementioned maximum score estimation, utility maximization under binary action, and treatment choice problems, among others. While this may seem daunting, if not infeasible, to those familiar with the literature, we attempt to develop a formal framework that establishes the validity of the proposed Gaussian inference procedure and demonstrate its practical ease.

Our proposed method of inference is easy to implement and hence practical, despite the inherently difficult nature of the problem. Unlike the method of \citet{horowitz1992smoothed}, we do not require the choice of a tuning parameter to smooth the indicator. Unlike the method of \citet{delgado2001subsampling}, we do not require the choice of a subsampling size. Furthermore, our inference method is based on the Gaussian distributions, which also facilitates an easy-to-implement bootstrap procedure.

Given the difficulty of the problem with partially identified nonparametric optimal policy functions, readers may wonder how we could enable Gaussian inference. The main novelty of this paper lies in the ``identification'' of a representative optimal classification policy function from among an infinite set of optimal classification policy functions. This is accomplished through the use of a strictly convex surrogate loss in place of the 0--1 loss. The use of surrogate losses has been explored in the statistical learning literature \citep[e.g.,][]{lugosi2004bayes, zhang2004statistical, steinwart2005consistency, bartlett2006convexity, zhao2012estimating}, and \citet{kitagawa2023constrained} adopt and extend this approach in the welfare maximization context. The motivations of these existing papers differ from ours: the existing work primarily aims to facilitate computation by convexifying the objective function while guaranteeing risk equivalence, whereas our objective is to establish the aforementioned ``identification'' of a representative optimal policy. While some of the existing papers advocate the use of the hinge loss for their respective purposes, we explicitly rule out the hinge loss as a candidate surrogate loss for our purpose. In this sense, our use and objective of the surrogate loss is orthogonal to those of the seemingly related papers in the literature.

Once the representing nonparametric optimizer is ``identified,'' we can proceed with econometric analyses for point-identified functional parameters, even though the underlying model admits infinitely many parameters in the identified set. We estimate this nonparametric representative policy function using a sieve approach. Our procedure goes beyond standard sieve M-estimation by accommodating infinite-dimensional nuisance parameters, such as the propensity score and conditional mean functions arising in the augmented inverse propensity score weighting estimator. In this setting, we achieve the standard nonparametric rate of convergence. Moreover, we establish a strong Gaussian approximation for the representative policy by applying Yurinskii's coupling inequality \citep[Theorem 10]{pollard2002}, which in turn enables Gaussian inference for the optimal classification policy.

With our estimator for the optimal representative binary classification policy, we can estimate the optimal policy value by evaluating its (cross-fitted) empirical analog at our optimal policy estimator. We establish its root-$n$ consistency and limiting Gaussian distribution. This result is achieved through a combination of several techniques and notable properties of our problem to be outlined ahead, and it shares a similar spirit to \citet{chen2003estimation}, who derive root-$n$ semiparametric results in related settings with non-smooth objective functions. In our setting, we show that a margin condition \citep[see][]{kitagawa2018should,ponomarev2024lower} plays a crucial role in ensuring that the population policy value function is sufficiently smooth at the optimal policy, thereby laying the foundation for Gaussian inference for the optimal policy values. As we accommodate infinite-dimensional nuisance parameters in the estimation of a finite-dimensional parameter, our approach bears resemblance to double machine learning \citep[DML,][]{chernozhukov2018double}. However, the asymptotic analysis differs from DML because we additionally employ a nonparametrically estimated functional optimizer, i.e., the classification policy.

\citet{matzkin1992nonparametric} and \citet{mbakop2021model} introduce nonparametric classification policy functions for estimation and learning in the contexts of (A) maximum score estimation and (B) empirical welfare maximization, respectively. We contribute to this literature by further developing inference methods and establishing their theoretical properties. Table~\ref{tab:relation_to_literature} summarizes the gaps in the existing literature that are filled by the present paper.

\begin{table}[tb]
\centering
\scalebox{0.94}{
\begin{tabular}{|c|l|l|}
\multicolumn{3}{c}{(A) Maximum Score Estimation}\\
\hline
Function Class & Estimation & Inference\\
\hline
Parametric & \citet{manski1975maximum, manski1985semiparametric}, etc. \hspace{0.2cm} & \citet{kim1990cube} \& \citet{horowitz1992smoothed}, etc. \hspace{0.2cm} \\
\hline
Nonparametric & \citet{matzkin1992nonparametric} & --This Paper--\\
\hline
\end{tabular}
}
\scalebox{0.94}{
\begin{tabular}{|c|l|l|}
\multicolumn{3}{c}{(B) Empirical Welfare Maximization}\\
\hline
Function Class & Estimation & Inference\\
\hline
Parametric & \citet{kitagawa2018should} \& \citet{athey2021policy}, etc. & \citet{rai2018statistical}, etc. \\
\hline
Nonparametric & \citet{mbakop2021model} & --This Paper--\\
\hline
\end{tabular}
}
\caption{Summary of the relation to the econometric literature on cost-sensitive binary classification. Panel (A) summarizes the gaps this paper fills in the maximum score estimation literature, and Panel (B) summarizes the gaps it fills in the empirical welfare maximization literature.}
\label{tab:relation_to_literature}
\end{table}

As already noted, in the maximum score literature (A), \cite{kim1990cube} and \cite{horowitz1992smoothed} develop limit distributions of the orignal maximum score estimator and the smoothed maximum estimator, respectively. \cite{chen2018best} construct a best subset maximum score prediction rule that is linear in covariates by maximizing a penalized objective function. Recently, \cite{chen2025relu} reformulate the maximum score estimator by rectified linear unit (ReLU) functions to achieve computation advantage and convergence rate that is faster than cube-root-n. All these work consider a linear index of finite number of covariates.
Under the empirical welfare maximization framework (B), \citet{rai2018statistical} conduct inference on optimal policy functions and maximized welfare under finite VC dimension. \cite{chernozhukov2025policy}  introduce a framework for selecting policies that maximize a penalized empirical welfare function with the penalty being proportional to a consistent estimator of the risk of each candidate policy. They primarily focus on cases in which the set of policies has finite carnality, with brief discussion of extensions to infinity. We contribute to the both literature, (A) and (B), by allowing the VC dimension of the classes of classification policy functions to diverge to infinity, so as to accommodate nonparametric policy functions.

In the context of treatment assignment, several plug-in approaches to inference exist, leveraging nonparametric nuisance parameters, such as the conditional average treatment effects, that are directly estimable
\citep[e.g.,][]{bhattacharya2012inferring,armstrong2023inference,park2025debiased,chen2025inference}.
In contrast, our approach is aligned with empirical welfare maximization as a special case of cost-sensitive binary classification, and is therefore complementary to these plug-in methods.\footnote{See \citet[][Section 2.3]{hirano2020asymptotic} for details about the conceptual distinction between the plug-in approach and the empirical welfare maximization approach in statistical decisions. Our approach is also conceptually different from \cite{xu2025asymptotic}, who analyzes the optimal point decision from a continuum of alternatives in limit experiments.}
Furthermore, our general binary classification framework extends beyond treatment assignment settings to encompass maximum score and utility maximization frameworks.

Finally, it is worth noting several related results from the statistics and biostatistics literature. 
\citet{luedtke2016statistical} develop root-$n$ rate confidence intervals for the optimal welfare even when the parameter is not pathwise differentiable.
\citet{wu2021resampling} propose a smoothing-based procedure for model-free inference on the optimal treatment. 
\citet{liang2022estimation} employ surrogate convex losses to conduct inference on low-dimensional parameters. 
\citet{wu2023model} develop a uniform inference method by quantifying the unknown link function, thereby enabling flexible modeling of interactions between treatment and covariates. 
The present paper, which adopts a surrogate-loss approach to uniform inference for nonparametric policies, complements specific aspects of each of these contributions.

The paper is organized as follows.
Section \ref{sec_setup} introduces the setup and presents motivating examples.
Section \ref{sec_issue} discusses the key issues arising from the discontinuous nature of the loss function.
Section \ref{sec:identification} develops the ``identification'' of the representative classification policy function.
Section \ref{sec4} presents estimation and inference procedures for the representative classification policy function, 
and Section \ref{sec5} establishes estimation and inference results for the optimal policy value.
Section \ref{sec_sim} examines the finite-sample performance of the proposed methods through simulation studies.
Section \ref{sec_emp} provides an empirical analysis that demonstrate the application of our method.
Section \ref{sec_con} concludes. 

\section{Setup and Examples}\label{sec_setup}

Let \(Z_i\) and \(X_i\), $i=1,2,\cdots, n$, denote random vectors of unit $i$ observed by researchers or policymakers. Some variables may be common to both \(Z_i\) and \(X_i\). Let \(\mathcal{G}\) denote a class of functions. With these notations, a broad class of econometric frameworks can be represented as a cost-sensitive binary classification problem:
\begin{align}
&\min_{g \in \mathcal{G}} L(g), \quad \text{where}
\notag \\
&L(g) = \mathbb{E}\!\left[\,\psi^{+}(Z_i,\eta_0)\, \mathbf{1}\{g(X_i) \geq 0\} + \psi^{-}(Z_i,\eta_0)\, \mathbf{1}\{g(X_i) < 0\}\right].
\label{eq:loss1}
\end{align}
Here, \(\psi^{+}\) and \(\psi^{-}\) denote non-negative functions that may depend on a possibly infinite-dimensional nuisance parameter \(\eta\), whose true value is denoted by \(\eta_0\). This general risk-minimization framework is motivated by three examples. 
Sections~\ref{sec:max_score}, \ref{sec:utility_max}, and \ref{sec:welfare_max} below introduce the maximum score estimator, utility maximization problem, and welfare maximization framework, respectively.

\subsection{Example I: Maximum Score}\label{sec:max_score}

Let \( Y_i \) be a binary random variable and \( X_i \) a random vector such that
\[
Y_i=\mathbf{1}\{g(X_i)+v_i\geq 0\},
\]
where the error term $v_i$ satisfies $\text{Median}(v_i|X_i)=0$ almost surely. The maximum score estimator \citep{manski1975maximum, manski1985semiparametric} for $g$ is defined as the sample analogue of
\begin{align*}
&\arg\max_{g \in \mathcal{G}}
\mathbb{E}\left[ Y_i \cdot \mathbf{1}\{g(X_i)\geq0\} + (1-Y_i) \cdot \mathbf{1}\{g(X_i)<0\} \right],
\end{align*}
for some function class \( \mathcal{G} \). This can be equivalently reformulated as
\begin{align*}
\arg\min_{g \in \mathcal{G}}
\mathbb{E}\left[ {\psi^+(Z_i,\eta)} \cdot \mathbf{1}\{g(X_i)\geq0\} + {\psi^-(Z_i,\eta)}\cdot \mathbf{1}\{g(X_i)<0\} \right],
\end{align*}
where $Z_i := Y_i$, \( \psi^{+}(z,\eta) := 1-z\), and \( \psi^{-}(z,\eta) := z \).
Therefore, the maximum score estimation can be viewed as a special case of the general risk-minimization framework with the risk given in \eqref{eq:loss1}.
$\blacktriangle$

\subsection{Example II: Expected Utility Maximization with Binary Actions}\label{sec:utility_max}

\citet{elliott2013predicting} consider the problem of maximizing the expected utility:
$$
\max_{a(\cdot)}\mathbb{E}\left[U(a(X_i),Y_i,X_i)\right]
$$
under a $\{-1,1\}$-valued binary outcome $Y_i$ with respect to a binary action $a(x) \in \{-1,1\}$ for each $x$.
Imposing the condition that
$
U(1,1,x) > U(-1,1,x)
$
and
$
U(-1,-1,x) > U(1,-1,x)
$ for all $x$ in the support of $X$ to ensure correct classification yields higher utility,
\citet{elliott2013predicting} show that the problem can be reformulated as $\max_{g \in \mathcal{G}}V(g)$, where
$$
V(g) = \mathbb{E}[ b(X_i) [Y_i+1-2c(X_i)] \cdot \mathbf{1}\{g(X_i) \ge 0\}
-b(X_i) [Y_i+1-2c(X_i)] \cdot \mathbf{1}\{g(X_i)<0\}],
$$ 
$b(x) = U(1,1,x) - U(-1,1,x) + U(-1,-1,x) - U(1,-1,x)$, 
and
$c(x) = [U(-1,-1,x)-U(1,-1,x)] / b(x)$.
The optimal action $a^*(x)$ can then be recovered from the optimal classifier $g^\ast$ as $a^\ast(x) = \mathrm{sign}(g^*(x))$ for each $x$.
Like Example I (Section \ref{sec:max_score}), the above maximization problem can be further reformulated as a special case of the general risk-minimization framework with the risk given in \eqref{eq:loss1} by taking $Z_i:=(Y_i,X_i')'$, $\psi^{+}(z,\eta) := |b(x)(y+1-2c(x))|-b(x)(y+1-2c(x))$, and $\psi^{-}(z,\eta) := |b(x)(y+1-2c(x))|+b(x)(y+1-2c(x))$.
$\blacktriangle$

\subsection{Example III: Welfare Maximization}\label{sec:welfare_max}
Let \( Z_i = (Y_i, A_i, X_i')' \), where \( Y_i \) denotes a scalar outcome, \( A_i \) a binary treatment indicator, and \( X_i \) a vector of covariates supported on $\mathcal{X}$. Let \( Y_i(1) \) and \( Y_i(0) \) denote the potential outcomes under treatment and control, respectively. The observed outcome is generated according to
\[
Y_i = Y_i(1) \cdot A_i + Y_i(0) \cdot (1 - A_i).
\]
Define the propensity score function as \( \pi(x) = \mathbb{P}(A_i = 1 \mid X_i = x) \).

A policymaker is interested in selecting a policy function \( g \) from \( \mathcal{G} \) that maximizes welfare, measured by the mean outcome:
\begin{align}\label{eq:ate}
\mathbb{E}\left[Y_i(1)\cdot\mathbf{1}\{g(X_i)\geq 0\} + Y_i(0)\cdot \mathbf{1}\{g(X_i)<0\}\right].
\end{align}
To analyze this objective, the literature commonly imposes the following set of assumptions.
\begin{enumerate}[(WM.i)]
	\item \label{as11} {\bf Unconfoundedness:} $(Y_i(1), Y_i(0)) {\mathrel{\perp\!\!\!\perp}} A_i~\vert~X_i$;
	\item \label{as12} {\bf Moments:} $\mathbb{E}[\vert Y_{i}(a)\vert]<\infty$ for each $a=0, 1$ and;
	\item \label{as13} {\bf Overlap:}  
	There exists $\underline{C} \in (0,1/2)$ such that $\mathbb{P}(\underline{C}\leq \pi_{0}(x)\leq 1-\underline{C})=1$ for all $x\in\mathcal{X}$.
\end{enumerate}
	
Under these assumptions (WM.\ref{as11})--(WM.\ref{as13}), the mean outcome \eqref{eq:ate} can be identified by the inverse propensity score weighting (IPW) as in \citet{kitagawa2018should}, which in turn may be rewritten in the augmented-IPW (AIPW) as in \citet{athey2021policy}:\small
\begin{align*}
&V(g) :=
\mathbb{E}\left[Y_i(0)\right]
\\
&+\mathbb{E}\left[\left(\left(\frac{A_{i}(Y_{i}-\mu(1, X_{i}))}{\pi(X_{i})}+\mu(1, X_{i})\right)-\left(\frac{(1-A_{i})(Y_{i}-\mu(0, X_{i }))}{1-\pi(X_{i})}+\mu(0, X_{i})\right)\right)\cdot\mathbf{1}\{g(X_i)\geq0\}\right],
\end{align*}\normalsize
where $\mu(l, x)=\mathbb{E}[Y_{i} \vert A_{i}=l, X_{i}=x]$ for $l \in \{0,1\}$. Collecting the nuisance parameters \( \eta := (\mu(1,\cdot), \mu(0,\cdot), \pi(\cdot)) \), we express the \( g \)-relevant component of the identified objective, \( V(g) \), as
\begin{align}
&
V(g)=\mathbb{E}[\psi_0(Z_i,\eta)]+\mathbb{E}\left[\left(\psi_1(Z_i,\eta) - \psi_0(Z_i,\eta)\right) \cdot \mathbf{1}\{g(X_i)\geq 0\}\right],
\qquad\text{where}\label{eq:value}
\\
&
\psi_1(Z_i,\eta) := \frac{A_{i}(Y_{i}-\mu(1, X_{i}))}{\pi(X_{i})}+\mu(1, X_{i})
\text{ and }
\psi_0(Z_i,\eta) := \frac{(1-A_{i})(Y_{i}-\mu(0, X_{i }))}{1-\pi(X_{i})}+\mu(0, X_{i}).
\label{eq:psi1psi0}
\end{align}
Denote the positive components of $\psi_1(Z_i,\eta)$ and $\psi_0(Z_i,\eta)$ by $\psi_1^{+}(Z_i,\eta)$ and $\psi_0^{+}(Z_i,\eta)$, respectively, and similarly denote the negative parts of $\psi_1(Z_i,\eta)$ and $\psi_0(Z_i,\eta)$ by $\psi_1^{-}(Z_i,\eta)$ and $\psi_0^{-}(Z_i,\eta)$, respectively.
Then, the welfare \eqref{eq:value} can be rewritten as
\begin{align}
	{V}(g)
	&=\mathbb{E}\left[\psi_1(Z_i,\eta)\mathbf{1}\{g(X_i)\geq 0\}+\psi_0(Z_i,\eta)\mathbf{1}\{g(X_i)<0\}\right]\notag\\
	&=\mathbb{E}\left[(\psi_1^{+}(Z_i,\eta)-\psi_1^{-}(Z_i,\eta)) \mathbf{1}\{g(X_i)\geq 0\}+(\psi_0^{+}(Z_i,\eta)-\psi_0^{-}(Z_i,\eta)) \mathbf{1}\{g(X_i)<0\}\right]\notag\\
	&=\mathbb{E}\left[\psi_1^{+}(Z_i,\eta)\left(1-\mathbf{1}\{g(X_i)<0\}\right)-\psi_1^{-}(Z_i,\eta)\mathbf{1}\{g(X_i)\geq 0\} \right. \notag\\
	&
	\qquad\quad\left. +\psi_0^{+}(Z_i,\eta)\left(1-\mathbf{1}\{g(X_i)\geq 0\}\right)-\psi_0^{-}(Z_i,\eta)\mathbf{1}\{g(X_i)<0\} \right]\notag\\
	&=\mathbb{E}\left[\psi_1^{+}(Z_i,\eta)+\psi_0^{-}(Z_i,\eta)\right]-\underbrace{\mathbb{E}\left[{\psi}^{+}(Z_i,\eta) \cdot\mathbf{1}\{g(X_i)\geq 0\}+{\psi}^{-}(Z_i,\eta)\cdot\mathbf{1}\{g(X_i)<0\}\right]}_{=\text{\eqref{eq:loss1}}},\label{eq:welfare}
\end{align}
where $\psi^{+}(Z_i,{\eta}) := \psi_1^{-}(Z_i,\eta)+ \psi_0^{+}(Z_i,\eta)$ and $\psi^{-}(z,{\eta}) := \psi_1^{+}(Z_i,\eta)+ \psi_0^{-}(Z_i,\eta)$.
Note that the second expectation in the last line takes the form of \eqref{eq:loss1}. Thus, the welfare maximization problem can be viewed as a special case of the general risk-minimization framework with a risk of the form \eqref{eq:loss1}.
$\blacktriangle$

\section{Issues}\label{sec_issue}

As Sections \ref{sec:max_score}--\ref{sec:welfare_max} demonstrate, the cost-sensitive binary classification problem $\min_{g \in \mathcal{G}} L(g)$ with the risk function given by \eqref{eq:loss1} encompasses several important econometric frameworks. However, this problem is more challenging to address than standard econometric problems. The difficulty becomes even more pronounced when allowing for a nonparametric class $\mathcal{G}$ of policy functions. 

The primary challenge arises from the discontinuous nature in the sample analog of the objective function. In practice, this discontinuity induces computational difficulties: the sample objective function is non-convex, making numerical optimization difficult, if not infeasible. Beyond these practical concerns, the discontinuity also has adverse theoretical implications, giving rise to non-standard limiting behavior and non-identification, as detailed in the following two subsections.

\subsection{Non-Standard Limit under Identifying Normalizations}\label{sec:non_standard}

To simplify the discussion, suppose for the moment that \(g\) belongs to a parametric class specified as \(g(x) = x^\prime \gamma\), with appropriate normalizations. In this case, the problem \(\min_{g \in \mathcal{G}} L(g)\) reduces to an \(M\)-estimation of \(\gamma\) with a discontinuous sample objective function. In this setting, \citet{kim1990cube} show that the resulting \(M\)-estimator converges at the cube-root-\(n\) rate and admits a non-standard asymptotic distribution. Consequently, naive bootstrap procedures are invalid in this context \citep{abrevaya2005bootstrap,leger2006bootstrap}.

\subsection{Non-Identification in General}\label{sec:non_identification}

Although Section~\ref{sec:non_standard} focuses on a simple parametric class, it already highlights the inherent difficulty of the problem. When \(\mathcal{G}\) is a nonparametric class, the challenge becomes even more severe, as the optimal policy function \(g\) is no longer point-identified even after normalization. This is because only the sign of \(g\) affects the risk in \eqref{eq:loss1}. In particular, if \(g^{\ast} \in \mathcal{G}\) is a solution, then any \(g^{\ast\ast} \in \mathcal{G}\) such that \(\text{sign}(g^{\ast\ast}(x)) = \text{sign}(g^{\ast}(x))\) for all \(x\) is also a solution. In a nonparametric class \(\mathcal{G}\), one can readily construct infinitely many such functions \(g^{\ast\ast}\) that are sign-equivalent to \(g^{\ast}\), thereby rendering the optimal policy function unidentifiable. Theoretical analysis of these partially identified nonparametric models poses even greater challenges than those encountered in the parametric case of Section~\ref{sec:non_standard}.

\section{Representative Identification}\label{sec:identification}

The difficulties associated with \eqref{eq:loss1}, as discussed in the previous section, stem from the discontinuity of the 0–1 loss function. We remedy them by replacing the 0–1 loss with a strictly convex surrogate loss. This substitution not only yields a continuous and convex objective to ease computation, but also ensures point identification of the optimal policy function even when $\mathcal{G}$ is nonparametric.

Let $\phi: \mathbb{R} \rightarrow \mathbb{R}$ be a surrogate loss function, and consider the following modification of \eqref{eq:loss1}:
\begin{align}\label{eq:surrogate1}
L^\ast(g) := \mathbb{E}\left[\psi^+(Z_i,\eta_0)\cdot \phi(-g(X_i)) + \psi^{-}(Z_i,\eta_0)\cdot \phi(g(X_i))\right].
\end{align}
The following theorem presents the utility of using this function $\phi(\cdot)$ in place of the original 0-1 loss.

\begin{thm}[Identification]\label{thm1}
Suppose that $\mathbb{E}[\psi^+(Z_i,\eta_0)|X_i] -\mathbb{E}[\psi^-(Z_i,\eta_0)|X_i] \neq 0$ almost surely and $\mathcal{G}$ includes all measurable functions. Let $\phi(\cdot)$ be differentiable at zero and $\partial \phi(0)<0$.
	\begin{enumerate}[(i)]
	\item  If $\phi(\cdot)$ is convex, then any $g_\ast \in\arg\min_{g\in\mathcal{G}} L^\ast(g)$ satisfies $g_\ast\in \arg\min_{g \in \mathcal{G}} L(g)$.
	\item If $\phi(\cdot)$ is strictly convex and $\mathcal{G}$ is convex, then there is unique $g_\ast$ such that $g_\ast = \arg\min_{g\in\mathcal{G}} L^\ast(g)$.\footnote{The uniqueness is up to the equivalence class in $\mathcal{G}$ identified by the underlying probability measure.}
	\end{enumerate}
\end{thm}

Theorem~\ref{thm1}(i) guarantees that minimizing the surrogate risk \( L^\ast \) yields an optimal policy function with respect to the original risk \( L \), as established in the literature \citep{bartlett2006convexity}. Furthermore, Theorem~\ref{thm1}(ii) ensures that the surrogate risk admits a unique minimizer (up to \( L_1 \)-equivalence). In other words, by employing a strictly convex surrogate function \( \phi(\cdot) \), we achieve \emph{point identification} of a nonparametric solution \( g_\ast \) to the original problem \( \min_{g \in \mathcal{G}} L(g) \). This, in turn, facilitates the application of econometric methods based on point estimation and their associated asymptotic properties, which are typically more tractable than those relying on set identification.

The surrogate loss approach has been explored in the literature \citep[e.g.,][]{bartlett2006convexity,zhao2012estimating,kitagawa2023constrained}; however, the motivation behind our use of it is fundamentally different. While existing studies primarily leverage surrogate losses to enable numerical optimization by convexifying the originally non-convex objective, our primary aim is to establish \textit{identification} of the optimal policy function. Moreover, while \citet{zhao2012estimating} and \citet{kitagawa2023constrained} advocate the hinge loss for their purposes, we explicitly rule out the hinge loss in part (ii) of our Theorem~\ref{thm1} and instead promote alternative loss functions that better align with our identification objective -- see below for a concrete list. In this respect, both our goals and the methodological approaches are largely orthogonal to those in the literature. To the best of our knowledge, the use of surrogate losses for the purpose of identification (or representation), rather than computational convenience, is novel. 

Although we identify only one representative solution \( g_\ast \), this suffices for conducting inference on the optimal classification policy. Indeed, \( g_\ast \) is sign-equivalent to any solution of $\min_{g \in \mathcal{G}}L(g)$ up to a null set. Accordingly, the optimal classification rule defined by \(x \mapsto \mathbf{1}\{x \in \mathcal{X} \mid g_\ast(x) \geq  0\} \) delivers the same policy decisions as the rule \( x \mapsto \mathbf{1}\{x \in \mathcal{X} \mid g(x) \geq 0\} \) associated with any solution \( g \in \arg\min_{g \in \mathcal{G}} L(g) \), possibly except on a set of measure zero. Thus, we may perform inference on the optimal classification policy and the corresponding optimal policy value (e.g., welfare) using the representative, point-identified representative policy function \( g_\ast \), effectively treating it as the unique solution.

Various choices for the function \( \phi \) have been proposed in the literature: the hinge loss, the exponential loss, the logistic loss, the squared loss, and the sigmoid loss, among others. 
While part~(i) of Theorem~\ref{thm1} accommodates all of these examples, part~(ii) rules out the hinge loss.

The use of different loss functions leads to different representers $g_*$. 
However, as argued above, these representers yield the same optimal classification rule, since they agree in sign except on a null set. As such, the choice of loss function, and hence the resulting representer, may be viewed as analogous to selecting different parameter normalizations proposed by \citet{manski1975maximum, manski1985semiparametric} in the context of the maximum score framework (Example I) introduced in Section \ref{sec:max_score}.

The assumption that 
	\begin{align*}
	\mathbb{E}[\psi^+(Z_i,\eta)\mid X_i] - \mathbb{E}[\psi^-(Z_i,\eta)\mid X_i] \neq 0 \quad \text{a.s.}
	\end{align*}
	plays a crucial role. If there exists an event with positive probability on which 
	\begin{align*}
	\mathbb{E}[\psi^+(Z_i,\eta)\mid X_i] - \mathbb{E}[\psi^-(Z_i,\eta)\mid X_i] = 0,
	\end{align*}
	then the function $g$ could take either positive or negative values on that event, thereby failing to point-identify the solution. 
	This condition is essential not only for point identification but also for the inference on the optimal policy value; see, for example, \citet[][Theorem~1]{luedtke2016statistical} for related discussions.

\section{Estimation and Inference for the Representative Policy Function}\label{sec4}

Given the `identification' result from Section \ref{sec:identification}, one can obtain a point estimate of the representative classification policy function $g_\ast$. Although $g_\ast$ is only one among infinitely many solutions, it suffices for constructing the optimal classification rule $x \mapsto \mathbf{1}\{g_\ast(x) \geq 0\}$, since only the sign of $g_\ast$ matters and any other solution has an equivalent sign almost everywhere. 
Consequently, the optimal classification policy and the optimal policy value (e.g., welfare) can be estimated. 

\subsection{Estimation of the Representative Policy Function}\label{sec:estimate_g}

Let $\mathcal{X}$ denote the support of $X_i$, and $\{p_1(\cdot),p_2(\cdot),...\}$ be a sieve basis consisting of measurable functions $p_k: \mathcal{X} \rightarrow \mathbb{R}$, $k \in \mathbb{N}$.
For each $k \in \mathbb{N}$, let $\mathcal{G}_k = \{x \mapsto p_{1:k}(x)'\beta  \ : \ \beta \in \mathbb{R}^k\}$ denote the sieve space, where $p_{1:k}(\cdot) := (p_1(\cdot),...,p_k(\cdot))'$.
Let $g_k$ denote the $L_2(\mathbb{P})$-projection of $g_\ast \in \mathcal{G}$ onto $\mathcal{G}_k$, and let $r_g = g_\ast-g_k$ be its approximation error.
Denote the complexity measure by $\xi_k := \sup_{x \in \mathcal{X}}\|p_{1:k}(x)\|$.

We now introduce the cross-fitting method for estimating \(g_\ast\).
Partition the index set \([n] = \{1,\ldots,n\}\) into \(m\) folds \(I_{1}, \ldots, I_{m}\), and let \(I^{c}_{(\ell)} := [n] \setminus I_{(\ell)}\) denote the complement of the \(\ell\)-th fold. Assume that the folds are approximately balanced, so that 
$
n_{(1)} \asymp \cdots \asymp n_{(m)},
$
where \(n_{(\ell)} := |I_{(\ell)}|\), and the number of folds \(m\) is fixed.

For a given fold \(\ell \in [m] = \{1,\ldots,m\}\), obtain an estimator \(\widehat{\eta}_{(\ell)}\) of the nuisance parameter \(\eta_0\) using the complementary subsample \(I_{(\ell)}^c\). 
Then, estimate the representative policy function \(g_\ast\) on the held-out fold \(I_{(\ell)}\). 
Specifically, the fold-wise estimator of \(g_\ast\) is given by
\begin{align*}
&\widehat{g}_{(\ell)}(x) := p_{1:k}(x)^{\prime} \widehat{\beta}_{(\ell)},
\qquad\text{where} 
\\
&\widehat{\beta}_{(\ell)} 
	:= \arg\min_{b \in \mathbb{R}^k} 
	\mathbb{E}_{n,(\ell)}\!\left[
	\widehat{\psi}^{+}_{(\ell)}(Z_i)\, \phi\!\left(-p_{1:k}(X_i)^{\prime}b\right) 
	+ \widehat{\psi}^{-}_{(\ell)}(Z_i)\, \phi\!\left(p_{1:k}(X_i)^{\prime}b\right)
	\right],
\end{align*}
\(p_{1:k}(\cdot) := (p_1(\cdot), \ldots, p_k(\cdot))^\prime\), 
\(\widehat{\psi}^{\pm}_{(\ell)}(\cdot) := \psi^{\pm}(\cdot, \widehat{\eta}_{(\ell)})\) for shorthand, 
and \(\mathbb{E}_{n,(\ell)}[f] := \frac{1}{|I_{(\ell)}|} \sum_{i \in I_{(\ell)}} f(Z_i)\) denotes the sample mean over the \(\ell\)-th fold. 

Finally, aggregate the \(m\) fold-wise estimators to obtain the cross-fitted estimator
\begin{align}\label{esti:cf}
\widehat{g}(x) := \frac{1}{m} \sum_{\ell=1}^m \widehat{g}_{(\ell)}(x).
\end{align}

For later use, define \(\widehat{\eta}\) as the full-sample estimator of the nuisance parameter \(\eta_0\), and let \(\widehat{\psi}^\pm(\cdot) := \psi^\pm(\cdot, \widehat{\eta})\). 
This estimator will be employed for variance estimation and bootstrap inference. 
The conditions required for both the cross-fitted and full-sample estimators are presented in Section \ref{sec41}.

\subsection{Large Sample Theory for the Representative Policy Function Estimation}\label{sec41}

We next establish the uniform convergence rate of \(\widehat{g}\) toward \(g_\ast\) and prove a Gaussian coupling principle that characterizes the Gaussian approximation of its estimation error. 

We now introduce a set of regularity conditions.

\begin{as}[Sieves]\label{as:series}
Let the following conditions be satisfied:
	\begin{enumerate}[(i)]
		\item \textbf{(Bounded Moments)} \label{as:s0} $\mathbb{E}[\vert\psi^+(Z_i, \eta)\vert^3\vert X_i=x] < \infty$ and $\mathbb{E}[\vert\psi^-(Z_i, \eta)\vert^3\vert X_i=x] < \infty$ uniformly for all $x \in \mathcal{X}$.
		\item \textbf{(Eigenvalues)} \label{as:s1}
		For $Q := \mathbb{E}[p_{1:k}(X_i)(\psi^{+}(Z_i)\cdot \partial^2\phi(-g_\ast(X_i))+\psi^{-}(Z_i)\cdot \partial^2\phi(g_\ast(X_i))) p_{1:k}(X_i)^{\prime}]$ and $\Sigma := \mathbb{E}[p_{1:k}(X_i)(-\psi^{+}(Z_i)\cdot \partial\phi(-g_\ast(X_i))+\psi^{-}(Z_i)\cdot \partial\phi(g_\ast(X_i)))^2 p_{1:k}(X_i)^{\prime}]$, there exist constant values $C_{\min}$, $C_{\max} \in (0,\infty)$ such that $C_{\min} < \lambda_{\min}(Q) < \lambda_{\max}(Q) < C_{\max}$ and $C_{\min} < \lambda_{\min}(\Sigma) < \lambda_{\max}(\Sigma) < C_{\max}$  with $ \lambda_{\min}(\cdot)$ and $\lambda_{\max}(\cdot)$ indicating the minimum and maximum eigenvalues.
		\item \textbf{(Growth Condition)} \label{as:s2} The complexity measure $\xi_k$ grows sufficiently slowly such that it satisfies
		$\sqrt{\xi_k^2 \log(k) / n} \to 0$ and $ \log(\xi_k) \lesssim \log(k)$

		\item \textbf{(Approximation Error)} \label{as:s3}  
		There is a sequence $\{B_k\}_k$ of bounds on approximation errors associated with the sequence $\{\mathcal{G}_k\}_k$, such that $\sqrt{n}B_k\log(k)\rightarrow0$ and
			\begin{align*}
				\Vert r_g\Vert_{\mathbb{P},\infty}:=\sup _{x \in \mathcal{X}}\vert r_g(x)\vert \leq B_k,
			\end{align*}
	\end{enumerate}
\end{as}

Assumption~\ref{as:series} collects standard regularity conditions commonly used in the series-based nonparametric estimation literature and adapts them to our cost-sensitive binary classification problem. 
Assumption~\ref{as:series}\eqref{as:s0} requires the conditional moments of the weights to be bounded and serves multiple purposes. 
In particular, the existence of the third moment is needed to establish a coupling inequality.
Assumption~\ref{as:series}\eqref{as:s1}, which rules out collinearity in the population signal matrix, is a conventional requirement in nonparametric analysis \citep[e.g.,][]{newey1997convergence}. 
Assumptions~\ref{as:series}\eqref{as:s2} and \eqref{as:s3} control the complexity of the function class and the nonparametric approximation error, respectively.

To facilitate our subsequent writings, we define some short-hand notations:
\begin{align*}
	&\Psi(Z_i,\eta,g):=\psi^{+}(Z_i,\eta)\phi(-g(X_i))+\psi^{-}(Z_i,\eta)\phi(g(X_i)),\\
	&\Psi_1(Z_i,\eta,g):=-\psi^{+}(Z_i,\eta)\cdot\partial\phi(-g(X_i))+\psi^{-}(Z_i,\eta)\cdot\partial\phi(g(X_i)), \text{ and }\\
	&\Psi_2(Z_i,\eta,g_1,g_2):=\psi^{+}(Z_i,\eta)\cdot\partial^2\phi(-g_1(X_i))+\psi^{-}(Z_i,\eta)\cdot\partial^2\phi(g_2(X_i)),
\end{align*}
where $\partial\phi(g)$ and $\partial^2\phi(g)$ denote the first- and second-order derivatives of $\phi$, respectively.
With these notations, we state the following conditions on the estimators of the nuisance parameter that enter the weights.

\begin{as}[Nuisance Parameter Estimators]\label{as:nuisance}~
There exists a sequence $\mathcal{S}_n$ of shrinking subsets such that $g_\ast, g_k, \widehat g \in \mathcal{S}_n$ with probability at least $1-\Delta_n$ such that $\Delta_n \to 0$ and the following conditions are satisfied.

\begin{enumerate}[(i)]
	\item \textbf{(Cross-Fitting)}For each $\ell \in [m]$ and any $g_1, g_2 \in \mathcal{S}_n$, the cross-fitted nuisance parameter estimator satisfies
\begin{align}
	&\label{as:cross1}\left\Vert\mathbb{E}_{n,(\ell)} \left[\left(\Psi_1(Z_i, \widehat{\eta}_{(\ell)},g_1)- \Psi_1(Z_i,\eta_0,g_\ast)\right)p_{1:k}(X_i) \right]\right\Vert= o_p(\sqrt{1/(n\log(k)^2)}),\\
	&\label{as:cross2}\left\Vert\mathbb{E}_{n,(\ell)} \left[\left(\psi^{\pm}(Z_i, \widehat{\eta}_{(\ell)})- \psi^{\pm}(Z_i,\eta_0)\right)p_{1:k}(X_i)p_{1:k}(X_i)^\prime\right]\right\Vert= o_p(\xi_k\sqrt{\log(k)/n}),\\
	&\label{as:cross3}\left\Vert\mathbb{E}_{n,(\ell)} \left[\left(\Psi_2(Z_i, \widehat{\eta}_{(\ell)},g_1,g_2)- \Psi_2(Z_i,\eta_0,g_\ast,g_\ast)\right)p_{1:k}(X_i)p_{1:k}(X_i)^\prime\right]\right\Vert= o_p(\xi_k\sqrt{\log(k)/n}),\\
	&\label{as:cross4}\left\Vert\mathbb{E}_{n,(\ell)} \left[\left(\Psi_1(Z_i,\widehat{\eta}_{(\ell)},g_1)^2- \Psi_1(Z_i,\eta_0,g_\ast)^2\right)p_{1:k}(X_i)p_{1:k}(X_i)^\prime\right]\right\Vert= o_p(\xi_k\sqrt{\log(k)/n}).
\end{align}  
\item \textbf{(Full-Sample)} For any $g_1, g_2 \in \mathcal{S}_n$ of $g_\ast$, the full-sample nuisance parameter estimator satisfies
\begin{align}
	&\label{as:full1}\left\Vert\mathbb{E}_{n} \left[\left(\Psi_1(Z_i, \widehat{\eta},g_1)- \Psi_1(Z_i,\eta_0,g_\ast)\right)p_{1:k}(X_i) \right]\right\Vert= o_p(\sqrt{1/(n\log(k)^2)}),\\
	&\label{as:full2}\left\Vert\mathbb{E}_{n} \left[\left(\psi^{\pm}(Z_i, \widehat{\eta})- \psi^{\pm}(Z_i,\eta_0)\right)p_{1:k}(X_i)p_{1:k}(X_i)^\prime \right]\right\Vert= o_p(\xi_k\sqrt{\log(k)/n}),\\
	&\label{as:full3}\left\Vert\mathbb{E}_{n} \left[\left(\Psi_2(Z_i, \widehat{\eta},g_1,g_2)- \Psi_2(Z_i,\eta_0,g_\ast,g_\ast)\right)p_{1:k}(X_i)p_{1:k}(X_i)^\prime \right]\right\Vert= o_p(\xi_k\sqrt{\log(k)/n}),\\
	&\label{as:full4}\left\Vert\mathbb{E}_{n} \left[\left(\Psi_1(Z_i,\widehat{\eta},g_1)^2- \Psi_1(Z_i,\eta_0,g_\ast)^2\right)p_{1:k}(X_i)p_{1:k}(X_i)^\prime\right]\right\Vert= o_p(\xi_k\sqrt{\log(k)/n}).
\end{align} 
\end{enumerate}
\end{as}

Assumption~\ref{as:nuisance} imposes rate restrictions on the estimation error of the subsample and full-sample estimators for $\eta_0$.  Conditions \eqref{as:cross1} and \eqref{as:full1} are used for the Gaussian coupling of large-dimensional score vectors, Conditions \eqref{as:cross2}--\eqref{as:cross3} and \eqref{as:full2}--\eqref{as:full3} ensure the convergence rate of the sample Gram matrix, and Conditions \eqref{as:cross4} and \eqref{as:full4} guarantee the convergence rate of the variance estimator.\footnote{The sample Gram matrix and the variance estimator will be introduced later in \eqref{eq:sample_gram} and \eqref{eq:sample_variance} respectively.}

In the context of Example~I (Maximum Score; Section~\ref{sec:max_score}) and Example~II (Expected Utility Maximization; Section ~\ref{sec:utility_max}), there is no nuisance parameter \(\eta\), so Assumption~\ref{as:nuisance} is trivially satisfied. 
In the context of Example~III (Welfare Maximization; Section~\ref{sec:welfare_max}), one can adopt alternative estimators subject to suitable low-level conditions to ensure that Assumption~\ref{as:nuisance} holds. Such examples include the $\ell_1$-penalized and related methods in sparse models \citep{belloni2013inference} and other machine learning methods under appropriate conditions. For the current work, we focus on $\ell_1$-penalized methods for estimating our nuisance parameters using full-sample and sample-splitting techniques.

Finally, we impose a smoothness condition on the surrogate loss and the representative policy function to facilitate the derivation of their asymptotic properties.

\begin{as}[Surrogate Loss]\label{as:phi}
	The surrogate loss function $\phi$ is twice continuously differentiable.
\end{as} 

\begin{as}[Representative Policy]\label{as:holder} 
The representative classification policy function $g_\ast$ is uniformly bounded on compact $\mathcal{X}$ and belongs to a Hölder ball $\Lambda^h(L)$ of smoothness order $h> 1/2$ and radius $L \in (0,\infty)$.\footnote{See \eqref{class:holder} in the appendix for the definition of the Hölder ball $\Lambda^h(L)$ of smoothness order $h$ and radius $L$.}
\end{as}

For standardization, we introduce the variance process
$\sigma^2(x)=p_{1:k}(x)^{\prime}Q^{-1}\Sigma Q^{-1}p_{1:k}(x)$,
where
\begin{align*}
	&Q:=\mathbb{E}[(\psi^{+}(Z_i)\cdot \partial^2\phi(-g_\ast(X_i))+\psi^{-}(Z_i)\cdot \partial^2\phi(g_\ast(X_i))) p_{1:k}(X_i) p_{1:k}(X_i)^{\prime}] \quad\text{and}\\
	&\Sigma:=\mathbb{E}[(-\psi^{+}(Z_i)\cdot \partial\phi(-g_\ast(X_i))+\psi^{-}(Z_i)\cdot \partial\phi(g_\ast(X_i)))^2 p_{1:k}(X_i) p_{1:k}(X_i)^{\prime}].
\end{align*}
We may estimate it by the sample counterpart
$\widehat{\sigma}^2(x)=p_{1:k}(x)^{\prime}\widehat{Q}^{-1}\widehat{\Sigma}\widehat{Q}^{-1}p_{1:k}(x)$
where
\begin{align}
	&\widehat{Q}:=\mathbb{E}_n[(\widehat{\psi}^{+}(Z_i)\cdot \partial^2\phi(-\widehat{g}(X_i))+\widehat{\psi}^{-}(Z_i)\cdot \partial^2\phi(\widehat{g}(X_i))) p_{1:k}(X_i) p_{1:k}(X_i)^{\prime}] \quad\text{and} \label{eq:sample_gram}\\
	&\widehat{\Sigma}:=\mathbb{E}_n[(-\widehat{\psi}^{+}(Z_i)\cdot \partial\phi(-\widehat{g}(X_i))+\widehat{\psi}^{-}(Z_i)\cdot \partial\phi(\widehat{g}(X_i)))^2 p_{1:k}(X_i)p_{1:k}(X_i)^{\prime}]. \label{eq:sample_variance}
\end{align}

With the above definitions and notation in place, the following theorem establishes the uniform convergence rate and Gaussian coupling for the representative policy function estimator \(\widehat{g}\).

\begin{thm}\label{thm6}
	In addition to the conditions for Theorem \ref{thm1} (ii), suppose that Assumptions \ref{as:series}, \ref{as:nuisance}, \ref{as:phi} and \ref{as:holder} hold. Then, we have
\begin{align*}
		\sup_{x\in\mathcal{X}}\vert\widehat{g}(x)-g_\ast(x)\vert\lesssim_p \xi_{k} \sqrt{\log (k)/n}+B_{k}.
\end{align*}
In addition, letting $\mathcal{N}_{k}\sim\mathcal{N}(\mathbf{0}_{k},I_{k\times k})$, we have
\begin{align}\label{coupling_yurinskii}
		\sqrt{n}\frac{(\widehat{g}(x)-g_\ast(x))}{\Vert \widehat{\sigma}(x)\Vert}=_d\frac{\sigma(x)^{\prime}}{\Vert \sigma(x)\Vert}\mathcal{N}_k+o(1/\log(k))
\text{ uniformly for all $x\in\mathcal{X}$,}
\end{align}
provided that $\sup _{x \in \mathcal{X}} \sqrt{n}\vert r_g(x)\vert /\Vert \sigma(x)\Vert=o(1/\log(k))$ and $\log(k)^{6}k^4\xi^{2}_{k}\log(n)^2/n\rightarrow0$.
\end{thm}

A few remarks are in order. 
First, the point identification result from Section~\ref{sec:identification} is the key enabler of our asymptotic analysis and plays a crucial role in constructing valid inference procedures. 
Second, as a consequence, our estimator achieves the standard nonparametric rate of convergence established in the sieve literature, despite the originally non-standard problem involving partial identification. 
Third, in addition to the requirements stated in Theorem~\ref{thm1}(ii), Assumption~\ref{as:phi} requires the surrogate loss to be twice continuously differentiable. 
This condition still accommodates all of the aforementioned options, again except the hinge loss.

Since our nonparametric policy function estimator is strongly approximated by a Gaussian process whose distribution depends on the data only through the covariance, it is feasible to conduct uniform nonparametric inference using bootstrapped critical values. 
The details of this inference procedure for \(g(\cdot)\) are presented in Section \ref{sec42}.

\subsection{Bootstrap Inference for the Representative Policy Function}\label{sec42}

Given Theorem~\ref{thm6}, it is natural to consider the following \(t\)-statistic process:
\begin{align}\label{t5}
	t(x) := \frac{\widehat{g}(x) - g_\ast(x)}{\widehat{\sigma}(x)}, 
	\qquad x \in \mathcal{X}.
\end{align}

The sieve score bootstrap \citep{chen2015sieve} can be employed to compute the relevant critical values. 
Specifically, let \(\{\omega_{i}\}_{i}\) be i.i.d. standard Gaussian random variables, independent of the data. 
Consider the score-bootstrap \(t\)-statistic process
\begin{align}\label{bs2}
	t^{b}_{g}(x)
	:= \frac{
		p_{1:k}(x)^{\prime}\widehat{Q}^{-1}
		\Big\{\mathbb{G}_{n}\!\big[\omega_{i}\big(-\widehat{\psi}^{+}(Z_i)\,\partial\phi(-\widehat{g}(X_i))
		+ \widehat{\psi}^{-}(Z_i)\,\partial\phi(\widehat{g}(X_i))\big) p_{1:k}(X_i)\big]\Big\}
	}{\widehat{\sigma}(x)}
\end{align}
for \(x \in \mathcal{X}\), where \(\widehat{\psi}^\pm(\cdot)\) is defined in Section~\ref{sec:estimate_g}, \(\widehat{Q}\) and \(\widehat{\sigma}\) are defined in Section~\ref{sec41}, and $\mathbb{G}_n[f]=\sqrt{n}(\mathbb{E}_n[f]-\mathbb{E}[f])$
denotes the empirical process under the i.i.d. sampling scheme.

To compute the critical value \(cv_{n}(1-\alpha)\) of the supremum statistic \(\sup_{x \in \mathcal{X}} |t(x)|\) in \eqref{t5}, one can evaluate \(\sup_{x \in \mathcal{X}} |t^{b}_{g}(x)|\) based on a large number of independent draws of \(\{\omega_{i}\}_{i}\). 
The critical value \(cv_{n}(1-\alpha)\) is then approximated as 
\begin{align}\label{t10}
	cv^{b}_{n}(1-\alpha) := (1-\alpha)\text{-quantile of } 
	\sup_{x \in \mathcal{X}} \big| t^{b}_{g}(x) \big| 
	\;\text{conditional on the data}.
\end{align}

Construct the uniform confidence band (UCB) for the representative policy function \(g_\ast\) as
\begin{align}\label{band_3}
	[\widehat{g}^{b}_{l}(x), \widehat{g}^{b}_{u}(x)] 
	:= \Big[\,\widehat{g}(x) - cv^{b}_{n}(1-\alpha)\cdot \widehat{\sigma}(x),~
	\widehat{g}(x) + cv^{b}_{n}(1-\alpha)\cdot \widehat{\sigma}(x)\,\Big],
	\qquad x \in \mathcal{X},
\end{align}
where \(cv^{b}_{n}(1-\alpha)\) guarantees that \(g_\ast(x) \in [\widehat{g}^{b}_{l}(x), \widehat{g}^{b}_{u}(x)]\) uniformly for all \(x \in \mathcal{X}\) with asymptotic confidence level \(100(1-\alpha)\%\), as formally stated in the following theorem.

We introduce the notation $\Psi^\ast_1(Z_i,\omega_i,\eta,g):=\omega_i(-\psi^{+}(Z_i,\eta)\cdot\partial\phi(-g(X_i))+\psi^{-}(Z_i,\eta)\cdot\partial\phi(g(X_i)))$, and the following assumption on the nuisance parameter estimation analogously to Assumption \ref{as:nuisance}:
\begin{as}\label{as:nuisance1} 
For any $g$ belonging to a shrinking neighborhood $\mathcal{S}_n$ of $g_\ast$, the full-sample nuisance parameter estimator satisfies
\begin{align*}
	&\left\Vert\mathbb{E}_{n} \left[\left(\Psi_1^\ast(Z_i,\omega_i, \widehat{\eta},g)- \Psi_1^\ast(Z_i,\omega_i,\eta_0,g_\ast)\right)p_{1:k}(X_i) \right]\right\Vert= o_p(\sqrt{1/(n\log(k)^2)}).
\end{align*}
\end{as}
\noindent
The following theorem establishes the uniform probability coverage.

\begin{thm}[Uniform Inference via Two-Sided UCB]\label{thm11}
In addition to the conditions of Theorem \ref{thm6}, suppose that Assumption \ref{as:nuisance1} holds. With the critical value $cv^{b}_{n}\left(1-\alpha\right)$ constructed in \eqref{t10}, the uniform confidence band (UCB) defined in \eqref{band_3} satisfies
	\begin{align}
		&\label{kpi232}\mathbb{P}\left\{g_\ast(x)\in[\widehat{g}^{b}_{l}(x),\widehat{g}^{b}_{u}(x)]~\text{for all }x\in\mathcal{X}\right\}=1-\alpha+o\left(1\right).
	\end{align}
\end{thm}

There has been progress in developing inference methods for parametric policy functions \citep{wu2021resampling,liang2022estimation,cheng2024inference}. 
However, to the best of our knowledge, inference for nonparametric policy functions is first addressed in this paper, thereby filling an important gap in the literature.

To compute the one-sided critical values, one can evaluate \(\sup_{x \in \mathcal{X}} t^{b}_{g}(x)\) and \(\inf_{x \in \mathcal{X}} t^{b}_{g}(x)\) based on a large number of independent draws of \(\{\omega_{i}\}_{i}\), as before. 
The upper and lower critical values, \(cv_{n}^{1,b}(1-\alpha)\) and \(cv_{n}^{1,b}(\alpha)\), are then computed as 
\begin{align}
	\label{t101}
	cv^{1,b}_{n}(1-\alpha) &:= (1-\alpha)\text{-quantile of } 
	\sup_{x \in \mathcal{X}} t^{b}_{g}(x) 
	\;\text{conditional on the data and } \{\omega_{i}\} \qquad\text{and}\\
	\label{t11}
	cv^{1,b}_{n}(\alpha) &:= \alpha\text{-quantile of } 
	\inf_{x \in \mathcal{X}} t^{b}_{g}(x) 
	\;\text{conditional on the data and } \{\omega_{i}\}.
\end{align}

We construct the one-sided uniform confidence bands (UCBs) for testing the uniform null hypotheses 
\(\mathcal{H}_0: g_\ast(x) \leq 0 \;\;\forall x \in \mathcal{X}\) 
or 
\(\mathcal{H}_0: g_\ast(x) \geq 0 \;\;\forall x \in \mathcal{X}\): 
\begin{align}
	\label{band_31}
	\widehat{g}^{b}_{1,l}(x) 
	&:= \widehat{g}(x) - cv^{1,b}_{n}(1-\alpha)\cdot \widehat{\sigma}(x),
	\qquad x \in \mathcal{X},\\
	\label{band_4}
	\widehat{g}^{b}_{1,u}(x) 
	&:= \widehat{g}(x) - cv^{1,b}_{n}(\alpha)\cdot \widehat{\sigma}(x),
	\qquad x \in \mathcal{X}.
\end{align}

Here, \eqref{t101} and \eqref{band_31} ensure that 
\(\sup_{x \in \mathcal{X}} \widehat{g}^{b}_{1,l}(x) \leq 0\) 
with asymptotic confidence level \(100(1-\alpha)\%\) under 
\(\mathcal{H}_0: g_\ast(x) \leq 0 \;\;\forall x \in \mathcal{X}\). 
Similarly, \eqref{t11} and \eqref{band_4} ensure that 
\(\sup_{x \in \mathcal{X}} \widehat{g}^{b}_{1,u}(x) \geq 0\) 
with asymptotic confidence level \(100(1-\alpha)\%\) under 
\(\mathcal{H}_0: g_\ast(x) \geq 0 \;\;\forall x \in \mathcal{X}\). 
This result is formalized in the theorem below.

\begin{thm}[Uniform Inference via One-Sided UCB]\label{thm111}
In addition to the conditions of Theorem \ref{thm6}, suppose that Assumption \ref{as:nuisance1} holds. 
With the critical values $cv^{1,b}_{n}(1-\alpha)$ constructed in \eqref{t101}, the one-sided lower UCB defined in \eqref{band_31} satisfies
		\begin{align}
			&\label{kpi2321}\mathbb{P}\left\{\widehat{g}^{b}_{1,l}(x)\leq 0~\text{for all }x\in\mathcal{X}\right\}\geq 1-\alpha+o\left(1\right)~~~\text{under}~~\mathcal{H}_0:g_\ast(x)\leq 0 \text{ for all } x \in \mathcal{X}.
		\end{align}
Similarly, with the critical values $cv^{1,b}_{n}\left(\alpha\right)$ constructed in \eqref{t11}, the one-sided upper UCB defined in \eqref{band_4} satisfies
		\begin{align}
			&\label{kpi233}\mathbb{P}\left\{\widehat{g}^{b}_{1,u}(x)\geq 0~\text{for all }x\in\mathcal{X}\right\}\geq 1-\alpha+o\left(1\right)~~~\text{under}~~\mathcal{H}_0:g_\ast(x)\geq 0 \text{ for all } x \in \mathcal{X}.
		\end{align}
\end{thm}

\section{Estimation and Inference for the Optimal Policy Value}\label{sec5}

Suppose that a policymaker is interested in making inference on the optimal policy value of the form
\begin{align*}
V_0 &:= \sup_{g \in \mathcal{G}} V(g), \quad \text{where} 
\\
V(g) &:= \mathbb{E}\!\left[ \psi_1(Z_i,\eta_0)\, \mathbf{1}\{g(X_i)\geq 0\} 
+ \psi_0(Z_i,\eta_0)\, \mathbf{1}\{g(X_i)<0\} \right],
\end{align*}
with 
$
\psi_1(z,\eta) := \overline{\psi}(z,\eta) - \psi^{+}(z,\eta)
$
and
$
\psi_0(z,\eta) := \overline{\psi}(z,\eta) - \psi^{-}(z,\eta)
$
for some function \(\overline{\psi}\).

Note that under this setup, the welfare function \(V\) can be rewritten as
\begin{align*}
V(g) = \mathbb{E}[\overline{\psi}(Z_i,\eta_0)] - L(g),
\end{align*}
where $L$ is defined in \eqref{eq:loss1}.
Therefore, the representative classification policy function \(g_\ast\) identified via the surrogate loss approach attains the maximum level of welfare within \(\mathcal{G}\); that is, \(V_0 = V(g_\ast)\). 
Although the representative policy function \(g_\ast\) is obtained by minimizing the surrogate risk \(L^\ast\) in \eqref{eq:surrogate1} to ensure point identification, this surrogate risk \(L^\ast\) does not coincide with the \emph{value} of the \emph{original} risk \(L\) in \eqref{eq:loss1}. 
To estimate the optimal value \(V_0\), we need to evaluate the \emph{original} risk \(L\), rather than the surrogate risk \(L^\ast\), at an estimate of \(g_\ast\).

For instance, in the context of Example~III discussed in Section~\ref{sec:welfare_max}, 
\(V_0\) corresponds to the optimal welfare value in \eqref{eq:welfare}, 
with the functions \(\psi_1\) and \(\psi_0\) defined in \eqref{eq:psi1psi0} 
and \(\overline{\psi}(z,\eta) = \psi_1^{+}(z,\eta) + \psi_0^{-}(z,\eta)\). 

\subsection{Estimation of the the Optimal Policy Value}\label{sec:estimation_optimal_welfare}

Let \(\widehat{g}\) denote the estimator of the representative classification policy function \(g_\ast\), as obtained in Section~\ref{sec:estimate_g}. To account for the estimation of \(\eta\), we employ a cross-fitting procedure to estimate the optimal policy value. Partition \([n] = \{1,\ldots,n\}\) into \(m\) folds \(I_{1}, \ldots, I_{m}\). Assume that the folds are approximately balanced, i.e., \(n_{(1)} \asymp \cdots \asymp n_{(m)}\), where \(n_{(\ell)} := |I_{(\ell)}|\), and that \(m\) is fixed. For a given fold \(\ell \in [m] = \{1,\ldots,m\}\), obtain an estimator \(\widehat{\eta}_{(\ell)}\) of \(\eta_0\) using the complementary subsample \(I_{(\ell)}^c\). We then estimate the optimal policy value \(V_0\) via the plug-in cross-fitted estimator
\begin{align}\label{value1}
	\widehat{V}_n(\widehat{g})
	:= \frac{1}{n} \sum_{\ell=1}^m \sum_{i \in I_{(\ell)}}
	\Big[
	\widehat{\psi}_{1,(\ell)}(Z_i)\,\mathbf{1}\{-\widehat{g}(X_i)\leq 0\}
	+ \widehat{\psi}_{0,(\ell)}(Z_i)\,\mathbf{1}\{\widehat{g}(X_i)<0\}
	\Big],
\end{align}
where \(\widehat{\psi}_{1,(\ell)}(Z_i) := \psi_1(Z_i,\widehat{\eta}_{(\ell)})\) and 
\(\widehat{\psi}_{0,(\ell)}(Z_i) := \psi_0(Z_i,\widehat{\eta}_{(\ell)})\) for all \(i \in I_{(\ell)}\) and each \(\ell \in [m]\).

In the context of Example~III in Section~\ref{sec:welfare_max}, the estimated weights take the form
\begin{align*}
\widehat{\psi}_{1,(\ell)}(Z_i) 
&= \frac{A_i\big(Y_i - \widehat{\mu}_{(\ell)}(1, X_i)\big)}{\widehat{\pi}_{(\ell)}(X_i)} 
+ \widehat{\mu}_{(\ell)}(1, X_i) \qquad\text{and}\\
\widehat{\psi}_{0,(\ell)}(Z_i) 
&= \frac{(1-A_i)\big(Y_i - \widehat{\mu}_{(\ell)}(0, X_i)\big)}{1-\widehat{\pi}_{(\ell)}(X_i)} 
+ \widehat{\mu}_{(\ell)}(0, X_i),
\end{align*}
where \(\widehat{\eta}_{(\ell)} = \big(\widehat{\mu}_{(\ell)}(1,\cdot),\, \widehat{\mu}_{(\ell)}(0,\cdot),\, \widehat{\pi}_{(\ell)}(\cdot)\big)\) 
denotes the estimator of the conditional mean functions \(\mu(1,\cdot)\) and \(\mu(0,\cdot)\), 
as well as the propensity score function \(\pi(\cdot)\), 
constructed using the complementary subsample \(I_{(\ell)}^c\) for all \(i \in I_{(\ell)}\).

\subsection{Asymptotic Properties for the Optimal Policy Value}\label{sec51}

The optimal policy value estimator \eqref{value1} is constructed via a plug-in rule using our estimator $\widehat{g}$ of the representative policy function $g_\ast$, as given in Section \ref{sec:estimate_g}. For valid inference, $\widehat{g}$ must satisfy conditions ensuring that both its estimation error and its surrogate misspecification error are dominated and hence do not affect the limiting distribution of the policy value estimator. Theorem \ref{thm6} shows that our estimator $\widehat{g}$ converges at the rate $\xi_{k}\sqrt{\log(k)/n}$. We verify below that, under a set of mild conditions, this nonparametric rate suffices to ensure its impact on the distribution of the policy value estimator is asymptotically negligible.

\begin{as}[Margin Condition]\label{as:psi1}
The random variable $U_i := \mathbb{E}[\psi_1(Z_i,\eta_0)|X_i] -\mathbb{E}[\psi_0(Z_i,\eta_0)|X_i]$ admits a continuous probability density function $f_U$ over a small neighborhood $[-\delta,\delta]$ of zero.
\end{as}
  
Assumption~\ref{as:psi1} corresponds to the standard \emph{margin condition} widely used in the empirical welfare maximization literature (see, for example, \cite{kitagawa2018should} and \cite{ponomarev2024lower}). 
 	This assumption requires the data to exhibit a ``low-noise'' property. Specifically, if the random variable $U_i$ has a continuous density around zero, then $\mathbb{P}(|U_i|<t) = O(t)$ for sufficiently small $t>0$, which further implies $\mathbb{E}[|U_i|1\{|U_i|<t\}] = O(t^2)$. Intuitively, when $|U_i|$ is small---meaning that the welfare difference between the two actions is nearly tied---the estimated optimal rule is more likely to misclassify the optimal decision, but each such misclassification incurs only a small welfare loss. If these near-tie events occur infrequently, as ensured by the margin condition, the expected welfare loss regarding the optimal rule is even smaller. So, the margin condition ensures that the population welfare functional $V(g)$ is not too sensitive to small deviations of $g$ from the optimal policy $g_\ast$.\footnote{Assumption~\ref{as:psi1} corresponds to the margin condition in \cite{kitagawa2018should}, where $\mathbb{P}(|U_i|<t)=O(t^\alpha)$ with $\alpha=1$. Our results remain valid under any distribution of $U_i$ satisfying $\mathbb{P}(|U_i|<t)=O(t^\alpha)$ for $\alpha\ge 1$, but generally fail when $\alpha<1$.}

For convenience in stating the next assumption on nuisance parameter estimation, define the \(L_2(\mathbb{P})\) norm by $\| f \|_{\mathbb{P},2} := \big(\mathbb{E}[\| f(X) \|^2]\big)^{1/2}$ where the norm \(\|\cdot\|\) inside the expectation denotes the Euclidean norm.

\begin{as}
\label{as:nuisance_2} 
\begin{enumerate}[(i)]
\item\label{as:nuisance_21} \textbf{(Nuisance Parameter Estimation)} $\Vert \widehat{\eta}_{(\ell)}-\eta_0\Vert_{\mathbb{P},2}=o_p(n^{-1/4})$ for each $\ell\in[m]$. 
Furthermore, with probability at least $1-\epsilon_n$ for $\epsilon_n=o(1)$, $\widehat{\eta}_{(\ell)}$ belongs to a shrinking neighborhood $\mathcal{T}_n$ of $\eta_0$ for each $\ell \in[m]$, such that
the following convergence results hold:
\begin{align*}
	&\sqrt{n} \sup _{\eta \in \mathcal{T}_n}\left\vert \mathbb{E} \left[\left(\psi_1(Z_i,\eta)-\psi_1(Z_i,\eta_0)\right)\mathbf{1}(-g_k(X_i)<0)\right]\right\vert=o(1),\\
	&\sqrt{n} \sup _{\eta \in \mathcal{T}_n}\left\vert \mathbb{E} \left[\left(\psi_0(Z_i,\eta)-\psi_0(Z_i,\eta_0)\right)\mathbf{1}(g_k(X_i)<0)\right]\right\vert=o(1),\\
	&\sqrt{n}\sup _{\eta \in \mathcal{T}_n}\left(\mathbb{E}\left\vert \left[\left(\psi_1(Z_i,\eta)-\psi_1(Z_i,\eta_0)\right)\mathbf{1}(-g_k(X_i)<0)\right]\right\vert^2\right)^{1 / 2}=o(1), \text{ and }\\
	&\sqrt{n}\sup _{\eta \in \mathcal{T}_n}\left(\mathbb{E}\left\vert \left[\left(\psi_0(Z_i,\eta)-\psi_0(Z_i,\eta_0)\right)\mathbf{1}(g_k(X_i)<0)\right]\right\vert^2\right)^{1 / 2}=o(1).
\end{align*}
\item\label{as:nuisance_22} \textbf{(Boundedness)} For $j=0,1$, and uniformly over the shrinking neighborhood $\mathcal{T}_n$ of $\eta_0$$, \mathbb{E}[\vert\psi_j(Z_i, \eta)\vert^q]$ is bounded for some $q > 4$, and $\mathbb E[\psi_j(Z_i, \eta)|X_i=x]$ is bounded over the support of $X$. 
\end{enumerate}
\end{as}

Assumption \ref{as:nuisance_2}.\eqref{as:nuisance_21} requires that the cross-fitted estimator \(\widehat{\eta}\) of the nuisance parameter induces asymptotically negligible estimation errors. 
This condition has been also employed by \citet[][Assumption 2]{athey2021policy} in the context of Example III, and can be achieved by employing a Neyman-orthogonal score, which ensures that first-step estimation errors are asymptotically negligible provided that \(\widehat{\eta}_{(\ell)}\) converges at the \(n^{-1/4}\) rate, rather than the parametric \(n^{-1/2}\) rate. Such convergence rates are attainable with a variety of machine learning methods. For example, the Lasso achieves the \(n^{-1/4}\) rate under approximate sparsity and restricted eigenvalue conditions \citep{bickelritovtsybakov2009,belloni2015some}. Random forests and boosted trees are valid provided the regression function is sufficiently smooth and sample splitting is employed \citep{buehlmannYu2002,wagerAthey2018}. Deep neural networks with controlled depth and bounded weight norms can approximate smooth functions at nearly minimax rates \citep{chenwhite1999,yarotsky2017,schmidtHieber2020}, among others. Assumption \ref{as:nuisance_2}.\eqref{as:nuisance_22} further imposes uniform bounds on the (conditional) expectations of weight functions, $\psi_1$ and $\psi_0$. Note that this condition allows for unbounded outcome variable $Y$ in the welfare maximization (Example III) and is also satisfied in the expected utility maximization with binary action (Example II) under the Condition 1 of \citet{elliott2013predicting}. It holds trivially in the maximum score estimation (Example I).

Under these rate conditions and the boundedness conditions, we obtain the following asymptotic normality result for our estimator of the optimal policy value.

\begin{thm}\label{thm8} 
In addition to the conditions for Theorem \ref{thm6}, suppose that Assumptions \ref{as:psi1} and \ref{as:nuisance_2} hold, and the sieve order $k$ satisfies $k^{4+\varepsilon}/n\lesssim1$ for some $\varepsilon>0$. Then, for any sequence of data generating processes $\mathbb{P}_n$, we have
	\begin{align*}
	\sqrt{n}\{\widehat{V}_n(\widehat{g})-V_0\}\rightarrow_d\mathcal{N}(0,\sigma^2_v),
	\end{align*}
	where $\sigma^2_v=\mathbb{E}[s_L(Z_i, \eta_0,g_\ast)^2]$ is the variance of the score $$s_L(Z_{i},\eta_0,g_\ast)=\psi_{1}(Z_i,\eta_0)\cdot\mathbf{1}(g_\ast(X_i)\geq0)+\psi_{0}(Z_i,\eta_0)\cdot\mathbf{1}(g_\ast(X_i)<0)-V_0.$$
\end{thm}

The asymptotic Gaussianity established in Theorem~\ref{thm8} enables inference on the optimal policy value \(V_0\) in the standard manner, i.e., using normal critical values such as 1.96. 
To approximate the asymptotic variance \(\sigma_v^2\) in practice, we introduce a bootstrap procedure in Section~\ref{sec52}.

While the proof in the appendix formally establishes the results of Theorem \ref{thm8}, we provide the intuition here, particularly illustrating why the estimation effect of our classification policy function estimator \(\widehat{g}\) as given in Section \ref{sec:estimate_g} is asymptotically negligible despite the fact that it appears in the indicator function and converges at a slower rate than the root-$n$ rate. Consider the decomposition:
\begin{align}
\sqrt{n}\big(\widehat{V}_n(\widehat{g}) - V_0\big)
&= \sqrt{n}\big(\widehat{V}_n(\widehat{g}) - \widehat{V}_n(g_k)\big) 
\label{eq:estimation_error_g_hat}\\
&+ \sqrt{n}\big(\widehat{V}_n(g_k) - V(g_k)\big) 
\label{eq:estimation_error_nuisance}\\
&+ \sqrt{n}\big(V(g_k) - V(g_\ast)\big)
\label{eq:approximation_error_sieve}\\
&+ \sqrt{n}\big(V(g_\ast) - V_0\big),
\label{eq:ht8}
\end{align}
where the term in \eqref{eq:estimation_error_g_hat} captures the estimation error of \(\widehat{g}\), 
the term in \eqref{eq:estimation_error_nuisance} collects the sampling error from \(\mathbb{E}_n\) and the estimation error of \(\widehat{\eta}\), the term \eqref{eq:approximation_error_sieve} indicates the sieve approximation error of \(g_k\) for \(g_\ast\), 
and the term in \eqref{eq:ht8} measures the identification error.
Theorem~\ref{thm1} implies that the identification error in \eqref{eq:ht8} is zero. 
The component in \eqref{eq:estimation_error_nuisance} corresponds to the standard semiparametric part and is asymptotically Gaussian under standard assumptions, in particular under ours. 
The remaining task is to show that the first component in \eqref{eq:estimation_error_g_hat} is asymptotically negligible. 
Since this argument is not entirely obvious, we provide the intuition here in the main text, while leaving the formal proof to the appendix.

The component in \eqref{eq:estimation_error_g_hat} in question can be further decomposed as
\begin{align}
\eqref{eq:estimation_error_g_hat}
&= \sqrt{n}\Big[\widehat{V}_n(\widehat{g}) - \widehat{V}_n(g_k) - \{V(\widehat{g}) - V(g_k)\}\Big]+\sqrt{n}[V(g_\ast)-V(g_k)]
\label{eq:estimation_error_g_hat1}\\
& + \sqrt{n}\{V(\widehat{g}) - V(g_\ast)\}.
\label{eq:estimation_error_g_hat2}
\end{align}
The two bracket groups on the right-hand side of \eqref{eq:estimation_error_g_hat1} can be shown to be asymptotically negligible by applying the maximal inequality for VC classes \citep[Lemma~A.3]{massart2006risk}, Sauer's Lemma \citep{lugosi2002pattern}, and the dominated sieve approximation errors. 
The term in \eqref{eq:estimation_error_g_hat2} is more delicate, since \(\widehat{g}\) converges only at the nonparametric rate as obtained in Theorem \ref{thm6}.
Nevertheless, this term is also negligible because its first-order effect vanishes due to the first-order condition for the optimality of \(g_\ast\) with respect to \(V\). As a result, the second-order term becomes the leading component, and its existence is guaranteed by the margin condition (Assumption \ref{as:psi1}). We further illustrate this point through numerical simulations in Section~\ref{sec:sim_welfare}. In particular, we show that the plug-in estimator $\widehat{V}_n(\widehat{g})$ and the oracle estimator $\widehat{V}_n(g_\ast)$ yield nearly identical confidence intervals.

In summary, the component in \eqref{eq:estimation_error_nuisance} drives the asymptotic normality in the same way as in standard semiparametric analysis, while all other components are asymptotically negligible. 
See Appendix \ref{sec:auxiliary_related}--\ref{sec:proof:welfare} for further details.
A couple of remarks are in order.

\begin{rem}\label{remark_new} Although the plug-in value estimator in \eqref{value1} is constructed using the estimator $\widehat{g}$ described in Section \ref{sec:estimate_g}, our inferential results do not rely on the specific form of this estimator. What is required for valid second-stage inference is that the first-stage policy function estimator converges sufficiently fast so that its estimation error becomes asymptotically negligible. Indeed, with slight modifications on the proof, we can show that any estimator $\widetilde{g}$ would be admissible if it converges uniformly over $\mathcal X$ to a limit $g_{\ast\ast}$ satisfying $V(g_{\ast\ast}) = V_0$ at a rate faster than $o_p(n^{-1/4})$. That is, our framework accommodates a broad class of first-stage methods, including $\ell_1$-penalized estimators \citep{athey2021policy}, parametric approaches based on the 0--1 loss \citep[e.g.,][]{kitagawa2018should}, and nonparametric CATE-based methods 
\citep[e.g.,][]{bhattacharya2012inferring,armstrong2023inference,park2025debiased}, provided they achieve the uniform convergence rate $o_p(n^{-1/4})$ to $\mathbb{E}[\psi_1(Z,\eta_0)\mid X=x] - \mathbb{E}[\psi_0(Z,\eta_0)\mid X=x]$.
Thus, while our construction employs the estimator $\widehat g$ for the representative classification policy $g_\ast$, the resulting limit theory supports valid inference for a wide range of first-stage estimators used in the policy learning literature.
\end{rem}

\begin{rem}\label{remark4}
Our optimal policy value estimation is related to semiparametric estimation with a non-smooth criterion \citep{chen2003estimation}. 
Our framework extends this by allowing not only non-smooth but also discontinuous criteria. 
Even when the loss, and hence the sample criterion, is discontinuous, the asymptotically Gaussian term dominates provided that the population criterion is smooth, as illustrated in the intuition above. 
In particular, the first-order condition for the optimality of \(g_\ast\) plays a crucial role in ensuring that the potentially problematic term in \eqref{eq:estimation_error_g_hat2} is in fact negligible.
$\blacktriangle$
\end{rem}

\begin{rem}\label{remark5}
\citet[Section~2.4]{hirano2012impossibility} highlight that non-differentiability in a population criterion function prevents regular estimation. 
In particular, such an impossibility phenomenon arises in estimating \(\phi(\theta) = \max\{\theta_1, \theta_2\}\), which is non-differentiable at the kink point \(\theta_1 = \theta_2\). 
This kink causes the plug-in estimator \(\max\{\widehat{\theta}_1, \widehat{\theta}_2\}\) to have a non-Gaussian distribution when \(\theta_1 = \theta_2\), even if \((\widehat{\theta}_1,\widehat{\theta}_2)\) were Gaussian and centered at \((\theta_1,\theta_2)\).

In contrast, our welfare criterion can be rewritten as
\begin{align*}
\sup_{g \in \mathcal{G}} V(g)
&=
\sup_{g \in \mathcal{G}} 
\mathbb{E}\!\left[ \mathbb{E}[\psi_1(Z_i,\eta_0)\mid X_i] \cdot \mathbf{1}\{g(X_i)\geq0\} 
+ \mathbb{E}[\psi_0(Z_i,\eta_0)\mid X_i] \cdot \mathbf{1}\{g(X_i)<0\} \right] \\
&= \mathbb{E}\!\left[ \max \left\{ \vartheta_1, \vartheta_2 \right\} \right],
\end{align*}
rather than \(\max\{\theta_1,\theta_2\}\), where 
\(\vartheta_1 := \mathbb{E}[\psi_1(Z_i,\eta_0)\mid X_i]\) 
and 
\(\vartheta_2 := \mathbb{E}[\psi_0(Z_i,\eta_0)\mid X_i]\) 
are random variables.
Hence, the two problems differ mathematically. 
As emphasized in Remark~\ref{remark4}, given the smoothness of the population objective, asymptotic normality is attainable even with a non-smooth sample criterion, with the first-order condition for the optimality of \(g_\ast\) playing the crucial role.
$\blacktriangle$
\end{rem}

\subsection{Bootstrap Inference for the Optimal Policy Value}\label{sec52}
Theorem~\ref{thm8} facilitates standard inference based on Gaussian critical values, e.g., \(\approx 1.96\). 
We test the null hypothesis \(\mathcal{H}_0: V_0 = v_0\) using the statistic
\begin{align*}
	T_{g^{\ast},L} := \sqrt{n}\big(\widehat{V}_n(\widehat{g}) - v_0\big).
\end{align*}
To approximate the variance \(\sigma_v^2\) of its Gaussian limit, we employ the bootstrap counterpart
\begin{align*}
	\widetilde{Z}_n := \mathbb{G}_n \big[\,\widehat{s}_L(Z_i, \widehat{\eta}_{(\ell(i))}, \widehat{g}) \cdot \delta_i\,\big],
\end{align*}
where the estimated score takes the form of
\begin{align*}
\widehat{s}_L(Z_{i},\widehat{\eta}_{(\ell(i))},\widehat{g})=\psi_{1}(Z_i,\widehat{\eta}_{(\ell(i))})\cdot\mathbf{1}(\widehat{g}(X_i)\geq 0)+\psi_{0}(Z_i,\widehat{\eta}_{(\ell(i))})\cdot\mathbf{1}(\widehat{g}(X_i)<0)-\widehat{V}_n(\widehat{g}),
\end{align*}
\(\widehat{\eta}_{(\ell(i))}\) denotes the nuisance parameter estimator based on the complementary subsample 
\(I^c_{(\ell(i))}\) of the fold \(I_{(\ell(i))}\) that contains observation \(i\), and the perturbation process \(\{\delta_i\}_{i \in [n]}\) is a sequence of i.i.d. standard Gaussian bootstrap weights independent of the data.

The following theorem establishes the consistency of the score bootstrap method for approximating the limiting Gaussian distribution of the estimated optimal policy value.

\begin{thm}\label{thm9}
In addition to the conditions of Theorem \ref{thm8}, suppose that the perturbation process $\{\delta_i\}_{i\in[n]}$ is a sequence of i.i.d. standard Gaussian bootstrap weights independent of the data. Given the data generating process $\mathbb{P}_n$, the bootstrapped score converges in distribution to a normal distribution:
	\begin{align*}
	\widetilde{Z}_n\rightarrow_d\mathcal{N}(0,\sigma^2_v),
	\end{align*}
	where $\sigma^2_v$ was defined in Theorem \ref{thm8}.
\end{thm}

For a fixed \(\alpha \in (0,1)\), let \(\widehat{c}_{1-\alpha}\) denote the \((1-\alpha)\)-quantile of \(\widetilde Z_n\). 
A consequence of Theorem~\ref{thm9} is the asymptotic coverage property
\begin{align*}
	\lim_{n \to \infty} \mathbb{P}\!\left(\sqrt{n}\big(\widehat{V}_n(\widehat{g}) - v_0\big) \leq \widehat{c}_{1-\alpha}\right) = 1-\alpha,
\end{align*}
under the null hypothesis \(\mathcal{H}_0: V_0 \leq v_0\) against alternative hypothesis \(\mathcal{H}_1: V_0>v_0\).

\section{Simulation Studies}\label{sec_sim}

In this section, we use numerical simulations to evaluate the finite-sample performance of our proposed inference methods for cost-sensitive binary classification. 
As an illustrative example, we focus on the welfare maximization problem (Example~III) introduced in Section~\ref{sec:welfare_max}. 
Within this context, we examine both the performance of our proposed uniform inference procedure for the optimal treatment assignment (classification) policy and the inference method for the optimal welfare (policy value).

The outcome is modeled as
$
Y_i = A_i \cdot \Delta(X_i) + S(X_i) + u_i,
$
where the idiosyncratic error term is generated as \(u_i \stackrel{\text{i.i.d.}}{\sim} \mathcal{N}(0,1)\). 
The binary treatment assignment is given by
$
A_i = 2 \cdot \text{Bernoulli}(\pi(X_i)) - 1,
$
so that \(A_i \in \{-1,1\}\). 
The treatment effect is specified as
$
\Delta(X_i) = \tanh\!\big((1, X_i^\prime)\gamma\big),
$
the baseline function as
$
S(X_i) = \sin(X_i^\prime \beta_S),
$
and the propensity score as
$
\pi(X_i) = \frac{\exp(X_i^\prime \beta_\pi)}{1 + \exp(X_i^\prime \beta_\pi)}.
$

We consider a nonlinear design featuring two covariates.
The first covariate, \(X_{i1} \sim \text{Uniform}[0,1]\), represents normalized pre-treatment income. 
The second covariate, \(X_{i2}\), is constructed by rescaling a discrete variable that mimics years of education. 
Specifically, for \(X_{i2}\), we draw a categorical variable taking values in \(\{7, \ldots, 18\}\) with probabilities calibrated to the empirical distribution of education levels in the JTPA data, and divide by 18 to ensure support on the unit interval.

The parameter values of \(\gamma\) are set to be consistent with the null hypotheses under consideration. 
For the one-sided null hypothesis \(\mathcal{H}_0: g_\ast(x) \leq 0 \;\; \forall x \in \mathcal{X}\), we consider the setting 
\(\gamma = (0, -1/\sqrt{n}, -1/\sqrt{n})^\prime\), with \(\beta_S = (-1, 1)^\prime\) and \(\beta_\pi = (1, -1)^\prime\). 
For the null \(\mathcal{H}_0: g_\ast(x) \geq 0 \;\; \forall x \in \mathcal{X}\), on the other hand, we instead use the setting 
\(\gamma = (0, 1/\sqrt{n}, 1/\sqrt{n})^\prime\), with \(\beta_S = (-1, 1)^\prime\) and \(\beta_\pi = (1, -1)^\prime\). 
Extensive simulations are conducted with sample sizes \(n \in \{250, 500, 1000\}\), and each configuration is replicated \(S = 1,000\) times through Monte Carlo iterations. 
The number of the cross-fitting folds is set to $m=2$ both for estimating the policy and welfare.

For nuisance parameters, the simulation design incorporates the conditional mean outcome functions \(\mu(a; x)\) for \(a \in \{-1, 1\}\), estimated via Lasso, and the propensity score function \(\pi(x)\), estimated via \(\ell_1\)-penalized logistic regression. 
The tuning parameters for both procedures are chosen by five-fold cross-validation.

The representative classification policy function is estimated using a sieve approximation based on a tensor-product Legendre polynomial basis. 
Specifically, for \(k \in \{2,3,4\}\), we employ the tensor-product basis constructed from the orthonormal Legendre polynomials \(\{p_j(x)\}_{j=0}^\infty\) of $k$-dimensions defined on the support \(\mathcal{X}\).
The sieve basis is given by
$
p_{1:k}(x_1, x_2) = \big\{\, p_j(x_1)\, p_l(x_2) : j, l \in [k] \,\big\},
$
and the policy function is approximated as
$
g_k(x_1, x_2) = \sum_{j=1}^{k} \sum_{l=1}^{k} \beta_{jl}\, p_j(x_1) p_l(x_2),
$
or, equivalently,
$
g_k(x_1, x_2) = p_{1:k}(x_1, x_2)^\prime \beta.
$

To remain consistent with the theoretical conditions required for the asymptotic analysis, the simulation study focuses exclusively on twice continuously differentiable surrogate loss functions. 
We consider the following three examples:
\begin{enumerate}[(a)]
	\item Logistic loss: $\phi(x) = \log(1 + e^{-x})$;
	\item Exponential loss: $\phi(x) = e^{-x}$; \text{ and }
	\item Squared loss: $\phi(x) = (1 - x)^2$.
\end{enumerate}

\subsection{Uniform Inference for Optimal Policy}

This subsection presents the finite-sample performance of our proposed uniform inference method for the optimal policy rule. 
We consider the two null hypotheses: 
\(\mathcal{H}_0: g_{\ast}(x) \leq 0 \;\; \forall x \in \mathcal{C}\) 
and 
\(\mathcal{H}_0: g_{\ast}(x) > 0 \;\; \forall x \in \mathcal{C}\),  
where the set \(\mathcal{C}\) is defined as 
\(\mathcal{C} := [0.05, 0.95] \times \{10\}\).

\begin{table}[!t]
	\centering
	\scalebox{1}{
		\begin{tabular}{c|ccc|ccc|ccc}
			\multicolumn{10}{c}{(I) $\mathcal{H}_0$: $g_\ast(x)\leq 0$ for all $x\in\mathcal{X}$ with $\gamma = (0, -1/\sqrt{n}, -1/\sqrt{n})^\prime$}\\
			\hline
			\multirow{2}{*}{Sample Size}  & \multicolumn{3}{c|}{Logistic Loss} 
			& \multicolumn{3}{c|}{Exponential Loss} 
			& \multicolumn{3}{c}{Squared Loss} \\
			& $k=2$ & $k=3$ & $k=4$
			& $k=2$ & $k=3$ & $k=4$
			& $k=2$ & $k=3$ & $k=4$ \\
			\hline
			250  & 0.977 & 0.992 & 0.994 & 0.971 & 0.979 & 0.975 & 0.971 & 0.979 & 0.975 \\
			500  & 0.987 & 0.995 & 0.999 & 0.977 & 0.994 & 0.996 & 0.977 & 0.994 & 0.996 \\
			1,000 & 0.993 & 0.998 & 0.999 & 0.988 & 0.997 & 0.995 & 0.988 & 0.997 & 0.995 \\
			\hline
			\multicolumn{10}{c}{}
	\end{tabular}}
	
	\scalebox{1}{
		\begin{tabular}{c|ccc|ccc|ccc}
			\multicolumn{10}{c}{(II) $\mathcal{H}_0$: $g_\ast(x)\geq 0$ for any $x\in\mathcal{X}$ with $\gamma = (0, 1/\sqrt{n}, 1/\sqrt{n})^\prime$}\\
			\hline
		\multirow{2}{*}{Sample Size}  & \multicolumn{3}{c|}{Logistic Loss} 
		& \multicolumn{3}{c|}{Exponential Loss} 
		& \multicolumn{3}{c}{Squared Loss} \\
		& $k=2$ & $k=3$ & $k=4$ 
		& $k=2$ & $k=3$ & $k=4$ 
		& $k=2$ & $k=3$ & $k=4$ \\
		\hline
		250  & 0.977 & 0.989 & 0.995 & 0.965 & 0.971 & 0.974 & 0.965 & 0.971 & 0.974 \\
		500  & 0.992 & 0.995 & 0.996 & 0.985 & 0.988 & 0.992 & 0.985 & 0.988 & 0.992 \\
		1,000 & 0.992 & 0.997 & 0.998 & 0.991 & 0.997 & 0.995 & 0.991 & 0.997 & 0.995 \\
		\hline

	\end{tabular}}
	\caption{Empirical size performance of one-sided uniform confidence bands for the optimal policy rule.}
	\label{tab:ucb1}
\end{table}

Table~\ref{tab:ucb1} reports the finite-sample uniform coverage frequencies based on the one-sided tests across sample sizes \(n \in \{100, 250, 500\}\) and the three surrogate loss functions. 
The top panel (I) corresponds to the null hypothesis \(\mathcal{H}_0: g_{\ast}(x) \leq 0 \;\; \forall x \in \mathcal{C}\), 
while the bottom panel (II) corresponds to the null hypothesis \(\mathcal{H}_0: g_{\ast}(x) > 0 \;\; \forall x \in \mathcal{C}\). 

The results confirm that our bootstrap uniform confidence bands successfully control the test size. 
The size performance is slightly conservative, which is attributable to the fact that the true data-generating functions lie in the interior of the null, far from the least favorable boundary points of the null.

\subsection{Gaussian Limit and Inference for the Optimal Welfare}\label{sec:sim_welfare}

This subsection demonstrates the Gaussian limit of our estimated welfare estimator via numerical studies. 
In addition to (a) the plug-in estimated optimal policy rule, we also study the Gaussian limit theory of welfare under three comparative benchmarks: (b) an oracle treatment assignment rule, (c) a treat-everyone policy, and (d) a random-design policy. 
The random-design policy (d) assigns treatment according to an independent Bernoulli\((0.5)\) distribution.

\begin{figure}[!t] \centering 
	\subfigure[Estimated optimal welfare]{\label{fig_01a}
		\includegraphics[height=0.4\columnwidth]{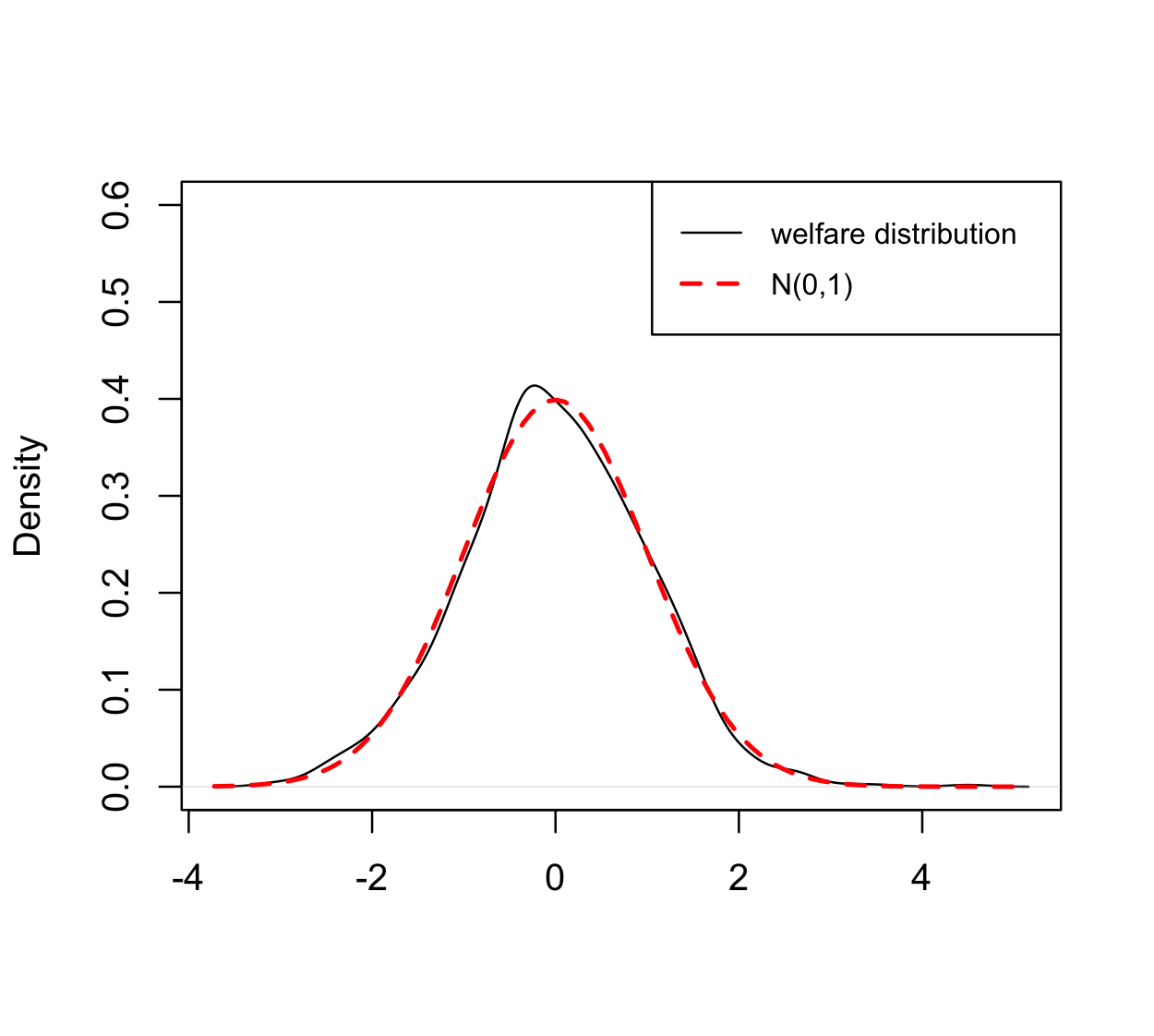} }
	\subfigure[Oracle optimal welfare]{
		\includegraphics[height=0.4\columnwidth]{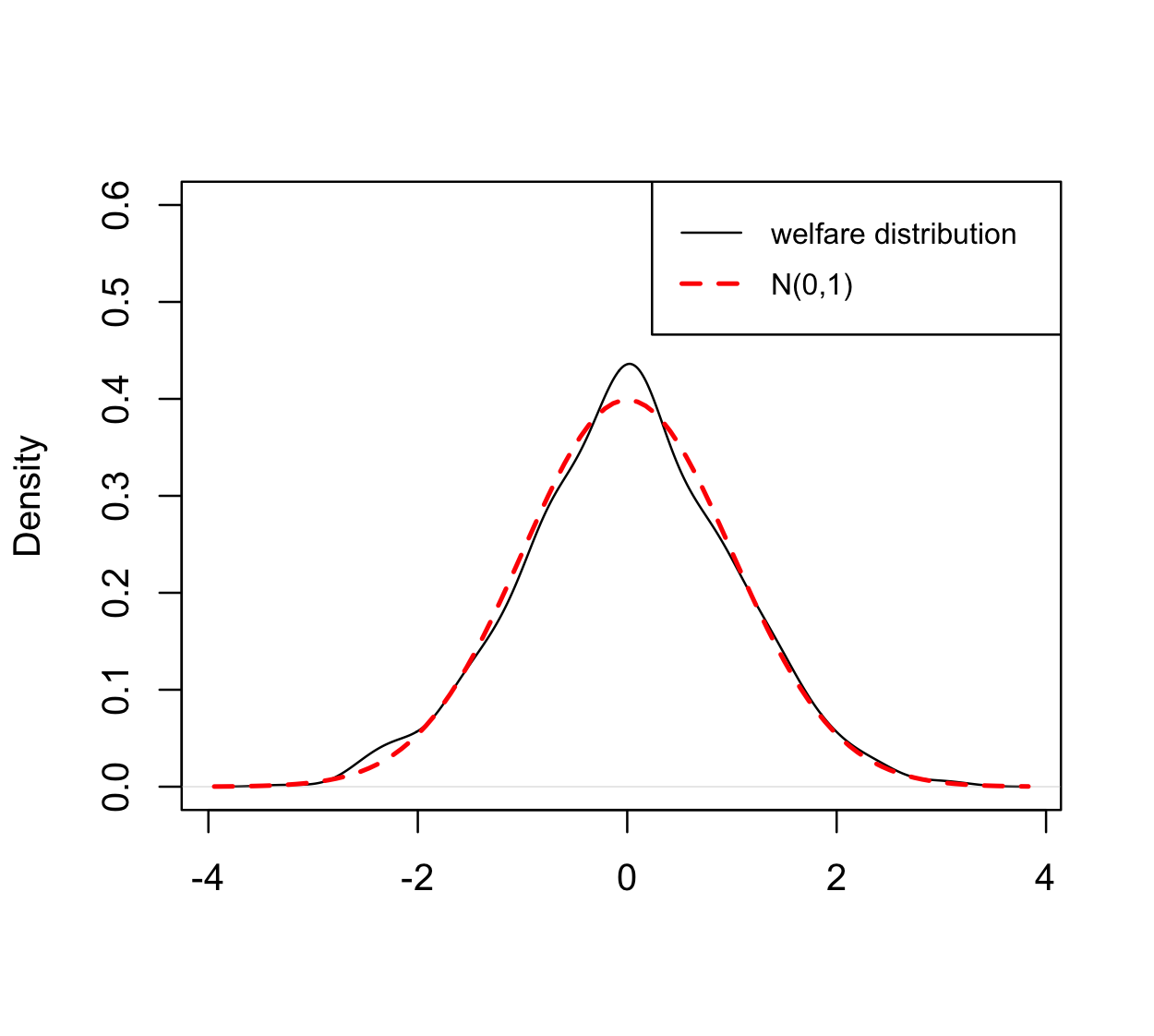}} \\
	\subfigure[Welfare with everyone treated]{
		\includegraphics[height=0.4\columnwidth]{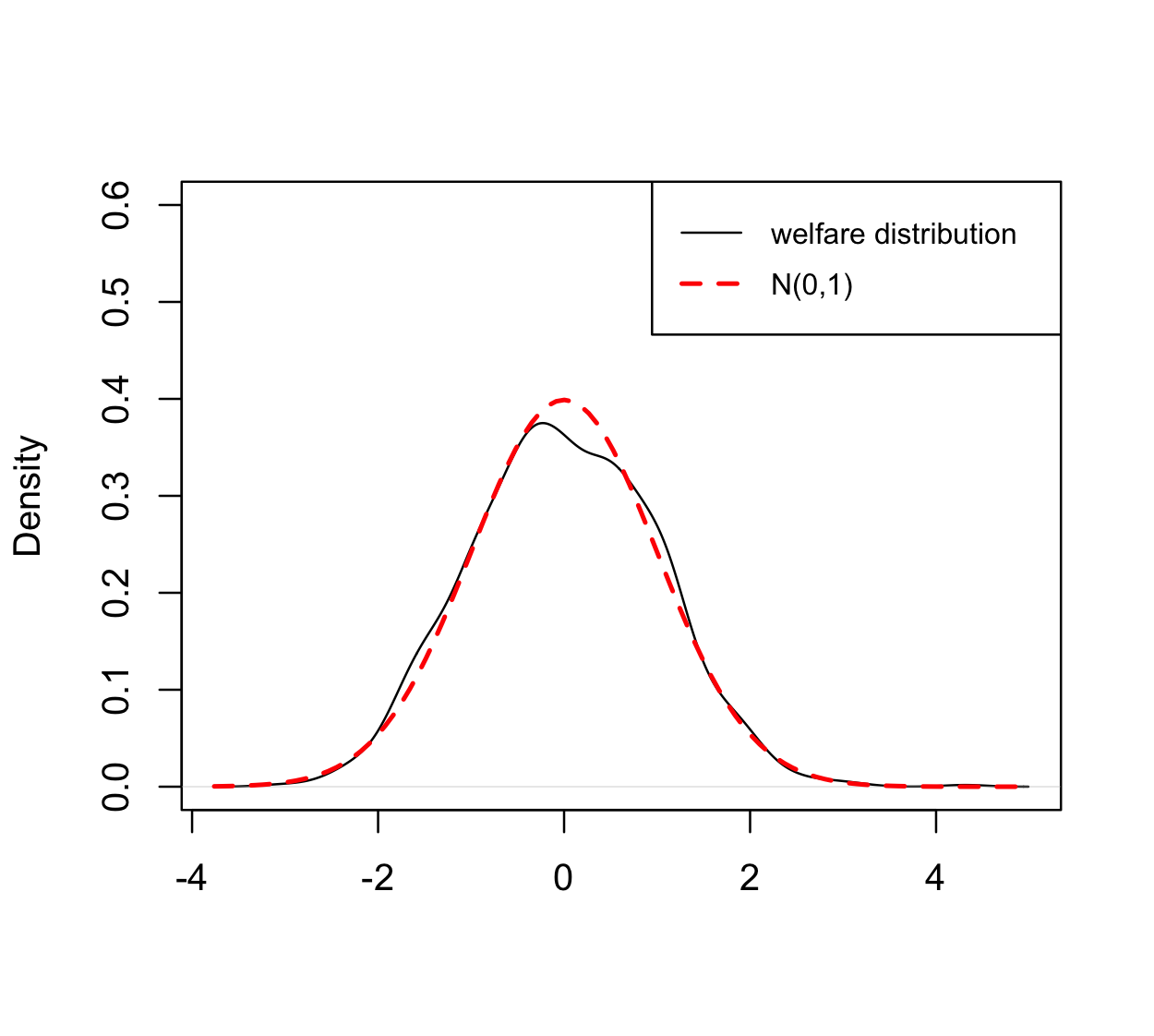}} 
	\subfigure[Welfare with random assign]{
		\includegraphics[height=0.4\columnwidth]{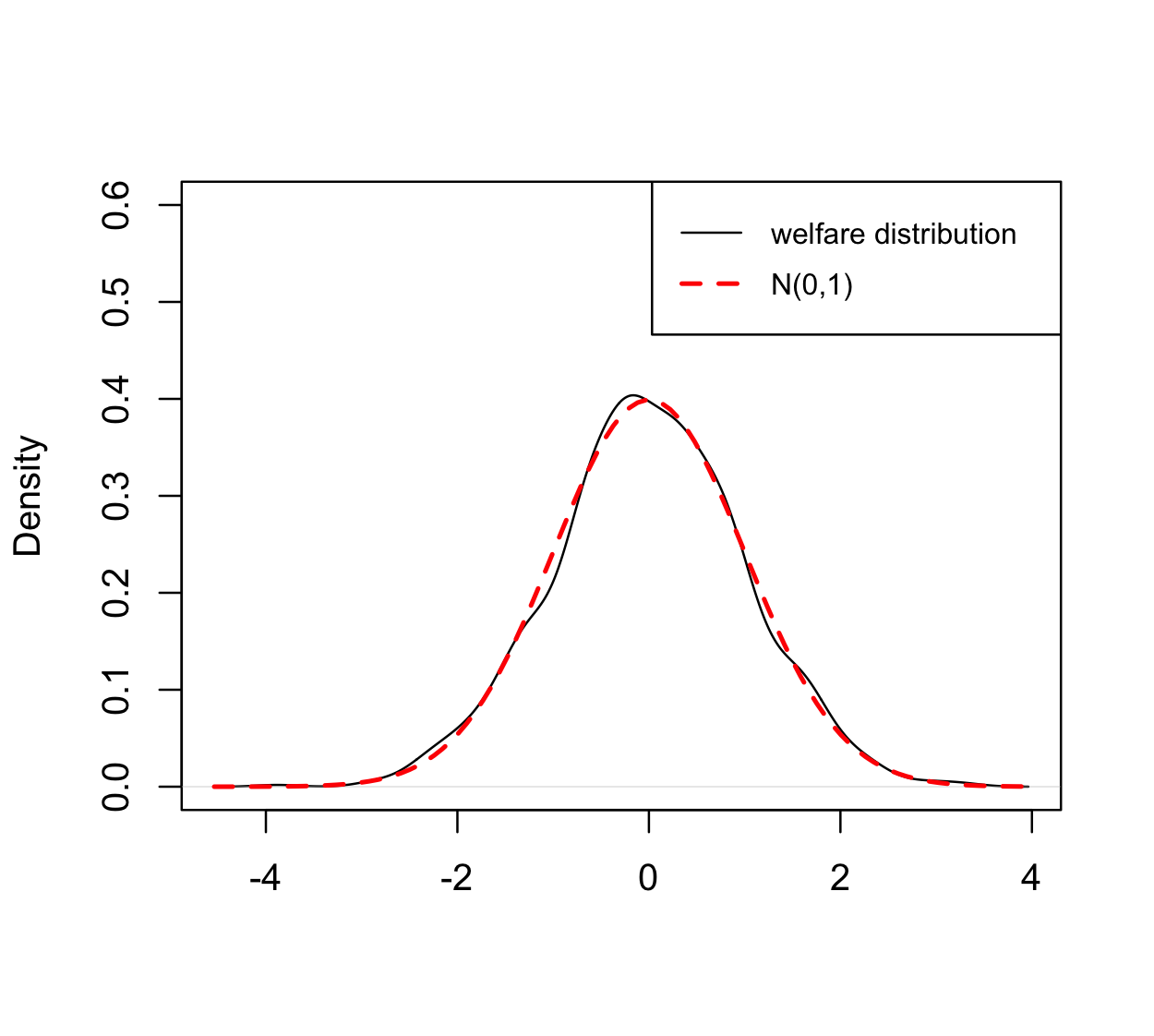} }
	\caption{Empirical densities of normalized welfare and standard Gaussianity ($n=500$ and $d_x=2$).}${}$ \label{fig_01}
\end{figure}		

For each of the four scenarios (a)--(d), we normalize the estimated optimal welfare using the Monte Carlo variance estimate computed across 1,000 simulation replications. 
We then compare the simulated densities of these normalized estimates with that of the standard normal distribution for each scenario. 
The sample size is fixed at \(n=500\) throughout. 
Figure~\ref{fig_01} displays the simulation density of the estimated optimal welfare in solid black, alongside the standard normal density in dashed red.

Each plotted density from (a) to (d) closely resembles the standard normal distribution.
In particular, the result in panel (a) for the plug-in estimated optimal welfare corroborates the theoretical property of Gaussian limit behavior established in Theorem~\ref{thm9}. 
Panel (a) is comparable to panel (b) for the oracle counterpart, and this observation is consistent with the asymptotic negligibility of the preliminary representative classification policy estimation emphasized in Section~\ref{sec51}. 
Although we report one set of results based on the logistic loss, these findings hold robustly across different surrogate loss functions.

\begin{figure}[!t] \centering 
	\includegraphics[height=0.5\columnwidth]{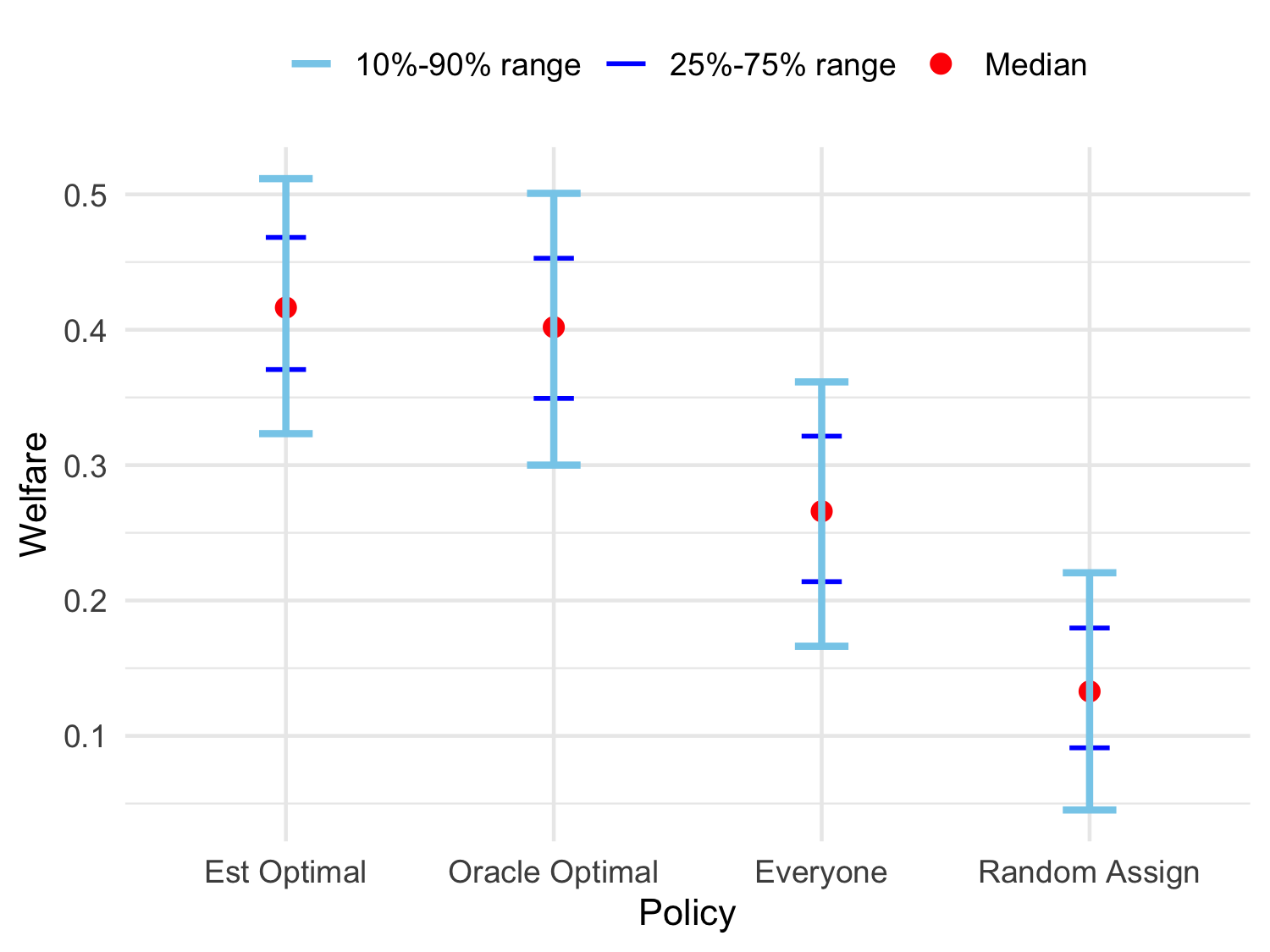} 
	\caption{Median, interquartile range, and interdecile range of the empirical welfare estimates 
under the estimated optimal, oracle optimal, treat-all, and random assignment rules.}
\label{fig_1}
\end{figure}

Figure~\ref{fig_1} reports the median, interquartile, and interdecile ranges of empirical welfare estimates under four treatment assignment rules:
	(a) the plug-in optimal policy estimated using the logistic loss (\emph{Est. Optimal});
	(b) the oracle assignment rule (\emph{Oracle Optimal});
	(c) the treat-everyone rule (\emph{Everyone}); and
	(d) the random assignment rule with estimated variance (\emph{Random Assign}).
The results are based on $n = 500$ observations and $S = 1{,}000$ Monte Carlo replications. 
The distribution under the estimated optimal policy closely resembles that under the oracle rule, 
corroborating the asymptotic negligibility of $\widehat g$ for 
$\sqrt{n}\bigl(\widehat V_n(\widehat g) - V_0\bigr)$ as argued in Section~\ref{sec51}. 
Moreover, the welfare estimates under both the estimated optimal policy and the oracle policy 
substantially exceed those attained under the treat-everyone and random assignment rules.

Finally, we conduct hypothesis tests to compare the welfare achieved under the estimated optimal policy to that under the benchmark policies $g_\dagger$.\footnote{See Appendix~\ref{ap:sec2} for details on this testing procedure.} Specifically, we perform a two-sided test of $\mathcal{H}_0: V_0 = V(g_\dagger)$ against $\mathcal{H}_1: V_0 \neq V(g_\dagger)$, as well as one-sided tests of $\mathcal{H}_0: V_0 \leq V(g_\dagger)$ versus $\mathcal{H}_1: V_0 > V(g_\dagger)$ and $\mathcal{H}_0: V_0 \geq V(g_\dagger)$ versus $\mathcal{H}_1: V_0 < V(g_\dagger)$. Table~\ref{tab:rej} reports the Monte Carlo rejection frequencies. Both the two-sided and right-sided tests are rejected in most Monte Carlo draws, indicating that the estimated optimal policy delivers a significantly higher level of welfare.

\begin{table}[!t]
	\centering
	\scalebox{1}{
		\begin{tabular}{lccc}
			\hline
			& \multicolumn{2}{c}{Benchmark policy $g_\dagger$}\\
			\cline{2-3}
			& Treat Everyone & Random Treat \\
			\hline
			Two-sided test & 0.790 & 0.990 \\
			Left-sided test & 0.000 & 0.000 \\
			Right-sided test & 0.881 & 0.996 \\
			\hline
	\end{tabular}}
	\caption{Monte Carlo rejection frequencies from hypothesis tests comparing the welfare under the optimal optimal policy to that under benchmark policies.}
	\label{tab:rej}
\end{table}

\section{Empirical Analysis}\label{sec_emp}

With the ``identification'', estimation, and inference methods developed in this paper, 
we now revisit the problem of policy learning using the experimental data from the National JTPA Study. 
While prior studies such as \citet{kitagawa2018should} and \citet{mbakop2021model} have analyzed this problem, 
they did not conduct formal inference for the optimal treatment assignment (classification) policy or its associated welfare (policy value). 
Our goal is to build on this literature and provide inference for these important features.

The JTPA study randomly assigned applicants to be eligible for a job training program for a period of 18 months. 
It collected background information on the applicants prior to random assignment, 
as well as administrative and survey data on their earnings over the 30 months following the assignment. 
The dataset consists of 11,008 observations from the adult sample (ages 22 and older).

Following benchmark studies in the literature, we adopt the following variable definitions. 
The outcome variable, \(Y_i\), is defined in two ways: 
(i) total individual earnings during the 30 months following program assignment, or 
(ii) total individual earnings during the 30 months following program assignment minus the average cost of services per treatment assignment (\$774). 
The covariates, \(X_i\), used to define the policy class are \emph{years of education} and \emph{pre-program earnings}.\footnote{For both the outcome variable (total individual earnings over the 30-month period) and the covariate (pre-program earnings), we divide their values by 100.}

We consider two types of policy functions: 
(a) a linear rule \(g(x) = \gamma^{\prime}x\); and 
(b) a class of nonlinear functions defined as follows:
Let \(\mathcal{X}_1\) denote the covariate space of \emph{years of education}, 
and let \(\mathcal{X}_2\) denote the covariate space of \emph{pre-program earnings}. 
The set of nonlinear assignment rules is given by
\[
\mathcal{G} = \bigl\{\, g(x_1, x_2) = f(x_1) - x_2 \;\big|\; f: \mathcal{X}_1 \to \mathcal{X}_2 \,\bigr\}.
\]

Linear and nonlinear rules have been studied in the literature \citep{kitagawa2018should, mbakop2021model}. 
In particular, while the nonlinear class follows \citet{mbakop2021model}, we allow for a more general specification without imposing their sign restriction ($f(x_1) \geq x_2$). 
We can accommodate this generality by leveraging our method’s ability to achieve nonparametric point ``identification,'' along with the associated estimation and inference procedures.
That said, the main contribution of this paper lies in its inference component, 
which has not been addressed by these preceding papers.

The optimal policy function is approximated using a sieve expansion based on tensor-product Legendre polynomial basis functions with sieve order \(k = 3\). 
The number of folds is set to \(m = 2\). 
The nuisance functions are estimated via machine learning methods: 
the propensity score is obtained using \(\ell_1\)-penalized logistic regression, 
and the conditional mean outcome is estimated using Lasso. 
In both cases, tuning parameters are selected via cross-validation. 
For inference, we implement a score-based bootstrap procedure with 1,000 repetitions 
to construct confidence bands for the optimal policy 
and a confidence interval for its corresponding welfare.

To ensure that the approximation error of Legendre polynomials diminishes in the \(L_\infty\) space, 
we normalize both covariates to lie within the unit interval \([0,1]\). 
Specifically, we define \(x_1\) as the \emph{years of education} divided by its maximum value, 
and \(x_2\) as normalized pre-treatment earnings, computed as
\[
x_2 = \frac{\text{pre-treatment earnings} - \min(\text{pre-treatment earnings})}
{\max(\text{pre-treatment earnings}) - \min(\text{pre-treatment earnings})}.
\]
Similarly, we normalize the outcome variable \(Y_i\) using the same min--max transformation:
\[
Y_i = \frac{\text{earnings}_i - \min(\text{earnings})}
{\max(\text{earnings}) - \min(\text{earnings})}.
\]

The surrogate loss function used in our analysis is the logistic loss. 
Our objective is to construct a left-sided uniform confidence band for \eqref{band_31} 
to test the null hypothesis 
\[
\mathcal{H}_0:\; g_\ast(x_1,x_2) \leq 0,
\]
over \(x_1 \in [7/18, 1]\), with \(x_2\) fixed at the 25th, 50th, and 75th quantiles of its empirical distribution.

\begin{figure}[!htbp] \centering 
	\subfigure[No Training Cost \& Linear Rule]{\label{fig_ucb_1}
		\includegraphics[height=0.4\columnwidth]{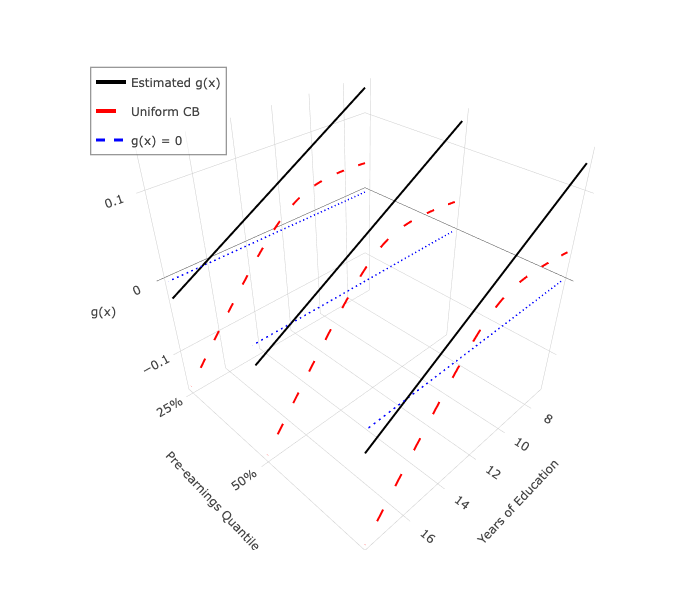} }
	\subfigure[No Training Cost \& Nonlinear Rule]{\label{fig_ucb_2}
		\includegraphics[height=0.4\columnwidth]{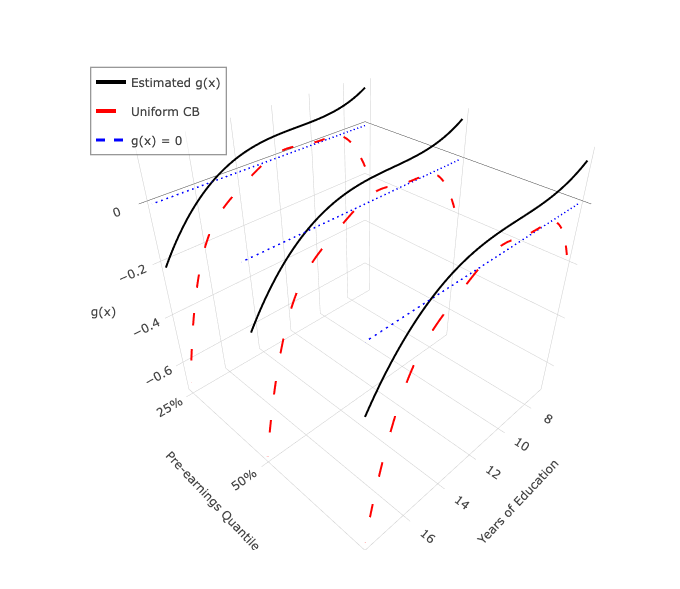} }
	\subfigure[Training Cost \& Linear Rule]{\label{fig_ucb_3}
		\includegraphics[height=0.4\columnwidth]{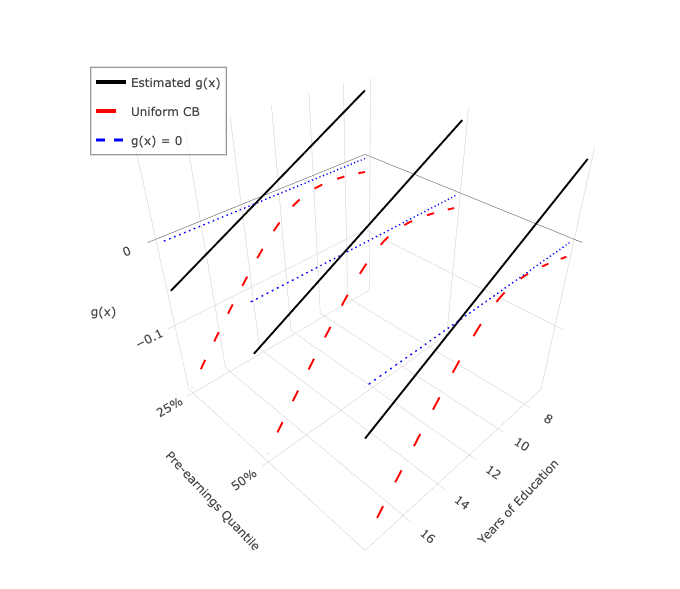} }
	\subfigure[Training Cost \& Nonlinear Rule]{\label{fig_ucb_4}
		\includegraphics[height=0.4\columnwidth]{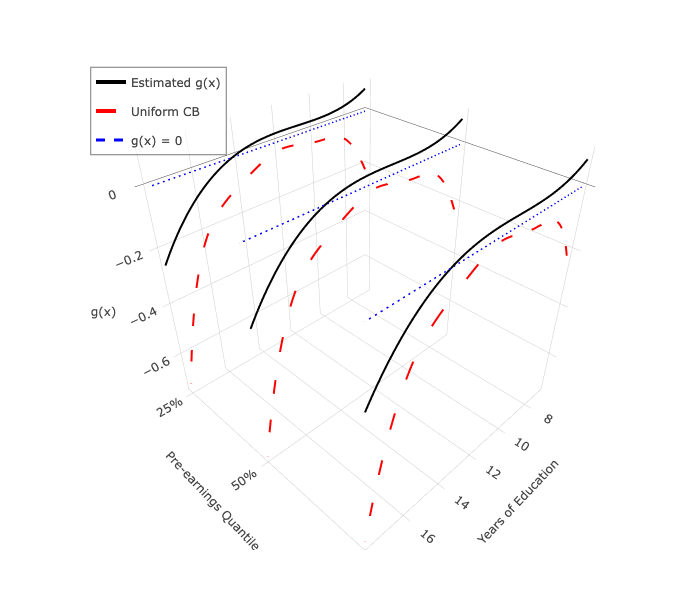} }
	\caption{$95\%$ uniform confidence bands for optimal policy rule $g_\ast(x)$. The estimated model is plotted in black dotted lines. The pointwise confidence sets that use the standard normal critical values are given in blue dotted lines. The bootstrapping uniform confidence bands are given in red dotted lines.} \label{fig_ucb}
\end{figure}	

First, we conduct inference results for the optimal policy rule. 
We compute the one-sided uniform confidence bands of \eqref{band_31} 
using the critical values obtained from the score bootstrap method described in Section \ref{sec4}. 
The sieve-based nonparametric estimators and their corresponding uniform confidence bands 
are computed for the four cases described above: 
(i) no training cost with a linear policy rule, 
(ii) no training cost with a nonlinear policy rule, 
(iii) training cost with a linear policy rule, and 
(iv) training cost with a nonlinear policy rule. 
For each case, we construct one-sided uniform confidence bands over \emph{years of education}, 
holding \emph{pre-program earnings} fixed at the 25th, 50th, and 75th quantiles.

Figure \ref{fig_ucb} presents the estimation and inference results for the optimal policy. 
When training costs are not considered and \emph{years of education} are below 12, 
Figures \ref{fig_ucb_1} and \ref{fig_ucb_2} show that the 95\% uniform confidence bands allow us 
to reject the uniform null hypothesis, indicating that assigning individuals in this range to the control group 
is statistically significant at the 5\% level. 
In contrast, when training costs are taken into account, this statistical significance disappears, 
as shown in Figures \ref{fig_ucb_3} and \ref{fig_ucb_4}. 
The economic intuition behind this finding is straightforward: 
as the cost of the National JTPA program increases, the net benefit of participation declines, 
reducing its appeal and thereby lowering the expected welfare for the representative individual. 
Overall, our inference method for the optimal policy rule complements 
the existing findings \citep[e.g.,][]{kitagawa2018should,mbakop2021model} 
by providing formal statistical inference as well as more general nonlinearity.

\begin{table}[!t]
	\centering
	\begin{tabular}{llccc}
		\hline
		& \multicolumn{4}{c}{\textbf{Outcome Variable: 30-Month Post-Program Earnings}} \\
		\cline{2-5}
		& & Optimal treat & Treat everyone & Random treat \\
		\hline
		\multicolumn{5}{l}{\textbf{No Treatment Cost}} \\
		Linear & Estimated Welfare Gain & \$1,285 & \$1,247 & \$502 \\
		& 95\% Confidence Interval & (\$901, \$1,668) & (\$867, \$1,626) & (\$16, \$988) \\
		Nonlinear & Estimated Welfare Gain & \$1,338 & \$1,247 & \$499 \\
		& 95\% Confidence Interval & (\$943, \$1,733) & (\$867, \$1,626) & (\$9, \$990) \\
		\hline
		\multicolumn{5}{l}{\textbf{\$774 Cost for Each Assigned Treatment}} \\
		Linear & Estimated Welfare Gain & \$533 & \$473 & \$111 \\
		& 95\% Confidence Interval & (\$120, \$946) & (\$93, \$852) & (-\$376, \$598) \\
		Nonlinear & Estimated Welfare Gain & \$566 & \$473 & \$109 \\
		& 95\% Confidence Interval & (\$160, \$972) & (\$93, \$852) & (-\$381, \$598) \\
		\hline
	\end{tabular}
	\caption{95\% confidence intervals of estimated welfare gain of treatment assignments}\label{tab:empirics}
\end{table}

We now report the estimation and inference results for (optimal) welfare. 
Table \ref{tab:empirics} presents the 95\% confidence intervals (CIs) for the estimated welfare under three different treatment assignment rules: 
(i) the estimated optimal policy rule based on the logistic surrogate loss, 
(ii) the policy that assigns all individuals to the treatment group, and 
(iii) the random assignment rule following a Bernoulli(0.5) distribution. 
Since Theorem \ref{thm9} establishes that the empirical welfare of a suitable plug-in treatment assignment rule is asymptotically normal, constructing CIs for welfare functions only requires appropriate variance estimates. 
Specifically, for the estimated optimal policy rule, we obtain the variance estimate using our score bootstrap procedure (see Section \ref{sec52}), while for the other two cases, we use the variance formula provided in Theorem \ref{thm8}. 

When linear rules are considered, the estimated welfare of the estimated optimal rule is 
\$1,285 in the case of no training cost and \$533 in the case of training costs. 
These findings are very similar to those reported by \citet[Table 1]{kitagawa2018should}, 
who obtain estimated welfare values of \$1,180 and \$404, respectively. 
While they do not develop a formal theory for inference, \citet[Table~1]{kitagawa2018should} also report confidence intervals based on bootstrap for this application.
Our confidence intervals for the two aforementioned cases, (\$901,~\$1{,}668) and (\$120,~\$946), are comparable to theirs but notably narrower, 
thereby offering greater precision and enhancing the reliability of statistical decisions.

Additionally, Table \ref{tab:empirics} compares the estimated optimal rule, 
the ``treat-everyone'' rule, and the random assignment rule across the cases. 
The estimated welfare of the random assignment rule lies outside the 95\% confidence intervals 
for the estimated optimal rule, clearly indicating that our estimated rule achieves higher welfare 
than the random assignment rule, which has been widely used in empirical economic research. 
Moreover, the estimated welfare of the optimal rule is consistently higher than that of the ``treat-everyone'' rule, 
further demonstrating the practical usefulness of our policy learning method.

\begin{table}[!t]
	\centering
	\scalebox{1}{
		\begin{tabular}{lccccc}
			\hline
			& \multicolumn{2}{c}{\ $k=1$} && \multicolumn{2}{c}{\ $k=3$} \\
			\cline{2-3}\cline{5-6}
			& Treat Everyone & Random Assign && Treat Everyone & Random Assign \\
			\hline
			\multicolumn{6}{l}{\textbf{No Treatment Cost}} \\
			Two-sided   & 0.514 & 0.000 && 0.282 & 0.000 \\
			Left-sided  & 0.766 & 1.000 && 0.877 & 1.000 \\
			Right-sided & 0.234 & 0.000 && 0.123 & 0.000 \\
			\hline
			\multicolumn{6}{l}{\textbf{\$774 Treatment Cost}} \\
			Two-sided   & 0.669 & 0.048 && 0.339 & 0.042 \\
			Left-sided  & 0.691 & 0.978 && 0.862 & 0.979 \\
			Right-sided & 0.309 & 0.022 && 0.138 & 0.021 \\
			\hline
	\end{tabular}}
	\caption{P-values from the hypothesis tests comparing the welfare under the optimal policy to that under benchmark policies.}
	\label{tab:pvalues}
\end{table}

Finally, we conduct hypothesis tests to compare the welfare achieved under the estimated optimal policy with that under the benchmark policies $g_\dagger \in \{\text{``Treat Everyone''}, \text{``Random Assign''}\}$.\footnote{See Appendix~\ref{ap:sec2} for details on this testing procedure.} Specifically, we perform a two-sided test of $\mathcal{H}_0: V_0 = V(g_\dagger)$ against $\mathcal{H}_1: V_0 \neq V(g_\dagger)$, as well as one-sided tests of $\mathcal{H}_0: V_0 \leq V(g_\dagger)$ versus $\mathcal{H}_1: V_0 > V(g_\dagger)$ and $\mathcal{H}_0: V_0 \geq V(g_\dagger)$ versus $\mathcal{H}_1: V_0 < V(g_\dagger)$. We consider two settings, depending on whether treatment costs are incorporated. Figure~\ref{tab:pvalues} summarizes the resulting $p$-values.

Several findings emerge. First, when comparing against the random assignment rule, both the two-sided and right-sided tests are rejected at the $5\%$ significance level, providing strong empirical evidence that the optimal policy rule achieves higher welfare than the random assignment rule. Second, when compared with the ``treat-everyone'' rule, the optimal policy rule does not provide sufficiently strong evidence to demonstrate higher welfare. This result holds robustly across the cases with and without treatment costs. This finding is unsurprising, as assigning all individuals to treatment can raise the overall level of welfare, especially among less educated individuals. However, when highly educated individuals represent only a small share of the population, the difference between the optimal treatment rule and the ``treat-everyone'' rule can be small, making it difficult to reject the null hypothesis that treats the two rules as equivalent.

\section{Conclusion}\label{sec_con}

In this paper, we have developed a toolkit for nonparametric uniform inference in cost-sensitive binary classification, a broad framework that encompasses maximum score estimation, utility-maximizing choice prediction, policy learning, and related problems. These settings are well known for slow convergence and nonstandard limiting behavior under parametric classification rules. Furthermore, nonparametric formulations of classification policy functions introduce the additional challenge of failures of identification.

To address these difficulties, we proposed a strictly convex surrogate loss that point-identifies a representative nonparametric classification policy function. Estimating this representative classification policy function provides a unified path to inference for both the optimal classification rule and its associated policy value. In particular, this strategy delivers a Gaussian approximation for the optimal classification policy and a Gaussian limit for the optimal policy value, enabling standard inferential procedures and substantially simplifying empirical implementation despite the intrinsic difficulty of the underlying problem.

Simulation evidence corroborates the theoretical results, showing that the proposed methods perform well in finite samples. Finally, our application to the National JTPA Study demonstrates how the procedure can be used to recover an optimal treatment assignment rule and to evaluate its welfare implications relative to alternative policies at hand.

While the primary contributions of this paper lie in developing inference methods for nonparametric classification policies, our approach also provides, as a by-product, practical benefits for estimating these policies. This improvement arises from two sources. First, the use of a strictly convex surrogate loss facilitates routine numerical optimization for sieve estimation. Second, and more importantly, our point identification of the representative policy function enables a straightforward estimation of the resulting classification rules.

\bigskip\bigskip\bigskip\bigskip
	
\newpage

\appendix
\section*{Appendix}

This appendix contains technical details. We use the same notations as in the main text unless otherwise noted.

\section{Proofs}\label{proofs}

\subsection{Proof for Section \ref{sec:identification}: Surrogate Identification}

\noindent\textbf{Proof of Theorem \ref{thm1}:} 
The proof consists of two parts, (i) and (ii).
\bigskip\\
{
{\bf Proof of Part (i):} We are going to prove that any $g_\ast \in \arg\min_{g\in\mathcal G} L^\ast(g)$ satisfies $g_\ast\in \arg\min_{g \in \mathcal{G}} L(g)$. If a policy $g$ minimizes $L(\cdot)$ in \eqref{eq:loss1}, it also minimizes the conditional risk
\begin{align*}
R(g(x);x) = C_1(x) \mathbf{1}\{-g(x)\leq 0\} + C_0(x) \mathbf{1}\{g(x)<0\}
\end{align*}
pointwisely for every $x$ possibly except on a set of probability measure zero, 
where 
\begin{align*}
C_1(x) = \mathbb{E}[\psi^+(Z_i,\eta_0)|X_i=x]
\quad\text{and}\quad
C_0(x) = \mathbb{E}[\psi^{-}(Z_i,\eta_0)|X_i=x].
\end{align*}
When $g(x)>0$, $R(g(x)) = C_1(x)$. 
Likewise, when $g(x)<0$, $R(g(x)) = C_0(x)$. 
Hence, any element $g^\dagger$ in the set of optimal policy functions satisfies
\begin{align*}
	g^\dagger(x)=
	\left\{\begin{matrix}
		\text{any positive value or }+\infty, & \text{if} & C_0(x)>C_1(x),\\
		\text{any negative value or }-\infty, & \text{if} & C_0(x)<C_1(x),\\
		\text{any value or }\pm\infty, &\text{if} & C_0(x) = C_1(x),
	\end{matrix}\right.
\end{align*}
possibly except on a set of probability measure zero.
The last case with $C_0(X_i) = C_1(X_i)$ is an event of probability measure zero by the statement of the theorem.
}

Now, consider the surrogate risk minimization problem
\begin{align*}
\min _{g\in\mathcal{G}} \mathbb{E}[\psi^+(Z_i,\eta) \cdot \phi(-g(X_i))+\psi^-(Z_i,\eta) \cdot \phi(g(X_i))].
\end{align*}
Similarly to the indicator case, we consider the conditional counterpart:
\begin{align*}
R^\ast(g(x);x)=C_1(x)\cdot\phi(-g(x))+C_0(x)\cdot\phi(g(x)).
\end{align*}
Since $\phi$ is differentiable at $0$ in the statement of the theorem, we can take the derivative
\begin{align*}
\left.\frac{\partial R^*(\bar g;x)}{\partial \bar g}\right|_{\bar g=0}=\phi^{\prime}(0)[-C_1(x)+C_0(x)].
\end{align*}
Since $\phi^{\prime}(0)<0$ in the statement of the theorem, the sign of this derivative equals
\begin{align*}
C_1(x)-C_0(x).
\end{align*}
Thus, if $C_0(x)>C_1(x)$, $R^*(g(x);x)$ decreases as $g(x)$ increases in a neighborhood of $0$. It follows from the convexity of $\phi(\cdot)$ that $R^\ast(g(x);x)$ is minimized by some $g(x)\in(0, \infty]$ at any $x\in\mathcal{X}$ with $C_0(x)>C_1(x)$. 
To see this, note that for any $x\in\mathcal{X}$ with $C_0(x)>C_1(x)$, there exists some $\check{g}(x)>0$ such that
\begin{align*}
R^\ast(\check{g}(x);x) \leq R^\ast(0;x)+\frac{\check{g}(x)}{2}\cdot\left.\frac{\partial R^*(\bar g;x)}{\partial \bar g}\right\vert_{\bar g=0}.
\end{align*}
Furthermore, the convexity of $\phi$ implies that for all $g(x)$,
\begin{align*}
R^\ast(g(x);x) \geq R^\ast(0;x)+g(x) \cdot\left.\frac{\partial R^*(\bar g;x)}{\partial \bar g}\right\vert_{\bar g=0}.
\end{align*}
In particular, if $g(x) \leq \check{g}(x) / 4$, then
\begin{align*}
R^\ast(g(x);x) \geq R^\ast(0;x)+\frac{\check{g}(x)}{4} \cdot\left.\frac{\partial R^*(\bar g;x)}{\partial \bar g}\right\vert_{\bar g=0}>R^\ast(0;x)+\frac{\check{g}(x)}{2}\cdot\left.\frac{\partial R^*(\bar g;x)}{\partial \bar g}\right\vert_{\bar g=0} \geq R^\ast(\check{g}(x);x).
\end{align*}
Hence, a minimizer $g_\ast(x)$ must be greater than $\check{g}(x)/4$, therefore should be positive as is $g^\dagger(x)$.

{If $C_1(x)>C_0(x)$, on the other hand, then $R^*(g ; x)$ increases as $g$ increases in a neighborhood of $0$. By a similar argument as above, a minimizer $g^*(x)$ should be negative as is $g^\dagger(x)$.}

From the above analysis, the sign of the minimizer $g^*(x)$ of the conditional surrogate risk $R^\ast(g(x))$ stays consistent with that of conditional risk $R(g(x))$ possibly except a set of probability measure zero. Therefore, it follows that any $g_\ast \in \arg\min L^\ast(g)$ satisfies $g_\ast\in \arg\min_{g \in \mathcal{G}} L(g)$. The proof is then complete. 

\bigskip\noindent
{{\bf Proof of Part (ii):} First, we prove that the solution to $\min_{g\in\mathcal{G}} L^\ast(g)$ is unique up to the equivalence class in $\mathcal{G}$ identified by the underlying probability measure, when $\phi$ is strictly convex.
Suppose that $g_1, g_2 \in \arg\min_{g \in \mathcal{G}}L^\ast(g)$ and $g_1 \neq g_2$ (up to the equivalence). For $\lambda \in(0,1)$, define the strict linear combination
\begin{align*}
	g_\lambda=\lambda g_1+(1-\lambda) g_2.
\end{align*}
Note that $g_\lambda \in \mathcal{G}$ because $\mathcal{G}$ is convex.
Since $\phi$ is strictly convex, $\mathbb{E}[\psi^{+}(Z_i,\eta_0)|X_i]>0$ a.s., and $\mathbb{E}[\psi^{-}(Z_i,\eta_0)|X_i]> 0$ a.s., 
\begin{align*}
		&\mathbb{E}[\psi^{+}(Z_i,\eta_0)|X_i] \cdot \phi(-g_\lambda(X_i))+\mathbb{E}[\psi^{-}(Z_i,\eta_0)|X_i] \cdot \phi(g_\lambda(X_i)) \\
		\leq& \lambda[\mathbb{E}[\psi^{+}(Z_i,\eta_0)|X_i] \cdot\phi(-g_1(X_i))+\mathbb{E}[\psi^{-}(Z_i,\eta_0)|X_i]\cdot\phi(g_1(X_i))]\\
		&+(1-\lambda)[\mathbb{E}[\psi^{+}(Z_i,\eta_0)|X_i] \cdot\phi(-g_2(X_i))+\mathbb{E}[\psi^{-}(Z_i,\eta_0)|X_i]\cdot\phi(g_2(X_i))],
\end{align*}
	where the inequality `$\leq$' is satisfied with the strict inequality `$<$' with probability greater than one as $g_1 \neq g_2$ (up to the equivalence).
	Thus, integrating both sides and applying the law of iterated expectations give
\begin{align*}
	L^\ast(g_\lambda)<\lambda \cdot L^\ast(g_1)+(1-\lambda) L^\ast(g_2),
\end{align*}
	showing a contradiction with $g_1,g_2 \in \arg\min_{g \in \mathcal{G}}L^\ast(g)$. This implies the uniqueness (up to the equivalence class) stated in part (ii) of the statement of the theorem.}
$\blacksquare$

\subsection{Proofs for Section \ref{sec4}: Estimation \& Inference for the Surrogate Policy}

\noindent\textbf{Proof of Theorem \ref{thm6}}: We first derive the convergence rate, and then establish the strong Guassian approximation.
\bigskip\\
{\bf Convergence Rate:}
For simplicity of writings, we focus on the case with $m=2$.
Thus, we simplify the notations, $\widehat{\psi}^{+}_{(\ell)}(Z_i)$, $\widehat{\psi}^{-}_{(\ell)}(Z_i)$ and $\mathbb{E}_{n,(\ell)}[\cdot]$, as $\widehat{\psi}^{+}_{i}$, $\widehat{\psi}^{-}_{i}$, and $\mathbb{E}_{n}[\cdot]$, respectively. We also simplify the notation of $\widehat{\eta}_{(\ell)}$, that use the complementary subsample $I_{(\ell)}^c$, as $\widehat{\eta}$. The subscripts of $\widehat{\beta}_{(\ell)}$ and $\widehat{g}_{(\ell)}$ are omitted similarly as $\widehat{\beta}$ and $\widehat{g}$, respectively. 

We split the proof into two steps: (a) we first apply \citet[Theorem 3.2]{chen2007large} and \citet{nair2005regularization} to prove a rough convergence rate; and (b) we then sharpen this rate to obtain a sharper bound.  

Recall the short-hand notation $\Psi(Z_{i},\eta,g):=\psi^{+}_i\phi(-g(X_i))+\psi^{-}_i\phi(g(X_i))$. First, when $\Vert g-g_{\ast}\Vert_{\mathbb{P},2}\leq \delta$ with all small enough $\delta>0$, by \citet[][Theorem 3]{chen1998sieve}, there exist some proxy functions $g_{\delta,1}$ and $g_{\delta,2}$ on the line segment connecting $g$ and $g_\ast$ such that
\begin{align}
&\text{Var}(\Psi(Z_{i},\eta,g)-\Psi(Z_{i},\eta,g_{\ast}))\notag\\
\leq& 4\mathbb{E}[(\psi^{+}_i)^2\partial\phi(-g_{\delta,1}(X_i))^2(g(X_i)-g_\ast(X_i))^2]+4\mathbb{E}[(\psi^{-}_i)^2\partial\phi(g_{\delta,2}(X_i))^2(g(X_i)-g_\ast(X_i))^2]\notag\\
\leq& 4\mathbb{E}[\mathbb{E}[(\psi^{+}_i)^2\vert X_i](g(X_i)-g_\ast(X_i))^2]\left(\sup_{x}\partial\phi(-g_{\delta,1}(x))\right)^2
\notag\\
&+4\mathbb{E}[\mathbb{E}[(\psi^{-}_i)^2\vert X_i](g(X_i)-g_\ast(X_i))^2] \left(\sup_{x}\partial\phi(g_{\delta,2}(x))\right)^2\notag\\
\leq& C_+\Vert g-g_\ast\Vert_{\mathbb{P},2}+C_-\Vert g-g_\ast\Vert_{\mathbb{P},2}\leq C_1 \delta^2,\label{cond3.7} 
\end{align}
where 
the first inequality is due to the functional mean value expansion in Banach space \citep[Chapter 4]{zeidler1995} around $g_\ast$ and the twice continuously Fréchet differentiable functional $g \mapsto \mathbb{E}[(\Psi(Z_i,\eta,g)-\Psi(Z_i,\eta,g_\ast))^2]$ \citep{zeidler1995} with Assumptions \ref{as:phi} and \ref{as:holder}, the second inequality uses the law of iterated expectations, and 
the last inequality follows from (i) $\mathbb{E}[(\psi^{+}_i)^2\vert X_i=x]$, $\mathbb{E}[(\psi^{-}_i)^2\vert X_i=x]<\infty$ uniformly for all $ \in \mathcal{X}$ under Assumption \ref{as:series}.\eqref{as:s0} and (ii) $\sup_{x\in\mathcal{X}}\partial\phi(g_\ast(x))<\infty$, which in turn follows from Assumptions \ref{as:phi} and \ref{as:holder}. Here, $C_+$, $C_-$ and $C_1$ are finite constants independent of $n$ and $k$. 
Specifically, the boundedness for $\sup_{x}\partial\phi(-g_{\delta,1}(x))<\infty$ and $\sup_{x}\partial\phi(g_{\delta,2}(x))<\infty$) in \eqref{cond3.7} holds by Assumption \ref{as:phi} and since
\begin{align*}
	\Vert g_{\delta,1}\Vert_{\mathbb{P},2}\leq t\Vert g_\ast- g_{\delta,1}\Vert_{\mathbb{P},2}+\Vert g_{\ast}\Vert_{\mathbb{P},2}<\infty~~\text{and}~~\Vert g_{\delta,2}\Vert_{\mathbb{P},2}\leq t\Vert g_\ast- g_{\delta,2}\Vert_{\mathbb{P},2}+\Vert g_{\ast}\Vert_{\mathbb{P},2}<\infty
\end{align*}
for some $t\in[0,1]$,
which accompanied by Lemma \ref{lm:holder} further yield $\Vert g_{\delta,1}\Vert_{\mathbb{P},\infty}<\infty$ and $	\Vert g_{\delta,2}\Vert_{\mathbb{P},\infty}<\infty$.

Hence, Condition 3.7 of \citet{chen2007large} is satisfied. 

Similarly, when $\Vert g-g_{\ast}\Vert_{\mathbb{P},2}\leq \delta \in (0,\infty)$, we have
\begin{align*}
	\vert \Psi(Z_{i},\eta,g)-\Psi(Z_{i},\eta,g_{\ast})\vert
	&\leq \Vert g-g_{\ast}\Vert_{\mathbb{P},\infty} \cdot2\vert (-\psi^{+}_i\partial\phi(- \check{g}^{(1)}(X_i))+\psi^{-}_i\partial\phi(\check{g}^{(2)}(X_i)))\vert\\
	&\leq \Vert g-g_{\ast}\Vert^{2h/(2h+d_x)}_{\mathbb{P},2}\cdot2\vert (-\psi^{+}_i\partial\phi(- \check{g}^{(1)}(X_i))+\psi^{-}_i\partial\phi(\check{g}^{(2)}(X_i)))\vert,
	\end{align*}
where
the first inequality is due to the definition of $\Vert\cdot\Vert_{\mathbb{P},\infty}$-norm and the second inequality follows by Lemma \ref{lm:holder}. 
Since $\mathbb{E}[ \vert (-\psi^{+}_i\partial\phi(- \check{g}^{(1)}(X_i))+\psi^{-}_i\partial\phi(\check{g}^{(2)}(X_i)))\vert^2]\leq C_2 \in(0,\infty)$ follows similarly to \eqref{cond3.7} and $(2h/(2h+d_x))\in(0,2)$, Condition 3.8 of \citet{chen2007large} is satisfied. 

Define the $L_2(\mathbb{P})$ covering number with bracketing, $\mathcal{N}_{[~]}( \mathcal{F}_n,\Vert\cdot\Vert_{\mathbb{P},2},\varepsilon)$, for the class $$\mathcal{F}_n=\{(\psi^+_{i}\phi(-g(X_i))+\psi^-_{i}\phi(g(X_i)))-(\psi^+_{i}\phi(-g_\ast(X_i))+\psi^-_{i}\phi(g_\ast(X_i))):\Vert g-g_\ast\Vert_{\mathbb{P},2} \leq \delta,~g\in\mathcal{G}_k\}.$$ 
For a constant $b>0$, the rate $\delta_n$ solves
\begin{align}
	\frac{1}{\sqrt{n} \delta_n^2} \int_{b \delta_n^2}^{\delta_n} \sqrt{\log(\mathcal{N}_{[~]}( \mathcal{F}_n,\Vert\cdot\Vert_{\mathbb{P},2},\epsilon))} d \varepsilon
	&\leq  \frac{1}{\sqrt{n} \delta_n^2} \int_{b \delta_n^2}^{\delta_n} \sqrt{k \cdot \log \left(1+\frac{k}{\varepsilon}\right)} d \varepsilon \label{entropy:2}
	\\
	&\leq \frac{1}{\sqrt{n} \delta_n^2}\sqrt{k\log(k)} \cdot \delta_n \leq \text { constant, }\notag
\end{align}
where $\log \mathcal{N}(\mathcal{F}_n,\Vert\cdot\Vert_{\mathbb{P},2},\varepsilon) \lesssim \operatorname{dim}(\mathcal{G}_k) \log (\operatorname{dim}(\mathcal{G}_k)/\varepsilon)$ for the finite-dimensional linear sieves \citep[Page 311, Cases 1.1, 1.2 and 1.3]{chen1998sieve}.

The solution is given by $\delta_n \asymp \sqrt{k\log(k)/n}$. Theorem 3.2 of \citet{chen2007large} along with Lemma \ref{lm:holder} yields $\Vert\widehat{g}-g_{k}\Vert_{\mathbb{P},2}\asymp\sqrt{k\log(k)/n}$ and $\Vert\widehat{g}-g_{k}\Vert_{\mathbb{P},\infty}\asymp(k\log(k)/n)^{h/(2h+d_x)}$. 
Thus, a rough convergence rate is obtained as
\begin{align*}
	\Vert \widehat{g}-g_\ast\Vert_{\mathbb{P},\infty}=(k\log(k)/n)^{h/(2h+d_x)}+B_k.
\end{align*}

Next, we sharpen the convergence rate. The first-order condition for the sieve estimator yields
\begin{align*}
	& \mathbb{E}_n\left[\left(-\widehat{\psi}^{+}_i\partial\phi(-p_{1:k}(X_i)^{\prime} \widehat{\beta})+\widehat{\psi}^{-}_i \partial\phi(p_{1:k}(X_i)^{\prime} \widehat{\beta})\right) p_{1:k}(X_i)\right]=\mathbf{0}_{k \times 1}.
\end{align*}
Then, the mean value theorem yields the coordinate-wise asymptotic representation
\begin{align*}
	\mathbf{0}_{k \times 1}&=\mathbb{E}_n[(-\widehat{\psi}^{+}_i \partial\phi(-p_{1:k}(X_i)^{\prime} \beta_{0})+\widehat{\psi}^{-}_i \partial\phi(p_{1:k}(X_i)^{\prime} \beta_{0})) p_{1:k}(X_i)]\\
	&+\mathbb{E}_n[(\widehat{\psi}^{+}_i \partial^2\phi(-p_{1:k}(X_i)^{\prime}  \check{\beta}^{(1)})+\widehat{\psi}^{-}_i \partial^2\phi(p_{1:k}(X_i)^{\prime} \check{\beta}^{(2)})) p_{1:k}(X_i)p_{1:k}(X_i)^{\prime}](\widehat{\beta}-\beta_{0}),
\end{align*}
where $\check{\beta}^{(s)}$ (with $s=1,2$) lies between $\widehat{\beta}$ and $\beta_{0}$.
(These values vary by coordinates, but we use the single notation for simplicity of writing.)

For the score vector, the following decomposition holds:
\begin{align}
	&\mathbb{E}_n[(-\widehat{\psi}^{+}_i \partial\phi(-p_{1:k}(X_i)^{\prime} \beta_{0})+\widehat{\psi}^{-}_i \partial\phi(p_{1:k}(X_i)^{\prime} \beta_{0})) p_{1:k}(X_i)]\notag\\
	&={\mathbb{E}_n[(-\psi^{+}_i \partial\phi(-p_{1:k}(X_i)^{\prime} \beta_{0})+\psi^{-}_i \partial\phi(p_{1:k}(X_i)^{\prime} \beta_{0})) p_{1:k}(X_i)]}\label{terma1}\\
	&+{\mathbb{E}_n[(-(\widehat{\psi}^{+}_i-\psi^{+}_i) \partial\phi(-p_{1:k}(X_i)^{\prime} \beta_{0})+(\widehat{\psi}^{-}_i-\psi^{-}_i)\partial\phi(p_{1:k}(X_i)^{\prime} \beta_{0})) p_{1:k}(X_i)]}.\label{terma2}
\end{align}

We are going to find bounds for each of the two terms, \eqref{terma1} and \eqref{terma2}, on the right-hand side.
Note the fact that $\{[-\psi^{+}_i\partial\phi(-p_{1:k}(X_i)^{\prime} \beta_0)+\psi_i^{-}\partial\phi(p_{1:k}(X_i)^{\prime} \beta_0)] p_{1:k}(X_i)\}_{i\in[n]}$ are mean-zero $k$-dimensional vectors. Consider the function set $\mathcal{F}=\{f_1,..., f_k\}$ where 
$$f_k(Z_i;X_i)=[-\psi^{+}_i\partial\phi(-p_{1:k}(X_i)^{\prime} \beta_0)+\psi_i^{-}\partial\phi(p_{1:k}(X_i)^{\prime} \beta_0)]{p_{s}(X_i)}$$
for $s\in[k]$ with the envelope  
\begin{align*}
F_i=&\max _{s \in[k]}\vert [-\psi^{+}_i\partial\phi(-p_{1:k}(X_i)^{\prime} \beta_0)+\psi_i^{-}\partial\phi(p_{1:k}(X_i)^{\prime} \beta_0)]{p_{s}(X_i)}\vert \qquad\text{and}
\\
M=&\max _{s \in[k], i \in[n]}\vert[-\psi^{+}_i\partial\phi(-p_{1:k}(X_i)^{\prime} \beta_0)+\psi_i^{-}\partial\phi(p_{1:k}(X_i)^{\prime} \beta_0)]{p_{s}(X_i)}\vert.
\end{align*}
Then, $\mathcal{N}(\mathcal{F}, e_Q, \varepsilon\Vert F\Vert_{Q, 2}) \leq k$ and the uniform entropy satisfies
\begin{align*}
	J(\delta, \mathcal{F}, F)=\int_0^\delta \sup _Q \sqrt{1+\log (\mathcal{N}(\mathcal{F}, e_Q, \varepsilon\Vert F\Vert_{Q, 2}))} d \varepsilon \lesssim \sqrt{\log (k)} \delta.
\end{align*}
Note that $\max _{s \in[K]} \mathbb{E}\{[-\psi^{+}_i\partial\phi(-p_{1:k}(X_i)^{\prime} \beta_0)+\psi_i^{-}\partial\phi(p_{1:k}(X_i)^{\prime} \beta_0)]{p_{s}(X_i)}\}^2 \leq C^2$ and $\Vert M\Vert_{\mathbb{P}, 2} \leq(n k)^{1/q}$. Define $\delta=C /\Vert F\Vert_{\mathbb{P}, 2}$ and apply the maximal inequality \citep[Theorem 5.2]{chernozhukov2014gaussian} to obtain the upper bound of the first term \eqref{terma1} as
\begin{align}
	\mathbb{E}\left\Vert \mathbb{E}_n\left[[-\psi^{+}_i\partial\phi(-p_{1:k}(X_i)^{\prime} \beta_0)+\psi_i^{-}\partial\phi(p_{1:k}(X_i)^{\prime} \beta_0)] p_{1:k}(X_i)\right] \right\Vert_{\infty}
	\notag\\
	\lesssim
	\sqrt{\log (k) / n}+\frac{\Vert M\Vert_{\mathbb{P}, 2} \log (k)}{n}
	\lesssim \sqrt{\log (k) / n},
	\label{eq:ht1}
\end{align}
where we use the fact that, as $q>4$
\begin{align*}
	\frac{\Vert M\Vert_{\mathbb{P}, 2} \log (k)}{n} \lesssim \frac{(n k)^{1 / q} \log (k)}{n} \lesssim \sqrt{\log (k) / n}.
\end{align*}
By Assumption \ref{as:nuisance}(i), the second term \eqref{terma2} can be bounded above as
\begin{align}
	\left\Vert\mathbb{E}_n\left[[-(\widehat{\psi}^{+}_{i}-\psi^{+}_i)\partial\phi(-p_{1:k}(X_i)^{\prime} \beta_0)+(\widehat{\psi}^{-}_{i}-\psi^{-}_i)\partial\phi(p_{1:k}(X_i)^{\prime} \beta_0)] p_{1:k}(X_i)\right] \right\Vert_{\infty}
\notag\\
	=o_p(\sqrt{1/(n\log(k))}). 
	\label{eq:ht1_1}
\end{align}
For the sample Gram matrix, we have a similar decomposition:
\begin{align*}
	&\Vert\mathbb{E}_n[(\widehat{\psi}^{+}_i \partial^2\phi(-p_{1:k}(X_i)^{\prime} \check{\beta}^{(1)})+\widehat{\psi}^{-}_i \partial^2\phi(p_{1:k}(X_i)^{\prime} \check{\beta}^{(2)})) p_{1:k}(X_i)p_{1:k}(X_i)^{\prime}]\Vert\\
	&\lesssim_p\Vert\mathbb{E}_n[(\widehat{\psi}^{+}_i +\widehat{\psi}^{-}_i) p_{1:k}(X_i)p_{1:k}(X_i)^{\prime}]\Vert\\ 
	&={\Vert\mathbb{E}_n[(\psi^{+}_i +\psi^{-}_i) p_{1:k}(X_i)p_{1:k}(X_i)^{\prime}]\Vert}+{\Vert\mathbb{E}_n[(\widehat{\psi}^{+}_{i}-\psi^{+}_i +\psi^{-}_i-\psi^{-}_i ) p_{1:k}(X_i)p_{1:k}(X_i)^{\prime}]\Vert}.
\end{align*}
The first inequality is from the fact $\sup_{x\in\mathcal{X}}\partial^2\phi(p_{1:k}(x)^{\prime} \check{\beta}^{(s)})$ is bounded a.s. (with $s\in\{1,2\}$) since  
\begin{align*}
	\sup_{x\in\mathcal{X}}\vert p_{1:k}(x)^{\prime}\check{\beta}^{(s)}\vert\lesssim_p \sup_{x\in\mathcal{X}}\vert g_k(x)\vert+\sup_{x\in\mathcal{X}}\vert p_{1:k}(x)\cdot (\widehat{\beta}-\beta_{0})\vert\lesssim_p2\sup_{x\in\mathcal{X}}\vert g_k(x)\vert<\infty,
\end{align*}
where $\sqrt{k^2\log(k)/n}=o(1)$ and $\widehat{\psi}^{+}_i$ and $\widehat{\psi}^{-}_i$ are non-negative. Rudelson’s LLN for Matrices given in \citet[Lemma 6.2]{belloni2015some}, with Assumptions \ref{as:nuisance}-\ref{as:phi}, yields 
\begin{align}
	\left\Vert (\mathbb{E}_n-\mathbb{E})[(\psi^{+}_i +\psi^{-}_i)  p_{1:k}(X_i)p_{1:k}(X_i)^{\prime}]\right\Vert&\lesssim_p\sqrt{\xi_{k}^2\log(k)/n} \quad\text{ and}\label{eq:gh0}\\
	\left\Vert (\mathbb{E}_n-\mathbb{E})[(\widehat{\psi}^{+}_{i}-\psi^{+}_i +\widehat{\psi}^{-}_i-\psi^{-}_i ) p_{1:k}(X_i)p_{1:k}(X_i)^{\prime}]\right\Vert&\lesssim_p\xi^2_{k}\log(k)/n.\label{eq:gh1} 
\end{align}
Combining the results of the score vector and sample matrix, we have
\begin{align}\label{rate}
	\Vert\widehat{\beta}-\beta_{0}\Vert_\infty\lesssim_p\sqrt{\log(k)/n},
\end{align}
$\Vert\widehat{\beta}-\beta_{0}\Vert_2\lesssim_p\sqrt{k\log(k)/n}$, and $\Vert\widehat{\beta}-\beta_{0}\Vert_1\lesssim_p\sqrt{k^2\log(k)/n}$. 

\bigskip

Consider a copy of the first-order approximation to our score vector:
\begin{align*}
	\frac{1}{\sqrt{n}}\sum^{n}_{i=1}\xi_{i},~~~\xi_{i}:=\Sigma^{-1/2}[(-\psi^{+}_i \partial\phi(-p_{1:k}(X_i)^{\prime} \beta_{0})+\psi^{-}_i \partial\phi(p_{1:k}(X_i)^{\prime} \beta_{0})) p_{1:k}(X_i)].
\end{align*}
Observe that the upper bound of the third moment is given by
\begin{align}
	\mathbb{E}\left\Vert \xi_{i}\right\Vert^3&\lesssim\mathbb{E}\left\Vert\left[(-\psi^{+}_i \partial\phi(-p_{1:k}(X_i)^{\prime} \beta_{0})+\psi^{-}_i \partial\phi(p_{1:k}(X_i)^{\prime} \beta_{0})) p_{1:k}(X_i)\right]\right\Vert^3\notag\\
	&\leq 4 \mathbb{E}\left[\left(\vert \psi^{+}_i \partial\phi(-g_k(X_i))\vert^3+\vert\psi^{-}_i \partial\phi(g_k(X_i))\vert^3\right)\Vert p_{1:k}(X_i)\Vert^3\right]\notag\\
	&\leq 4\sup_{x\in\mathcal{X}}\left(\vert \partial \phi(-g_k(x))\vert^3\vee\vert \partial \phi(g_k(x))\vert^3\right)\cdot\mathbb{E}\left[\mathbb{E}\left[(\vert\psi^{+}_{i}\vert^3+\vert\psi^{-}_{i}\vert^3)\vert X_i\right]\cdot \Vert\left[p_{1:k}(X_i)\right]\Vert^3\right]\notag\\ 
	&\lesssim\mathbb{E}\left\Vert\left[p_{1:k}(X_i)\right]\right\Vert^3\lesssim k\xi_{k},\label{eq:yurinskii_3rd_moment}
\end{align}
where 
the first inequality uses Assumption \ref{as:series}\eqref{as:s1}, 
the second inequality follows from the triangle inequality and the arithmetic inequality $(a+b)^3 \leq 4(a^3+b^3)$ for any non-negative $a$ and $b$, 
the third inequality employs the law of iterated expectations, 
the fourth inequality holds by Assumption \ref{as:series}\eqref{as:s0} and the bounds, $\sup_x\vert\partial\phi(g_k(x))\vert\lesssim1$ and $\sup_x\vert\partial\phi(-g_k(x))\vert\lesssim1$, and 
the last equality is due to the definition of $\xi_k$. 
In particular,  $\sup_x\vert\partial\phi(g_k(x))\vert\lesssim1$ holds since
\begin{align*}
	\sup_x\vert\partial\phi(g_k(x))\vert&\leq \sup_x\vert\partial\phi(g_\ast(x))\vert+\sup_x \vert\partial^2\phi(g_\ast(x))\vert\cdot\Vert g_k-g_\ast\Vert_{\mathbb{P},\infty}+o(B_{k}),\\
	&\lesssim \sup_x\vert\partial\phi(g_\ast(x))\vert+C_0\sup_x\vert\partial^2\phi(g_\ast(x))\vert<\infty
\end{align*}
holds by the Taylor theorem in Banach sapce \citep[Chapter 4]{zeidler1995} and  Assumptions \ref{as:series}\eqref{as:s3}, \ref{as:phi}, and \ref{as:holder}. Similarly, we have $\sup_x\vert\partial\phi(-g_k(x))\vert\lesssim1$.

Therefore, Lemma \ref{lm:yurinskii} yields
\begin{align}\label{eq:re1}
	\mathbb{P}\left\{\left\Vert \frac{\sum^{n}_{i=1}\xi_{i}}{\sqrt{n}}-\mathcal{N}_{k}\right\Vert\geq 3\delta (\log(k))^{-1}\right\}&\lesssim\frac{\log (k)^{3}nk^{2}\xi_{k}}{\left(\delta \sqrt{n}\right)^3}\left(1+\frac{\log \left(k^2\xi_{k}\right)}{k}\right)\\
	&\lesssim\frac{\log (k)^{3}k^{2}\xi_{k}}{\delta^3\sqrt{n}}\left(1+\frac{\log (n)}{k}\right)\rightarrow0,
\end{align}  
where the convergence follows from the rate restriction $\log(k)^{6} k^4\xi^{2}_{k}\log^2n/n\rightarrow0$ stated in the theorem.

Next, we are going to bound the errors of the covariance and moment estimators.
Observe that
\begin{align}
	&\Vert \widehat{Q}-Q\Vert\notag\\
	\leq&\Vert\mathbb{E}_n[((\widehat{\psi}^{+}_i \partial^2\phi(-\widehat{g}(X_i))+\widehat{\psi}^{-}_i \partial^2\phi(\widehat{g}(X_i)))-(\psi^{+}_i \partial^2\phi(-g_\ast(X_i))+\psi^{-}_i \partial^2\phi(g_\ast(X_i))))p_{1:k}(X_i) p_{1:k}(X_i)^{\prime}]\Vert\label{rr1}\\
	&+\Vert(\mathbb{E}_n-\mathbb{E})[(\psi^{+}_i \partial^2\phi(-g_\ast(X_i))+\psi^{-}_i \partial^2\phi(g_\ast(X_i)))p_{1:k}(X_i) p_{1:k}(X_i)^{\prime}]\Vert.\label{rr2}
\end{align}
The second term \eqref{rr2} on the right-hand side can be bounded by Rudelson’s LLN for Matrices \citep[Lemma 6.2]{belloni2015some} as
\begin{align*} 
	\mathbb{E}\Vert(\mathbb{E}_n-\mathbb{E})[(\psi^{+}_i \partial^2\phi(-g_\ast(X_i))+\psi^{-}_i\partial^2\phi(g_\ast(X_i)))p_{1:k}(X_i) p_{1:k}(X_i)^{\prime}]\Vert
\\
\lesssim\frac{\xi_{k}^2 \log (k)}{n}+\sqrt{\frac{\xi_{k}^2\Vert Q\Vert \log (k)}{n}}\lesssim(1/\log(k)).
\end{align*}
Hence, Markov's inequality yields that
\begin{align*}
	\Vert(\mathbb{E}_n-\mathbb{E})[(\psi^{+}_i \partial^2\phi(-g_\ast(X_i))+\psi^{-}_i\partial^2\phi(g_\ast(X_i)))p_{1:k}(X_i) p_{1:k}(X_i)^{\prime}]\Vert
	\\
	\lesssim_p\frac{\xi_{k}^2 \log (k)}{n}+\sqrt{\frac{\xi_{k}^2\Vert Q\Vert \log (k)}{n}}\lesssim(1/\log(k)).
\end{align*}
The first term \eqref{rr1} can be bounded with \eqref{as:full3} of Assumption \ref{as:nuisance} as
\begin{align*}
\Vert\mathbb{E}_n[((\widehat{\psi}^{+}_i \partial^2\phi(-\widehat{g}(X_i))+\widehat{\psi}^{-}_i \partial^2\phi(\widehat{g}(X_i)))-(\psi^{+}_i \partial^2\phi(-g_\ast(X_i))+\psi^{-}_i \partial^2\phi(g_\ast(X_i))))p_{1:k}(X_i) p_{1:k}(X_i)^{\prime}]\Vert\\
=o_p(\xi_k\sqrt{\log(k)}/n^{1/2}).
\end{align*}
Combining the results, we obtain $\Vert \widehat{Q}-Q\Vert\lesssim_p(1/\log(k))$. 
By using Rudelson’s LLN for Matrices and \eqref{as:full4} of Assumption \ref{as:nuisance}, we similarly obtain $\mathbb{E}\Vert\widehat{\Sigma}-\Sigma\Vert=o(1/\log(k))$.
Thus, it follows that 
\begin{align*}
	&\Vert\widehat{Q}^{-1}\widehat{\Sigma}\widehat{Q}^{-1}-Q^{-1}\Sigma Q^{-1}\Vert\\
	&\lesssim_p\Vert(\widehat{Q}^{-1}-Q^{-1}) \widehat{\Sigma} \widehat{Q}^{-1}\Vert+\Vert Q^{-1}(\widehat{\Sigma}-\Sigma) \widehat{Q}^{-1}\Vert+\Vert Q^{-1} \Sigma(\widehat{Q}^{-1}-Q^{-1})\Vert=o_p(1/\log(k)).
\end{align*}

All eigenvalues of $Q^{-1}\Sigma Q^{-1}$ are bounded away from zero by Assumption \ref{as:series}\eqref{as:s1}. Therefore, we have
\begin{align}
\left\vert\frac{\widehat{\sigma}(x)}{\sigma(x)}-1\right\vert \leq\left\vert\frac{\widehat{\sigma}(x)^2}{\sigma(x)^2}-1\right\vert=\frac{\left\Vert p_{1:k}(x)^{\prime}(\widehat{Q}^{-1}\widehat{\Sigma}\widehat{Q}^{-1}-Q^{-1}\Sigma Q^{-1}) p_{1:k}(x)\right\Vert}{\left\Vert p_{1:k}(x)^{\prime} (Q^{-1}\Sigma Q^{-1}) p_{1:k}(x)\right\Vert}
\notag\\
\lesssim_p\Vert\widehat{Q}^{-1}\widehat{\Sigma}\widehat{Q}^{-1}-Q^{-1}\Sigma Q^{-1}\Vert
\lesssim_po_p(1/\log(k)).
\label{eq:re2}
\end{align}
Define $\alpha_{1:k}(x)=p_{1:k}(x)/\Vert p_{1:k}(x)\Vert$, and then we have $\sup_{x}\Vert\alpha(x)\Vert\leq 1$. Combining \eqref{eq:ht1_1}--\eqref{eq:gh1}, \eqref{eq:re1}, \eqref{eq:re2} and Assumption \ref{as:series}.\eqref{as:s3}, we obtain
\begin{align*}
	& \left\Vert\sqrt{n} \alpha(x)^{\prime}(\widehat{\beta}-\beta)-\alpha(x)^{\prime} (Q^{-1}\Sigma Q^{-1})^{1 / 2} \mathcal{N}_k\right\Vert\\
	\leq&\left\Vert\frac{1}{\sqrt{n}} \sum_{i=1}^n \alpha(x)^{\prime} (Q^{-1}\Sigma Q^{-1})^{1 / 2} \xi_i-\alpha(x)^{\prime} (Q^{-1}\Sigma Q^{-1})^{1 / 2} \mathcal{N}_k\right\Vert+o_p(1/\log(k))+O_p(\sqrt{n}B_k)
	\\ 
	=&o_p(1/\log(k))
\end{align*}
uniformly over $x \in \mathcal{X}$. Since $\Vert \alpha(x)^{\prime}(Q^{-1}\Sigma Q^{-1})^{1 / 2}\Vert$ is bounded from below by Assumption \ref{as:series}.\eqref{as:s1}, we conclude that the Gaussian coupling principle \eqref{coupling_yurinskii} holds. $\blacksquare$

\bigskip

\noindent\textbf{Proof of Theorem \ref{thm11}}: 
Similarly to the proof of Theorem \ref{thm6}, we apply Yurinskii's coupling \citep[Theorem 4.4]{belloni2015some} and strongly approximate the leading term of the score vector. 

Consider the bootstrap counterpart of the first-order approximation to our score vector:
\begin{align*}
\frac{1}{\sqrt{n}}\sum^{n}_{i=1}\widetilde{\xi}_{i},~~~\widetilde{\xi}_{i}:=\Sigma^{-1/2}[\omega_i(-\psi^{+}_i \partial\phi(-p_{1:k}(X_i)^{\prime} \beta_{0})+\psi^{-}_i \partial\phi(p_{1:k}(X_i)^{\prime} \beta_{0})) p_{1:k}(X_i)].
\end{align*}
Observe that the upper bound of the third moment is given by
\begin{align}
\mathbb{E}\Vert \widetilde{\xi}_{i}\Vert^3&\lesssim\mathbb{E}\left\Vert[\omega_i(-\psi^{+}_i \partial\phi(-p_{1:k}(X_i)^{\prime} \beta_{0})+\psi^{-}_i \partial\phi(p_{1:k}(X_i)^{\prime} \beta_{0})) p_{1:k}(X_i)]\right\Vert^3\label{eq:boot1}\\
&\lesssim\mathbb{E}\left\Vert[(-\psi^{+}_i \partial\phi(-p_{1:k}(X_i)^{\prime} \beta_{0})+\psi^{-}_i \partial\phi(p_{1:k}(X_i)^{\prime} \beta_{0})) p_{1:k}(X_i)]\right\Vert^3\label{eq:boot2}\\
&\lesssim\mathbb{E}\left\Vert\left[p_{1:k}(X_i)\right]\right\Vert^3\lesssim k\xi_{k},\label{eq:boot3}
\end{align}
where 
the line \eqref{eq:boot1} holds by the bounded eigenvalues of $\Sigma$ under Assumption \ref{as:series}\eqref{as:s1},
the line \eqref{eq:boot2} is due to the independence between $\omega_i$ and $Z_i$, and 
the line\eqref{eq:boot3} can be proved identically to \eqref{eq:yurinskii_3rd_moment}.

Therefore,Yurinskii's coupling principle yields
\begin{align}\label{eq:re3}
\mathbb{P}\left\{\left\Vert \frac{\sum^{n}_{i=1}\widetilde{\xi}_{i}}{\sqrt{n}}-\mathcal{N}_{k}\right\Vert\geq 3\delta \log(k)^{-1}\right\}&\lesssim\frac{\log (k)^{3}nk^{2}\xi_{k}}{\left(\delta \sqrt{n}\right)^3}\left(1+\frac{\log \left(k^2\xi_{k}\right)}{k}\right)\notag\\
&\lesssim\frac{\log (k)^{3}k^{2}\xi_{k}}{\delta^3\sqrt{n}}\left(1+\frac{\log (n)}{k}\right)\rightarrow0,
\end{align}  
where the convergence follows from the rate restriction $\log(k)^{6} k^4\xi^{2}_{k}\log^2n/n\rightarrow0$.
Therefore, 
\begin{align}
	&\left\Vert\frac{1}{\sqrt{n}}\sum^{n}_{i=1}\Sigma^{-1/2}[\omega_i(-\widehat{\psi}^{+}_i \partial\phi(-\widehat{g}(X_i))+\widehat{\psi}^{-}_i \partial\phi(\widehat{g}(X_i))) p_{1:k}(X_i)]-\mathcal{N}_k\right\Vert\notag\\
	&\leq \left\Vert\frac{1}{\sqrt{n}}\sum^{n}_{i=1}\Sigma^{-1/2}[\omega_i(-\psi^{+}_i \partial\phi(-g_k(X_i))+\psi^{-}_i \partial\phi(g_k(X_i))) p_{1:k}(X_i)]-\mathcal{N}_k\right\Vert\notag\\
	&+\sqrt{n}\left\Vert\Sigma^{-1/2}\mathbb{E}_{n} \left[\left(\Psi_1^\ast(Z_i,\omega_i, \widehat{\eta},g_1)- \Psi_1^\ast(Z_i,\omega_i,\eta_0,g_\ast)\right)p_{1:k}(X_i) \right]\right\Vert\notag\\
	&=o_p(1/\log(k))\label{eq:nusiance}
\end{align}
for the numerator of the bootstrap process,
where the last line is due to Assumption \ref{as:nuisance1}, \eqref{eq:re3}, and the bounded eigenvalues of $\Sigma$ under Assumption \ref{as:series}\eqref{as:s1}. 
We thus obtain the strong Gaussian approximation
\begin{align*}
	\left\Vert\frac{1}{\sqrt{n}}\sum^{n}_{i=1}[\omega_i(-\widehat{\psi}^{+}_i \partial\phi(-\widehat{g}(X_i))+\widehat{\psi}^{-}_i \partial\phi(\widehat{g}(X_i))) p_{1:k}(X_i)]-\Sigma^{1/2}\mathcal{N}_k\right\Vert=o_p(1/\log(k)),
\end{align*}
for the bootstrapping statistic process.
It follows that
\begin{align}\label{qpa10}
\sup_{x\in\mathcal{X}}\left\vert t^{b}_{g}(x)-\frac{p_{1:k}(x)^{\prime}\widehat{Q}^{-1}\cdot\Sigma^{1/2}\mathcal{N}_{k}}{\widehat{\sigma}(x)}\right\vert=o_p(1/\log(k)).
\end{align}

Second, define the centered Gaussian process $x \mapsto t^{b,\ast}_{g}(x)$ by
\begin{align*}
t^{b,\ast}_{g}(x):=\frac{p_{1:k}(x)^{\prime}Q^{-1}\cdot\Sigma^{1/2}\mathcal{N}_{k}}{\sigma(x)}.
\end{align*}
By Theorem \ref{thm6}, which shows the consistency of $\widehat{\Sigma}$ and $\widehat{Q}$, we obtain
\begin{align}
\sup_{x\in\mathcal{X}}\left\vert t^{b,\ast}_{g}(x)-\frac{p_{1:k}(x)^{\prime}\widehat{Q}^{-1}\cdot\Sigma^{1/2}\mathcal{N}_{k}}{\widehat{\sigma}(x)}\right\vert=&\sup_{x\in\mathcal{X}}\left\vert \frac{p_{1:k}(x)^{\prime}Q^{-1}\cdot\Sigma^{1/2}\mathcal{N}_{k}}{\sigma(x)}-\frac{p_{1:k}(x)^{\prime}\widehat{Q}^{-1}\cdot\Sigma^{1/2}\mathcal{N}_{k}}{\widehat{\sigma}(x)}\right\vert
\notag\\
=&o_p(1/\log(k)).
\label{qpa20}
\end{align}

Third, combining \eqref{qpa10} and \eqref{qpa20}, we have
\begin{align}
\sup_{x\in\mathcal{X}}\left\vert t^b_{g}(x)-t^{b,\ast}_{g}(x)\right\vert
\leq&
\sup_{x\in\mathcal{X}}\left\vert t^{b}_{g}(x)-\frac{p_{1:k}(x)^{\prime}\widehat{Q}^{-1}\cdot\Sigma^{1/2}\mathcal{N}_{k}}{\widehat{\sigma}(x)}\right\vert+ \sup_{x\in\mathcal{X}}\left\vert t^{b,\ast}_{g}(x)-\frac{p_{1:k}(x)^{\prime}\widehat{Q}^{-1}\cdot\Sigma^{1/2}\mathcal{N}_{k}}{\widehat{\sigma}(x)}\right\vert
\notag\\
=&o_p(1/\log(k)).
\label{qpa3}
\end{align}
Based on \eqref{qpa3} and the proof of \citet[Theorem 5.5]{belloni2015some}, we obtain
\begin{align}\label{cssci2}
\sup_{s\in\mathbb{R}}\left\vert \mathbb{P}\left(\sup_{x\in\mathcal{X}}\left\vert t^{b}_{g}(x)\right\vert\leq s\right)-\mathbb{P}\left(\sup_{x\in\mathcal{X}}\left\vert t^{b,\ast}_{g}(x)\right\vert\leq s\right)\right\vert=o(1).
\end{align}
Also, observe that
\begin{align}\label{cssci3}
\mathbb{P}\left(\sup_{x\in\mathcal{X}}\left\vert t^{b,\ast}_{g}(x)\right\vert\leq s\right)-\mathbb{P}\left(\sup_{x\in\mathcal{X}}\left\vert {t}(x)\right\vert\leq s\right)
=
o(1)
\end{align}
as $n\rightarrow\infty$
for any $s\in\mathbb{R}$. 

Combining \eqref{cssci2} and \eqref{cssci3}, we obtain
\begin{align*}
	&\sup_{s\in\mathbb{R}}\left\vert \mathbb{P}\left(\sup_{x\in\mathcal{X}}\left\vert t(x)\right\vert\leq s\right)-\mathbb{P}\left(\sup_{x\in\mathcal{X}}\left\vert t^{b}_{g}(x)\right\vert\leq s\right)\right\vert\\
	&\leq\sup_{s\in\mathbb{R}}\left\vert \mathbb{P}\left(\sup_{x\in\mathcal{X}}\left\vert t(x)\right\vert\leq s\right)-\mathbb{P}\left(\sup_{x\in\mathcal{X}}\left\vert t^{b,\ast}_{g}(x)\right\vert\leq s\right)\right\vert +\sup_{s\in\mathbb{R}}\left\vert \mathbb{P}\left(\sup_{x\in\mathcal{X}}\left\vert t^{b}_{g}(x)\right\vert\leq s\right)-\mathbb{P}\left(\sup_{x\in\mathcal{X}}\left\vert t^{b,\ast}_{g}(x)\right\vert\leq s\right)\right\vert\\
	&=o_p(1),
\end{align*}
and prove the claim. $\blacksquare$

\subsection{An Auxiliary Result Related to Section \ref{sec5}}\label{sec:auxiliary_related}

To facilitate our discussions, we write \begin{align*}
		C_1(x) = \mathbb{E}[\psi^+(Z_i,\eta_0)|X_i=x]
		\quad\text{and}\quad
		C_0(x) = \mathbb{E}[\psi^{-}(Z_i,\eta_0)|X_i=x].
\end{align*}
\begin{lm}\label{margin for gstar}
Suppose Assumptions \ref{as:phi}, \ref{as:psi1}, \ref{as:nuisance_2}.(\ref{as:nuisance_22}), and conditions of Theorem \ref{thm1}(ii) hold. Then, there exists small $\delta$ such that for any $\varepsilon<\delta$, $\mathbf{1}\{|g_\ast(x)|\leq \varepsilon\}\leq \mathbf{1}\{|C_1(x)-C_0(x)|\leq K_1 \varepsilon \}$ for a positive $K_1$ that does not depends on $x$ and $\varepsilon$.
\end{lm}

\noindent\textbf{Proof of Lemma \ref{margin for gstar}}: Fix an arbitrary $x\in\mathcal X$. By Assumption of Theorem \ref{thm1}, we can ignore the case in which $C_1(x)=C_0(x)=0$. 

Suppose first $C_0(x)\neq 0$. From the proof of Theorem \ref{thm1}, we know that for a given choice of smooth function $\phi$, the first order condition implies
\[
H(g_\ast(x)) = \frac{C_1(x)}{C_0(x)},
\]
where $H(g):= \frac{\phi'(g)}{\phi'(-g)}$ is differentiable  and a strictly decreasing in $g$, with $H(0)=1$. Let $K_\delta=\max_{t\in[-\delta,\delta]}|H'(t)| $. By Mean Value Theorem, it follows that $|g_\ast(x)|\leq \varepsilon$ implies that $|\frac{C_1(x)}{C_0(x)}-1|\leq K_\delta \varepsilon$ or $|C_1(x)-C_0(x)|\leq K_\delta|C_0(x)| \varepsilon$. 

If $C_0(x)=0$, then $C_1(x)\neq 0$ (since we have excluded the case that both are zeros), then we can define $\tilde H(g) = \frac{\phi'(-g)}{\phi'(g)}$. Following similar logic, we can conclude that $|g_\ast(x)|<\varepsilon$ implies that $|C_1(x)-C_0(x)|< \tilde K_\delta|C_1(x)| \varepsilon$, where $\tilde K_\delta=\max_{t\in[-\delta,\delta]}|\tilde H'(t)| $. 

Finally, let $K_1=\sup_{x\in\mathcal X}\max\{K_\delta C_1(x),\tilde K_\delta C_0(x)\}$, then we can conclude that $|g_\ast(x)|\leq \varepsilon$ implies that $|C_1(x)-C_0(x)|\leq K_1 \varepsilon$, where $K_1$ does not depends on $x$ or $\varepsilon$. $\blacksquare$

The next proposition bounds the estimation error of the estimated welfare \eqref{value1} and establihes its oracle inequality.
\begin{pp}\label{thm7}
In addition to the conditions for Theorem \ref{thm6}, suppose that that Assumptions \ref{as:psi1} and \ref{as:nuisance_2} hold and the sieve order for the estimator $\widehat{g}$ satisfies $k^{4+\varepsilon}/n\lesssim1$ for some $\varepsilon>0$. Then, we have
	\begin{align*}
		\mathbb{E}\left[\left\vert\widehat{V}_n(\widehat{g})-\sup_{g \in \mathcal{G}} V(g)\right\vert\right] \lesssim \inf _k\left[\left\vert V_0-V(g_k)\right\vert+C_v\sqrt{\frac{\log(k)}{n}}\right],
	\end{align*}
where $C_v$ is a finite universal constant and $g_k:=\arg\min_{g\in\mathcal{G}_k}L^\ast(g)$. 
\end{pp}

\noindent\textbf{Proof of Proposition \ref{thm7}}: 
The estimated welfare follows the decomposition:
\begin{align*}
		\widehat{V}_n(\widehat{g})-V_0&=(\widehat{V}_n(\widehat{g})-\widehat{V}_n(g_k))+(\widehat{V}_n(g_k)-V(g_k))+(V(g_k)-V(g_\ast))+\underbrace{(V(g_\ast)-V_0)}_{=0}\\
		&=\underbrace{(\widehat{V}_n(\widehat{g})-\widehat{V}_n(g_k))}_{\text{term A}}+\underbrace{(\widehat{V}_n(g_k)-V(g_k))}_{\text{term B}}+\underbrace{(V(g_k)-V(g_\ast))}_{\text{term C}}.
\end{align*}
In what follows, we will bound each of the three terms, A, B, and C, in the above expression. 
For the convenience of proofs, we define 
\begin{align*}
	&V_n(g):= \mathbb{E}_n[\psi_1(Z_i,\eta_0) \cdot \mathbf{1}\{g(X_i)\geq0\}+\psi_0(Z_i,\eta_0) \cdot \mathbf{1}\{g(X_i)<0\}]. 
\end{align*}

For term A, we adopt the further decomposition:
\begin{align*}
	\widehat{V}_n(\widehat{g})-\widehat{V}_n(g_k)&=[V_n(\widehat{g})-V_n(g_k)-(V(\widehat{g})-V(g_k))]+(V(\widehat{g})-V(g_k))\\
	&+[\widehat{V}_n(\widehat{g})-\widehat{V}_n(g_k)-(V_n(\widehat{g})-V_n(g_k))]\\
	&=[V_n(\widehat{g})-V_n(g_k)-(V(\widehat{g})-V(g_k))]+(V(\widehat{g})-V(g_k))+o_p(1/\sqrt{n}),
\end{align*}
where the last right-hand side term is $o_p(1/\sqrt{n})$ by Assumption \ref{as:nuisance_2}. 

We now analyze the second term on the right-hand side. Note $|V(\widehat{g})-V(g_k)|\leq |V(\widehat{g})-V(g_\ast)|+|V(g_\ast)-V(g_k)|$. Let $U_i := \mathbb{E}[\psi_1(Z_i,\eta_0)|X_i] -\mathbb{E}[\psi_0(Z_i,\eta_0)|X_i]$. Then, with some positive constant $K_1$,
\begin{align}
& |V(\widehat{g})-V(g_\ast)|\notag\\
\leq &\mathbb E[|U_i|\mathbf{1}\{|g_\ast(X_i)|\leq |\widehat g(X_i)-g_*(X_i)|\}]\notag\\
\leq &\mathbb E[|U_i|\mathbf{1}\{|g_\ast(X_i)|\leq\sup_{x} |\widehat g(x)-g_*(x)|\}]\notag\\
\leq &\mathbb E[|U_i|\mathbf{1}\{|U_i|\leq K_1\sup_{x}|\widehat g(x)-g_*(x)|\}\notag\\
 \lesssim_p& K_1 \sup_{x}|\widehat g(x)-g_*(x)| \mathbb P(|U_i|\leq \sup_{x}|\widehat g(x)-g_*(x)|\})\notag\\
 \lesssim_p &K_1 \left(\sup_{x}|\widehat g(x)-g_*(x)|\right)^2 =o_p\left(n^{-\frac{1}{2}}\right),\label{eq0}
\end{align}
where the first inequality holds because $\widehat g(X_i)$ and $g_\ast(X_i)$ disagree implies that $|g_\ast(X_i)|\leq |\widehat g(X_i)-g_*(X_i)|$, and when they disagree, the absolute difference $|V(\widehat{g})-V(g_\ast)|$ is bounded by $|U_i|$. The third inequality holds by Lemma \ref{margin for gstar}, and the fourth inequality holds by the margin condition in Assumption \ref{as:psi1}. The last equality holds by Theorem \ref{thm6} because the condition $k^{4+\varepsilon}/n\lesssim1$ implies that the rate restriction $\xi_k\sqrt{\log(k)/n}\lesssim 1/n^{1/4}$. Also, Assumtion \ref{as:series}\eqref{as:s3} implies $B_k\lesssim (\log(k))^{-1}n^{-1/2}=o(n^{-1/2})$. Similarly, using $g_k$ in the place of $\widehat g$, we can show that $|V(g_k)-V(g_\ast)|=o_p(n^{-\frac{1}{2}})$ because $\sup_{x}|g_k(x)-g_\ast(x)|\leq B_k=o(n^{-1/2})$ again by Assumption \ref{as:series}(iv). Therefore, we can conclude $|V(\widehat{g})-V(g_k)|\leq |V(\widehat{g})-V(g_\ast)|+|V(g_\ast)-V(g_k)| = o_p(n^{-1/2})$.

Since the smoothness of $V(\cdot)$ yields the upper bound of the envelope function $\sigma^2\lesssim (1/n^{1/4})$, the maximal inequality of VC-class (Lemma \ref{lm:vc}), Sauer's Lemma (Lemma \ref{lm:sauer}), and \citet[Lemma A.1]{kitagawa2018should} yield the upper bound: 
\begin{align}
	& \left[V_n(\widehat{g})-V_n(g_k)-\{V(\widehat{g})-V(g_k)\}\right] \notag\\
	& \leq \mathbb{E}\left[\sup_{\sup_{x\in\mathcal{X}}\vert g(x)-g_k(x)\vert \leq \check{c}_2 /n^{1/4}}\left\vert V_n(g)-V_n(g_k)-\{V(g)-V(g_k)\}\right\vert\right] \notag\\
	& \leq C(1/\sqrt{n})\sqrt{(1/n^{1/4})k\log(n/k)}\lesssim1/n^{1/2},\label{eq01}
\end{align}
where $\check{c}_2$ is a generic constant such that $\sup_x\vert\widehat{g}(x)-g_k(x)\vert\leq\check{c}_2 \cdot(1/n)^{1/4}$. Combine both \eqref{eq0} and \eqref{eq01}, we can conclude that the ``term A'' is of order $o_p(n^{-1/2})$.

For term B, we adopt the further decomposition:
\begin{align}\label{kk01}
	\widehat{V}_n(g_k)-V(g_k) & =(\widehat{V}_n(g_k)-V_n(g_k))+(V_n(g_k)-V(g_k)),
\end{align}
whose first term can be in turn decomposed as:
\begin{align*}
	& (\widehat{V}_n(g_k)-V_n(g_k))\\ 
	=&\underbrace{-\frac{1}{n} \sum_{\ell=1}^m \sum_{i \in I_{(\ell)}}\left[\left(\psi_1(Z_i,\widehat{\eta}_{(\ell)})-\psi_1(Z_i,\widehat{\eta}_{0})\right)\mathbf{1}\{g_k(X_i)\geq 0\}\right]}_{\text{term (R1)}}
	\\ 
	&-\underbrace{\frac{1}{n} \sum_{\ell=1}^m \sum_{i \in I_{(\ell)}}\left[\left(\psi_0(Z_i,\widehat{\eta}_{(\ell)})-\psi_0(Z_i,\widehat{\eta}_{0})\right) \mathbf{1}\{g_k(X_i)<0\}\right]}_{\text{term (R2)}},
\end{align*}
where each of the two terms, (R1) and (R2), converges in probability to zero uniformly as
\begin{align*}
\sup _{g \in \mathcal{G}_k}\vert \text{(R1)}\vert=o_p((\log(k)/n)^{1/2})
\quad\text{and}\quad
\sup _{g \in \mathcal{G}_k}\vert \text{(R2)}\vert=o_p((\log(k)/n)^{1/2}),
\end{align*}
under Assumption \ref{as:nuisance_2}. 

To bound the second term of \eqref{kk01}, consider the function class
\begin{align*}
	\mathcal{F}:=\left\{\left.f_k\right\vert f_k(Z_i):=\psi_1(Z_i,\eta_{0})\cdot\mathbf{1}(g_k(X_i)\geq 0)+\psi_0(Z_i,\eta_{0})\cdot\mathbf{1}(g_k(X_i)<0)\right\}
\end{align*}
with envelope $F_i=\max_{g_k\in\mathcal{G}_k}\vert f_k(Z_i)\vert$ and $M=\max_{i\in[n],g_k\in\mathcal{G}_k}\vert f_k(Z_i)\vert$.
We have
\begin{align*}
	M&\leq \max_{i\in[n],g_k\in\mathcal{G}_k}\left\vert\psi_1(Z_i,\eta_{0})\cdot\mathbf{1}(g_k(X_i)\geq 0)+\psi_0(Z_i,\eta_{0})\cdot\mathbf{1}(g_k(X_i)<0)\right\vert\\
	&\leq \max_{i\in[n],g_k\in\mathcal{G}_k}\left\vert\psi_1(Z_i,\eta_{0})\cdot\mathbf{1}(g_k(X_i)\geq 0)-\psi_1(Z_i,\eta_{0})\cdot\mathbf{1}(g_k(X_i)<0)\right\vert\\
	&+\max_{i\in[n],g_k\in\mathcal{G}_k}\left\vert\psi_1(Z_i,\eta_{0})\cdot\mathbf{1}(g_k(X_i)<0)-\psi_0(Z_i,\eta_{0})\cdot\mathbf{1}(g_k(X_i)<0)\right\vert\\
	&\lesssim_p n^{1/q}.
\end{align*}
Similarly, we have $\Vert F_i\Vert_{\mathbb{P}_n,2}\lesssim1$ and $\sigma^{2}_{n}:=\sup_{g_k\in\mathcal{G}_k}(1/n)\sum_{i\in[n]}f_k(Z_i)^2\lesssim_p1$ under Assumption \ref{as:psi1}. The entropy integral inequality \citep[Theorem 2.14.1]{vandervaart1996weak} yields 
\begin{align*}
	&\sqrt{n}\mathbb{E}\left[\sup_{g_k\in\mathcal{G}_k}\left\vert(\mathbb{E}_n-\mathbb{E})\left[\psi_1(Z_i,\eta_{0})\cdot\mathbf{1}(g_k(X_i)\geq 0)+\psi_0(Z_i,\eta_{0})\cdot\mathbf{1}(g_k(X_i)<0)\right]\right\vert\right]\\
	& \lesssim \int_0^{\sigma_n} \sqrt{1+\log \mathcal{N}(\mathcal{F}, e_{\mathbb{P}_n}, \varepsilon)} d \varepsilon \lesssim\Vert F_i\Vert_{\mathbb{P}_n, 2} \int_0^{\sigma_n /\Vert F_i\Vert_{\mathbb{P}_n, 2}} \sqrt{1+\log \mathcal{N}(\mathcal{F}, e_{\mathbb{P}_n},\Vert F_i\Vert_{\mathbb{P}_n, 2} \varepsilon)} d \varepsilon\\
	&\lesssim \sqrt{\log(k)},
\end{align*}
where $e_{\mathbb{P}_n}(f^{(1)}_k,f^{(2)}_{k}):=\Vert f^{(1)}_k-f^{(2)}_{k}\Vert_{\mathbb{P}_n,2}$. 

Finally, for term C, we have
\begin{align*}
	\left\vert V(g_k)-V(g_\ast)\right\vert=\left\vert V(g_k)-\sup_{g\in\mathcal{G}}V(g)\right\vert,
\end{align*}
where $g_k=\arg\min_{g\in\mathcal{G}_k}L^\ast(g)$. 

Combining the three terms together proves the claimed statement
$\blacksquare$

\subsection{Proofs for Section \ref{sec5}: Estimation and Inference for the Optimal Welfare}\label{sec:proof:welfare}

\noindent\textbf{Proof of Theorem \ref{thm8}}: 
As in Proposition \ref{thm7}, the difference between $\widehat{V}_n(\widehat{g})$ and $V_0$ can be decomposed as
\begin{align}
\sqrt{n}(\widehat{V}_n(\widehat{g})-V_0)
=&\sqrt{n}(\widehat{V}_n(\widehat{g})-\widehat{V}_n(g_k))
\label{eq:proof:welfare1}\\
&+\sqrt{n}(\widehat{V}_n(g_k)-V(g_k))
\label{eq:proof:welfare2}\\
&+\sqrt{n}(V(g_k)-V(g_\ast))
\label{eq:proof:welfare4}\\
&+
\sqrt{n}(V(g_\ast)-V_0),
\label{eq:proof:welfare3}
\end{align}
where
\eqref{eq:proof:welfare1} measures the estimation error of the approximate policy;
\eqref{eq:proof:welfare2} contains the estimation error of nuisance parameters and sample moments, \eqref{eq:proof:welfare4} measures sieve approximation error; and
\eqref{eq:proof:welfare3} is the identification error by the surrogate loss.

Since the identified representative classification policy function $g_\ast$ belongs to the set of optimal policy functions by Theorem \ref{thm1}, the identification error \eqref{eq:proof:welfare3} of the surrogate loss approach is exactly zero. \eqref{eq:proof:welfare1} and \eqref{eq:proof:welfare4} are negligible by the analysis on term A in Proposition \ref{thm7}, respectively. It remains to further analyze the second right-hand side term \eqref{eq:proof:welfare2}, which can be decomposed as
\begin{align}\label{kk1}
	\sqrt{n}\{\widehat{V}_n(g_k)-V(g_k)\} & =\sqrt{n}\{\widehat{V}_n(g_k)-V_n(g_k)+V_n(g_k)-V(g_k)\}\notag \\
	& =o_p(1)+\sqrt{n}\{V_n(g_k)-V(g_k)\},
\end{align}
where the second inequality follows from Assumption \ref{as:nuisance_2}.\eqref{as:nuisance_21}. 
The Lindeberg-L\'evy central limit theorem yields the Gaussian limit theory for the second term $\sqrt{n}\{V_n(g_k)-V(g_k)\}$ in \eqref{kk1}. 

Combining all these components proves the claimed statement. $\blacksquare$
\bigskip 

\noindent\textbf{Proof of Theorem \ref{thm9}}: 
Similar steps as in those in the proof of Theorem \ref{thm8} applied to $\widetilde{Z}_n$ yields
\begin{align*}
	\widetilde{Z}_n=\mathbb{G}_n [s_L(Z_i, \eta_0, g_k) \delta_i]+o_p(1),
\end{align*}
where the Lindeberg-L\'evy central limit theorem implies $\mathbb{G}_n [s_L(Z_i, \eta_0, g_k) \delta_i]\rightarrow_d\mathcal{N}(0,\sigma^2_v)$.
$\blacksquare$

\section{Auxiliary Results from the Literature}\label{auxiliary}

\begin{lm}[\citealp{chen1998sieve}, Lemma 2]\label{lm:holder}
	Let $h=s+\gamma$ where $s \in \mathcal{Z}_+$ and $\gamma \in (0,1]$. Define the Hölder space with smoothness order $p>1/2$ and radius $L\in(0,\infty)$ by 
	\begin{align}\label{class:holder}
	\Lambda^h(L)=\left\{f:\mathcal{X} \rightarrow \mathbb{R},~\Vert f\Vert_\mathcal{H}=\sup _{x, y \in\mathcal{X},~x \neq y} \frac{\vert f^{(s)}(x)-f^{(s)}(y)\vert}{\vert x-y\vert^\gamma} \leq L\right\},
	\end{align}
	where $\mathcal{X}\subseteq \mathbb{R}^{d_x}$ is compact. Then, for any $f\in\Lambda^h(L)$, we have $$\Vert f\Vert_{\mathbb{P},\infty} \leq 2\Vert f\Vert_{\mathbb{P},2}^c L^{1-c},$$ where $c=2(s+\gamma)/(2(s+\gamma)+d_x)$.
\end{lm}

Consider a class $\mathcal{A}$ of subsets of $\mathbb{R}^d$, and let $x_1,..., x_n \in \mathbb{R}^d$ be arbitrary points. Recall that the finite set $\mathcal{A}\left(x_1^n\right) \subset\{0,1\}^n$ defined by
\begin{align*}
	\mathcal{A}(x_1^n) =\left\{b=(b_1,..., b_n) \in\{0,1\}^n:~b_i =\mathbb{I}_{[x_i \in A]}, i=1,..., n \text { for some } A \in \mathcal{A}\right\}
\end{align*}
plays an essential role in bounding uniform deviations of empirical measures. 
In particular, the maximal cardinality of $\mathcal{A}(x_1^n)$ denoted by
\begin{align*}
\mathbb{S}_{\mathcal{A}}(n)=\max _{x_1,..., x_n \in \mathbb{R}^d}\vert\mathcal{A}(x_1^n)\vert,
\end{align*}
i.e., the shatter coefficient, yields simple bounds via the Vapnik-Chervonenkis inequality.

\begin{lm}[Sauer's Lemma; \citealp{lugosi2002pattern}, Corollary 1.3]\label{lm:sauer}
Let $\mathcal{A}$ be a class of sets with VC dimension $V<\infty$. Then, for all $n$,
\begin{align*}
	\mathbb{S}_{\mathcal{A}}(n) \leq(n+1)^V,
\end{align*}
and for all $n \geq V$,
\begin{align*}
	\mathbb{S}_{\mathcal{A}}(n) \leq\left(\frac{n e}{V}\right)^V .
\end{align*}
\end{lm}
\begin{lm}[Maximal Inequality of VC Class; \citealp{massart2006risk}, Lemma A.3]\label{lm:vc}
Let $\mathcal{F}$ be a countable VC-class with dimension not larger than $V \geq 1$ and assume that $\sigma>0$ is such that $\mathbb{E}(f(Z_i)) \leq \sigma^2$ for every $f \in \mathcal{F}$. Let
	\begin{align*}
	W_{\mathcal{F}}^{+}=\sup _{f \in \mathcal{F}} \mathbb{G}_n(f)~~\text{and}~~W_{\mathcal{F}}^{-}=\sup _{f \in \mathcal{F}}-\mathbb{G}_n(f),
	\end{align*}
and $H_{\mathcal{F}}=\log \mathbb{S}_{\mathcal{F}}(n)$. Then, there exists an absolute constant $K$ such that
\begin{align*}
\sqrt{n}(\mathbb{E}[W_{\mathcal{F}}^{-}] \vee \mathbb{E}[W_{\mathcal{F}}^{+}]) \leq \frac{K}{2} \sigma \sqrt{\mathbb{E}[H_{\mathcal{F}}]},
\end{align*}
provided that $\sigma \geq K \sqrt{\mathbb{E}[H_{\mathcal{F}}]/n}$.
\end{lm}

\begin{lm}[\citealp{pollard2002}, Theorem 10]\label{lm:yurinskii}
	Let $\zeta_1,..., \zeta_n$ be independent $k$-vectors with $\mathbb{E}[\zeta_i]=0$ for each $i$, and $\Delta:=\sum_{i=1}^n \mathbb{E}\Vert \zeta_i\Vert^3$ finite. Let $S$ denote a copy of $\zeta_1+...+\zeta_n$ on a sufficiently rich probability space $(\Omega, \mathcal{A}, \mathbb{P})$. For each $\delta>0$, there exists a random vector $T$ in this space with a $\mathcal{N}(0, \operatorname{var}(S))$ distribution such that
	\begin{align*} 
		\mathbb{P}\{\Vert S-T\Vert>3 \delta\} \leq C_y B\left(1+\frac{|\log (1 / B)|}{k}\right) \text { where } B:=\Delta k \delta^{-3},
	\end{align*}
	for some generic constant $C_y\in(0,\infty)$.
\end{lm}

\section{Testing Optimal Welfare Against Benchmark Designs}\label{ap:sec2}
This section proposes tests that compare the welfare of the estimated optimal policy to that of benchmark policies. We consider two benchmark designs for $g_\dagger$: (i) treating everyone; and (ii) assigning treatment at random. Formally, we consider the null hypothesis
	\begin{align*}
		\mathcal{H}_0: V_0 = V(g_{\dagger}),
	\end{align*}
	where $V(g_{\dagger})$ denotes the welfare associated with the benchmark policy $g_{\dagger}$. The test statistic is defined as
	\begin{align}
		T_{g_\ast,g_{\dagger}} := \sqrt{n}\,\big(\widehat{V}_n(\widehat{g}) - \widehat{V}_n(g_{\dagger})\big),
	\end{align}
	where $\widehat{g}$ is the estimated optimal policy and $\widehat{V}_n(\cdot)$ is defined in \eqref{value1}.

We partition $[n]$ into $m$ approximately balanced folds $I_{1}, \ldots, I_{m}$, so that $n_{(1)} \asymp \cdots \asymp n_{(m)}$. For each fold $\ell \in [m] = \{1,\ldots,m\}$, we have a nuisance parameter estimator $\widehat{\eta}_{(\ell)}$ of $\eta_0$ using the complementary subsample $I_{(\ell)}^c$. The critical values of $T_{g_\ast,g_{\dagger}}$ are obtained via the score bootstrap,
\begin{align}\label{ap:bootstrap}
	\widetilde{Z}_n := \mathbb{G}_n \Big[\widehat{s}_L(Z_i,\widehat{\eta}_{(\ell(i))},\widehat{g},g_\dagger) \cdot \delta_i \Big],
\end{align}
where $\{\delta_i\}_{i\in[n]}$ are i.i.d.\ $\mathcal{N}(0,1)$ random variables independent of the observed data and $\widehat{\eta}_{(\ell(i))}$ denotes the nuisance estimator calculated on the complementary subsample $I^c_{(\ell(i))}$. The score function is 
\begin{align}
	&\widehat{s}_L(Z_i,\widehat{\eta}_{(\ell(i))},\widehat{g},g_\dagger) 
	\notag\\
	=& (\psi_{1}(Z_i,\widehat{\eta}_{(\ell(i))})-\psi_{0}(Z_i,\widehat{\eta}_{(\ell(i))}))\Big(\mathbf{1}\{g_\ast(X_i)\geq 0\}-\mathbf{1}\{g_{\dagger}(X_i)\geq 0\}\Big)-\Big(\widehat{V}_n(\widehat{g})-\widehat{V}_n(g_\ast)\Big).
	\label{ap:score}
\end{align}
For a given significance level $\alpha \in (0,1)$, let $\widehat{c}_{1-\alpha}$ denote the $(1-\alpha)$ quantile of $\lvert \widetilde{Z}_n \rvert$. Then, under the null hypothesis $\mathcal{H}_0: V_0 = V(g_\dagger)$, we have
\begin{align*}
	\lim_{n \to \infty} \mathbb{P}\!\left(\lvert T_{g_\ast,g_{\dagger}} \rvert \leq \widehat{c}_{1-\alpha}\right) = 1-\alpha,
\end{align*}
and the theoretical validity of this result follows by arguments analogous to those used in the proof of Theorem~\ref{thm9}.

The implementation details for conducting this test are summarized as follows:
\begin{enumerate}[Step (i)]
	\item \textbf{Test statistic.}  
	Construct the statistic
	\begin{align*}
	T_{g_\ast,g_{\dagger}} := \sqrt{n}\,\big(\widehat{V}_n(\widehat{g}) - \widehat{V}_n(g_{\dagger})\big),
	\end{align*}
	where $\widehat{g}$ is the estimated policy and $\widehat{V}_n(\cdot)$ is defined in \eqref{value1}.
	\item \textbf{Bootstrap process.}  
	Generate the score bootstrap process
	\begin{align*}
	\widetilde{Z}_n := \mathbb{G}_n \!\left[\, \widehat{s}_L(Z_i,\widehat{\eta}_{(\ell(i))},\widehat{g}) \cdot \delta_i \right],
	\end{align*}
	following \eqref{ap:bootstrap} and \eqref{ap:score}, where $\{\delta_i\}_{i\in[n]}$ are i.i.d.\ $\mathcal{N}(0,1)$ variates independent of the data.
	\item \textbf{Critical value.}  
	For a given level $\alpha \in (0,1)$, let $\widehat{c}_{1-\alpha}$ denote the $(1-\alpha)$ quantile of $\lvert \widetilde{Z}_n \rvert$.
	\item \textbf{Decision rule.}  
	Reject the null hypothesis $\mathcal{H}_0: V_0 = V(g_\dagger)$ whenever
	\begin{align*}
	\lvert T_{g_\ast,g_{\dagger}} \rvert > \widehat{c}_{1-\alpha}.
	\end{align*}
	Then, under $\mathcal{H}_0: V_0 = V(g_\dagger)$,
	\begin{align*}
	\lim_{n \to \infty} \mathbb{P}\!\left( \lvert T_{g_\ast,g_{\dagger}} \rvert \leq \widehat{c}_{1-\alpha}\right) = 1-\alpha.
	\end{align*}
	\item \textbf{Bootstrap P-value.}  
	In practice, the P-value can be computed directly from the bootstrap distribution as
	\begin{align*}
	\text{P-value} = \mathbb{P}^\ast\!\left( \lvert \widetilde{Z}_n \rvert \geq \lvert T_{g_\ast,g_{\dagger}} \rvert~\big|~\{Z_i\}_{i\in[n]} \right),
	\end{align*}
	where $\mathbb{P}^\ast$ denotes the probability conditional on data. Equivalently, P-value is estimated as the proportion of bootstrap draws for which $\vert \widetilde{Z}_n \vert$ exceeds $\vert T_{g_\ast,g_{\dagger}} \vert$.
\end{enumerate}

\clearpage

\clearpage
	\renewcommand\bibname{\LARGE \textbf {Bibliography}}
	\bibliographystyle{chicago}
	\bibliography{learning}	

\end{document}